\documentclass[]{aa} 
\usepackage{txfonts}
\usepackage{natbib,graphicx,amssymb}
\usepackage[english]{babel}
\usepackage{url}
\newcommand{\kms}{km\,s$^{-1}$}

\begin{document}

\title{The molecular gas reservoir of 6 low-metallicity galaxies 
	from the {\it Herschel} Dwarf Galaxy Survey}
\subtitle{A ground-based follow-up survey of CO(1-0), CO(2-1), and CO(3-2)}

\author{
  D.~Cormier\inst{1}
  \and S.~C.~Madden\inst{2}
  \and V.~Lebouteiller\inst{2}
  \and S.~Hony\inst{3}
  \and S.~Aalto\inst{4}
  \and F.~Costagliola\inst{4,5}
  \and A.~Hughes\inst{3}
  \and A.~R{\'e}my-Ruyer\inst{2}
  \and N.~Abel\inst{6}
  \and E.~Bayet\inst{7}
  \and F.~Bigiel\inst{1}
  \and J.~M.~Cannon\inst{8}
  \and R.~J.~Cumming\inst{9}
  \and M.~Galametz\inst{10}
  \and F.~Galliano\inst{2}
  \and S.~Viti\inst{11}
  \and R.~Wu\inst{2}
}

\institute{ 
Institut f\"ur theoretische Astrophysik, 
Zentrum f\"ur Astronomie der Universit\"at Heidelberg, 
Albert-Ueberle Str. 2, D-69120 Heidelberg, Germany
\email{diane.cormier@zah.uni-heidelberg.de} 
\and
Laboratoire AIM, CEA/DSM - CNRS - Universit\'e Paris
  Diderot, Irfu/Service d'Astrophysique, CEA Saclay, 91191
  Gif-sur-Yvette, France 
\and
Max-Planck-Institute for Astronomy, K{\"o}nigstuhl 17, 69117 Heidelberg, Germany
\and
Department of Earth and Space Sciences, Chalmers University of Technology, Onsala Space Observatory, 439 92 Onsala, Sweden
\and
Instituto de Astrof{\'i}sica de Andaluc{\'i}a, Glorieta de la Astronom{\'i}a s/n, 18008 Granada, Spain
\and
 University of Cincinnati, Clermont College, Batavia, OH, 45103, USA
\and
 Sub-Dept. of Astrophysics, Dept. of Physics at University of Oxford, Denys Wilkinson Building, Keble Road, Oxford OX1 3RH, U.K. 
\and
 Department of Physics \& Astronomy, Macalester College, 1600 Grand Avenue, Saint Paul, MN 55105, USA
\and 
Onsala Space Observatory, Chalmers University of Technology, 439 92 Onsala, Sweden
\and
 Institute of Astronomy, University of Cambridge, Madingley Road, Cambridge CB3 0HA, UK
\and
 Department of Physics and Astronomy, University College London, Gower Street, London, WC1E 6BT, UK
}

%\date{draft after comments}
%\date{Submitted to A\&A}

\abstract
%{\it Context}: 
{Observations of nearby starburst and spiral galaxies have revealed 
that molecular gas is the driver of star formation. However, some 
nearby low-metallicity dwarf galaxies are actively forming stars, but CO, 
the most common tracer of this reservoir is faint, leaving us with a puzzle 
about how star formation proceeds in these environments. }
%
%{\it Aims}: 
{We aim to quantify the molecular gas reservoir in a subset 
of 6 galaxies from the \textit{Herschel} Dwarf Galaxy Survey 
with newly acquired CO data, and link this reservoir to the observed 
star formation activity. }
%
%{\it Method}: 
{We present CO(1-0), CO(2-1), and CO(3-2) observations obtained at 
the ATNF Mopra 22-m, APEX, and IRAM 30-m telescopes, as well as 
[C~{\sc ii}]~157$\mu$m and [O~{\sc i}]~63$\mu$m observations obtained 
with the \textit{Herschel}/PACS spectrometer 
in the 6 low-metallicity dwarf galaxies: 
Haro\,11, Mrk\,1089, Mrk\,930, NGC\,4861, NGC\,625, and UM\,311. 
We derive molecular gas mass from several methods including the use 
of the CO-to-H$_{\rm 2}$ conversion factor $X_{\rm CO}$ (both Galactic and 
metallicity-scaled values) and of dust measurements. 
The molecular and atomic gas reservoirs are compared 
to the star formation activity. 
We also constrain the physical conditions of the molecular clouds using 
the non-LTE code \textsc{RADEX} and the spectral synthesis code Cloudy. }
%
%{\it Results}: 
{We detect CO in 5 of the 6 galaxies, including first detections 
in Haro\,11 {(Z$\sim$0.4~Z$_{\odot}$)}, Mrk\,930 {(0.2~Z$_{\odot}$)}, 
and UM\,311 {(0.5~Z$_{\odot}$)}, but CO remains undetected in 
NGC\,4861 {(0.2~Z$_{\odot}$)}. 
The CO luminosities are low while [C~{\sc ii}] is bright in these galaxies, 
resulting in [C~{\sc ii}]/CO(1-0) $\ge$10\,000. 
Our dwarf galaxies are in relatively good agreement with 
the Schmidt-Kennicutt relation for total gas. They show short molecular 
depletion time scales, even when considering metallicity-scaled 
$X_{\rm CO}$ factors. 
Those galaxies are dominated by their H~{\sc i} gas, 
except Haro\,11 which has high star formation efficiency {and 
is dominated by} ionized and molecular gas. 
We determine the mass of each ISM phase in Haro\,11 using Cloudy 
and estimate an equivalent $X_{\rm CO}$ 
factor which is 10 times higher than the Galactic value. 
{Overall, our results confirm the emerging picture 
that CO suffers from significant selective photodissociation 
in low-metallicity dwarf galaxies.} 
}
{}

\titlerunning{Molecular gas in dwarf galaxies}
\authorrunning{Cormier et al.}
\maketitle

%%%%%%%%%%%%%%
\section{Introduction}
\label{sect:intro}
%%%%%%%%%%%%%%
On galactic scales, the star formation rate is observed to correlate 
with the {total (molecular and atomic)} gas reservoir, 
following the empirical Schmidt-Kennicutt law 
\citep[e.g.][]{schmidt-1959,kennicutt-1998}: 
\begin{equation}
\label{eq:kslaweq}
\Sigma_{SFR} \propto (\Sigma_{gas})^n, \mathrm{with}~n \simeq 1.4,
\end{equation}
where $\Sigma_{SFR}$ is the star formation rate surface density, and 
$\Sigma_{gas}$ is the gas surface density. 
There is evidence that the star formation law in galaxies is mostly regulated by 
the molecular gas rather than the {total gas} content, 
and that the timescale to convert molecular gas into stars is to first order constant 
for disk galaxies and around $\tau_{dep}$$\sim$2~Gyr 
\citep{bigiel-2008,leroy-2008,bigiel-2011,genzel-2012}. 
\cite{leroy-2005} analyzed the star formation law in non-interacting dwarf galaxies 
of the northern hemisphere with metallicities 
$12+\log(O/H) \ge 8.0$\footnote{The metallicity $12+\log(O/H)$ is denoted by $O/H$ throughout the paper.}. 
They find that the center of dwarf galaxies and more massive spiral galaxies 
follow the same relationship between molecular gas, 
measured by CO(1-0), and star formation rate (SFR), measured by 
the radio continuum, with a power-law index $n \simeq 1.2-1.3$. 
%

%%%%%
The tight correlation between star formation and molecular gas emission 
results from the conditions required for molecules to be abundant. 
High density is a prerequisite for star formation. In order to be protected 
against photodissociation, molecules also require a dense and shielded 
environment, where CO acts as a main coolant of the gas. 
At low metallicities, this correlation may not hold since other lines -- 
particularly atomic fine-structure lines such as the [C~{\sc ii}]~157$\mu$m 
line -- can also cool the gas efficiently enough to allow stars to form 
\citep{krumholz-2011,glover-2012}. 
The formation of H$_{\rm 2}$ on grain surfaces is also affected in these environments. 
At extremely low metallicities, below 1/100\,$Z_{\odot}$\footnote{The solar 
metallicity is taken as $(O/H)_{\odot}=8.7$ \citep{asplund-2009}.}, 
\cite{krumholz-2012b} demonstrates that the timescale to form molecules 
is larger than the thermal and free-fall timescales. As a consequence, 
star formation may occur in the cold atomic phase before the medium becomes 
fully molecular \citep[see also][]{glover-2012}.

%%%%%
On the observational side, many low-metallicity dwarf galaxies 
($1/40\,Z_{\odot} \le Z \le 1/2\,Z_{\odot}$) are forming stars but 
show little observed molecular gas as traced by CO emission, standing as 
outliers on the Schmidt-Kennicutt relation \citep[e.g.][]{galametz-2009,schruba-2012}. 
How star formation occurs in these environments is poorly known. 
Such a discrepancy with the Schmidt-Kennicutt relation  {may imply} either 
higher star formation efficiencies (SFE) than in normal galaxies, or 
larger total gas reservoirs than measured by CO, 
as favored by recent studies \citep[e.g.][]{schruba-2012,glover-2012}.

%%%%%
Most of the molecular gas in galaxies is in the form of cold H$_{\rm 2}$, 
which is not directly traceable {due to the lack of low energy transitions 
(no dipole moment), leaving the second most abundant molecule, CO, 
the most common molecular gas tracer (see \citealt{bolatto-2013} for a review 
on the CO-to-H$_{\rm 2}$ conversion factor). 
CO, however, is difficult to detect in low-metallicity dwarf galaxies.  
Does this imply that dwarf galaxies truly contain little molecular gas, or 
that CO becomes a poor tracer of H$_{\rm 2}$ in low-metallicity ISM?}
The dearth of {detectable CO in dwarf galaxies} can be a direct 
result of the chemical evolution (from lower C and O abundances), or 
due to the structure and conditions of the ISM. Modeling of 
photodissociation regions (PDRs) demonstrates the profound 
effects of the low metallicity on the gas temperatures, which often 
result in an enhanced CO emission from higher-J transitions, 
and on the structure of the PDR envelope \citep{roellig-2006}. 
{While H$_{\rm 2}$ is self-shielded from 
the UV photons and can survive in the PDR, CO survival depends 
on the dust shielding. CO is therefore more easily prone 
to photodestruction, especially in low-metallicity star-forming galaxies, 
which have both a hard radiation field and a low dust-to-gas mass ratio, 
leaving, if any, only small filling factor molecular clumps, which are difficult to detect 
with single-dish antennas due to severe beam dilution effects} 
\citep[e.g.][]{taylor-1998,leroy-2011,schruba-2012}. 
This molecular gas not traced by CO has been 
referred to as the ``dark gas'' \citep{wolfire-2010,grenier-2005}.

%%%%%
{The full molecular gas reservoir in dwarf galaxies should, 
therefore, include the molecular gas traced by CO, and the CO-dark gas 
that emits brightly in other emission lines. PDR tracers, such as [O~{\sc i}] and [C~{\sc ii}], 
which usually arise from the surface layer of CO clouds and hence co-exist 
with the CO-dark molecular gas \citep{poglitsch-1995,madden-1997,glover-2010}, 
can trace the CO-dark gas.}
The [C~{\sc ii}]~157$\mu$m far-infrared (FIR) fine-structure line is 
one of the most important coolants of the ISM \citep{tielens-1985,wolfire-1995}, 
and was first used in \cite{madden-1997} to quantify the total molecular gas 
reservoir in a dwarf galaxy. 
New evidence of the presence of a significant reservoir of CO-dark molecular gas, 
on the order of 10 to 100 times that inferred 
by CO \citep{madden-2000}, is suggested by the exceptionally high [C~{\sc ii}]-to-CO 
{ratios found in ``The Dwarf Galaxy Survey'' \citep[DGS;][]{madden-2013, cormier-2010}, 
a {\it Herschel} Key Program which has observed 50 nearby 
low-metallicity dwarf galaxies in the FIR/submillimeter (submm) dust bands 
and the FIR fine-structure lines of the ionized and neutral gas, including 
[C~{\sc ii}]~157$\mu$m and [O~{\sc i}]~63$\mu$m.} 
%

%%%%%
{In this paper, we present new CO observations of 6 dwarf galaxies from the DGS: 
Haro\,11, Mrk\,930, Mrk\,1089, NGC\,4861, NGC\,625, and UM\,311, with metallicities 
ranging from 1/6 to 1/2\,Z$_{\odot}$ (Table~\ref{table:subgeneral}).
Section~\ref{sect:obs} describes the observations and data reduction of 
the CO(1-0), CO(2-1), and CO(3-2) data sets, as well as the 
[C~{\sc ii}]~157$\mu$m and [O~{\sc i}]~63$\mu$m lines from \textit{Herschel} 
and the warm H$_{\rm 2}$ lines from {\it Spitzer}.} 
We quantify the physical conditions of the molecular phase in section~\ref{sect:prop}, 
using empirical diagnostics, the non-Local Thermal Equilibrium (non-LTE) 
code \textsc{RADEX}, and excitation diagrams for the warm H$_{\rm 2}$ gas. 
In particular, we focus our analysis on comparing the cold and warm molecular gas 
reservoirs that are inferred from several methods ($X_{\rm CO}$ conversion factor, dust, etc.). 
In addition, we apply a full radiative transfer modeling to the low-metallicity 
galaxy Haro\,11 as a case study in section~\ref{sect:cloudy}, 
in order to characterize properties of the CO-dark gas in the PDR. 
{We then discuss our results in the context of the overall 
star formation activity in these galaxies}, 
and investigate how the estimated amount of molecular gas relates 
to other galaxy properties (atomic reservoir, SFR, 
etc.; section~\ref{sect:sfr}).
Throughout the paper, the quoted molecular gas masses refer to 
H$_{\rm 2}$ masses, except in section~\ref{sect:sfr} where the masses 
are multiplied by a factor $1.36$ to include helium in the total gas reservoir.

%%%%%%%%%%%%%%%%%%%%%%%%
\section{Surveying the neutral gas in low-metallicity galaxies: [C~{\sc ii}], [O~{\sc i}], H$_{\rm 2}$, and CO data}
\label{sect:obs}
%%%%%%%%%%%%%%%%%%%%%%%%
%%%%%
\subsection{Target Selection}
%%%%%
Due to the intrinsically faint CO and instrumental sensitivity limitations, 
studies of CO in {low-metallicity environments} have essentially 
focused on Local Group galaxies, such as the Magellanic Clouds, 
IC\,10, NGC\,6822, or M\,33. 
The proximity of the Magellanic Clouds, for example, has allowed for detailed 
studies of molecular clouds, where the MAGMA survey achieves 11\,pc 
resolution with Mopra \citep{wong-2011}, and ALMA resolves even sub-pc 
structures \citep{indebetouw-2013}. Those Local Group galaxies 
however probe a limited range of physical conditions. 
{Outside of the Local Group, CO is detected in tidal dwarf galaxies 
\citep[e.g.][]{braine-2001} and Magellanic type spirals and irregular galaxies 
\citep{albrecht-2004,leroy-2005,leroy-2009a}, but detections are sparse 
in Blue Compact Dwarf (BCD) galaxies \citep[e.g.][]{israel-2005}. 
From the DGS, only the galaxies NGC\,1569, NGC\,6822 
(1/5-1/6\,$\rm{Z_{\odot}}$), Haro\,2, Haro\,3, II\,Zw\,40, IC\,10, NGC\,4214, 
NGC\,5253 (1/3\,$\rm{Z_{\odot}}$), and He\,2-10 (1/2\,$\rm{Z_{\odot}}$) 
have been observed in more than one CO transition 
\citep{sage-1992,kobulnicky-1995,taylor-1998,meier-2001,bayet-2004,israel-2005}. }
To date, CO has only been detected in dwarf galaxies with metallicities 
$\geq$\,1/6\,Z$_{\odot}$, with the exception of the close-by galaxy WLM 
\citep{elmegreen-2013}, which has a metallicity of $\sim$\,1/8\,Z$_{\odot}$. 

\begin{figure*}
\begin{minipage}{18cm}
\centering
\includegraphics[clip,trim=.1cm .1cm 1cm .4cm,width=5.7cm]{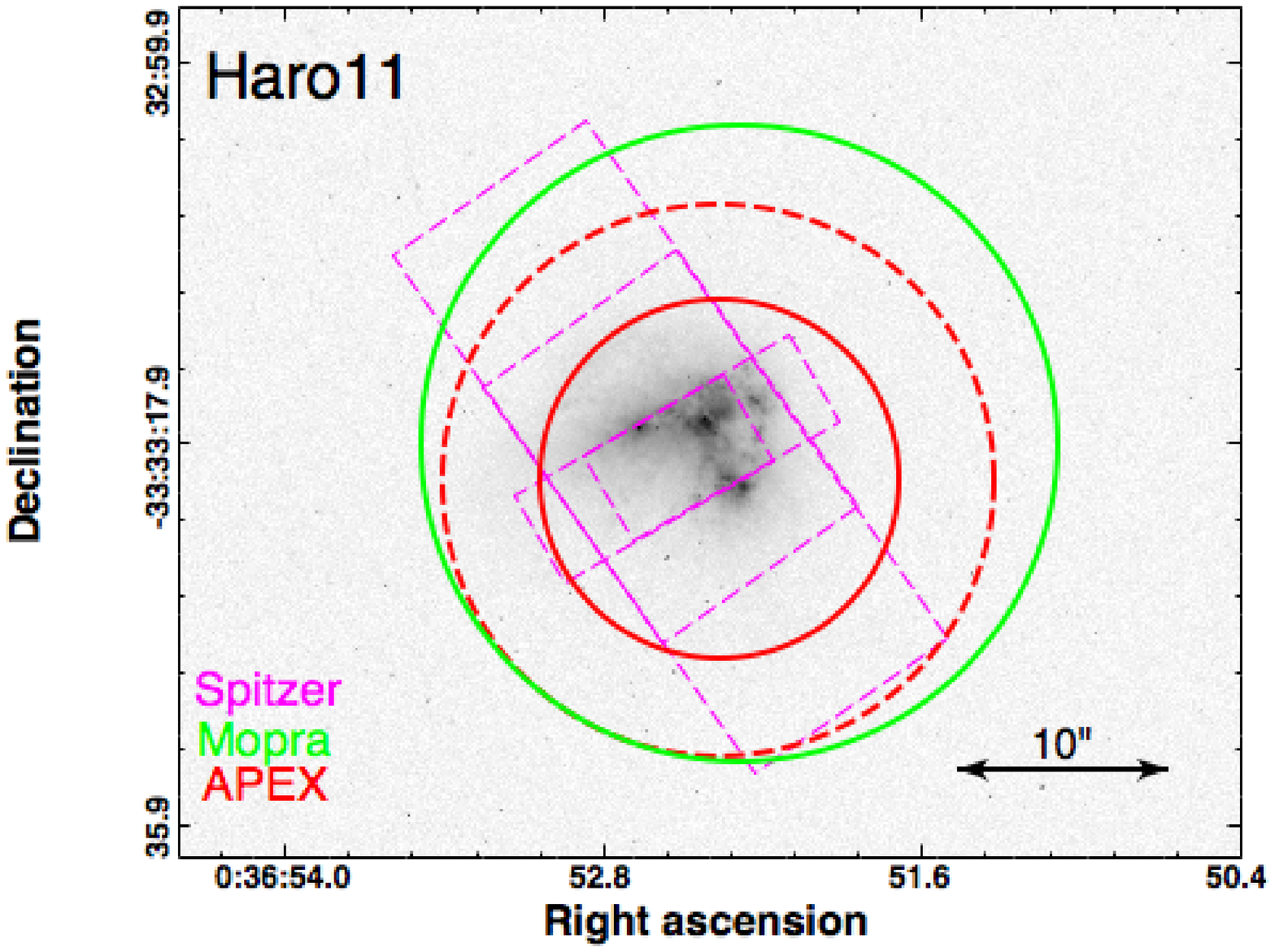}
\includegraphics[clip,trim=.6cm -.5cm 0 .8cm, width=6.1cm]{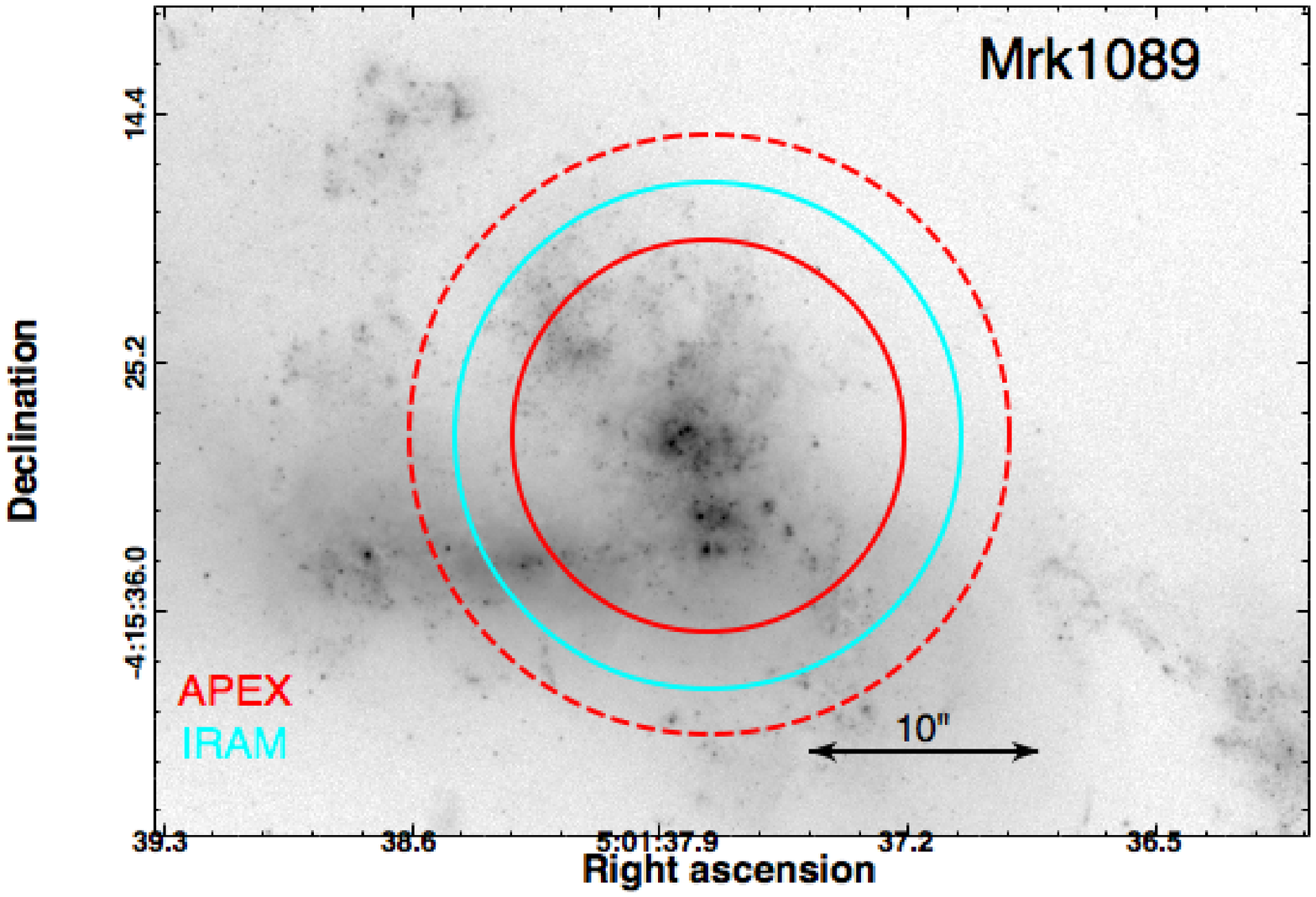}
\includegraphics[clip,trim=.5cm 0 0 0, width=6cm]{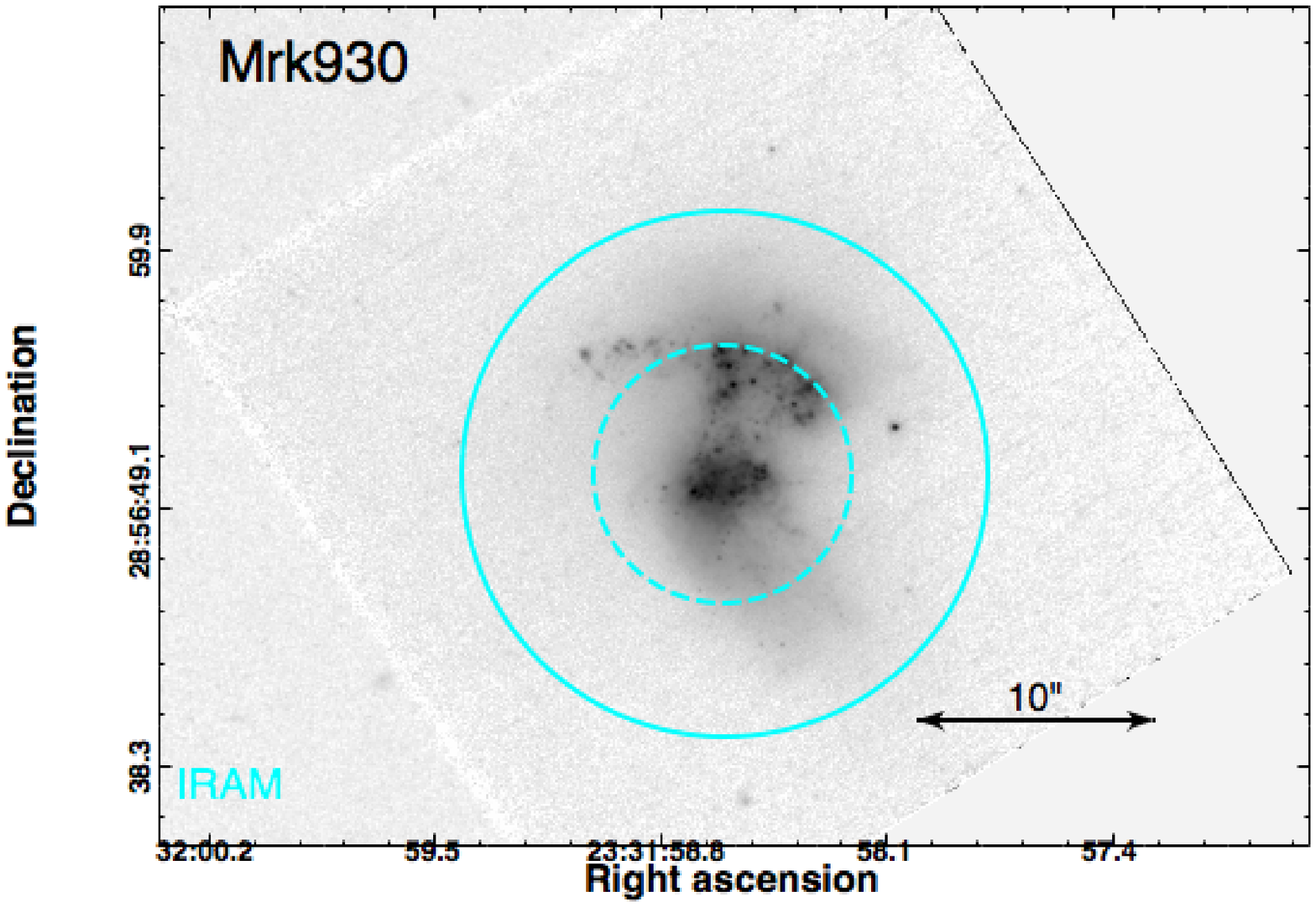}
\includegraphics[clip,trim=.2cm 0 .5cm 0, width=5.7cm]{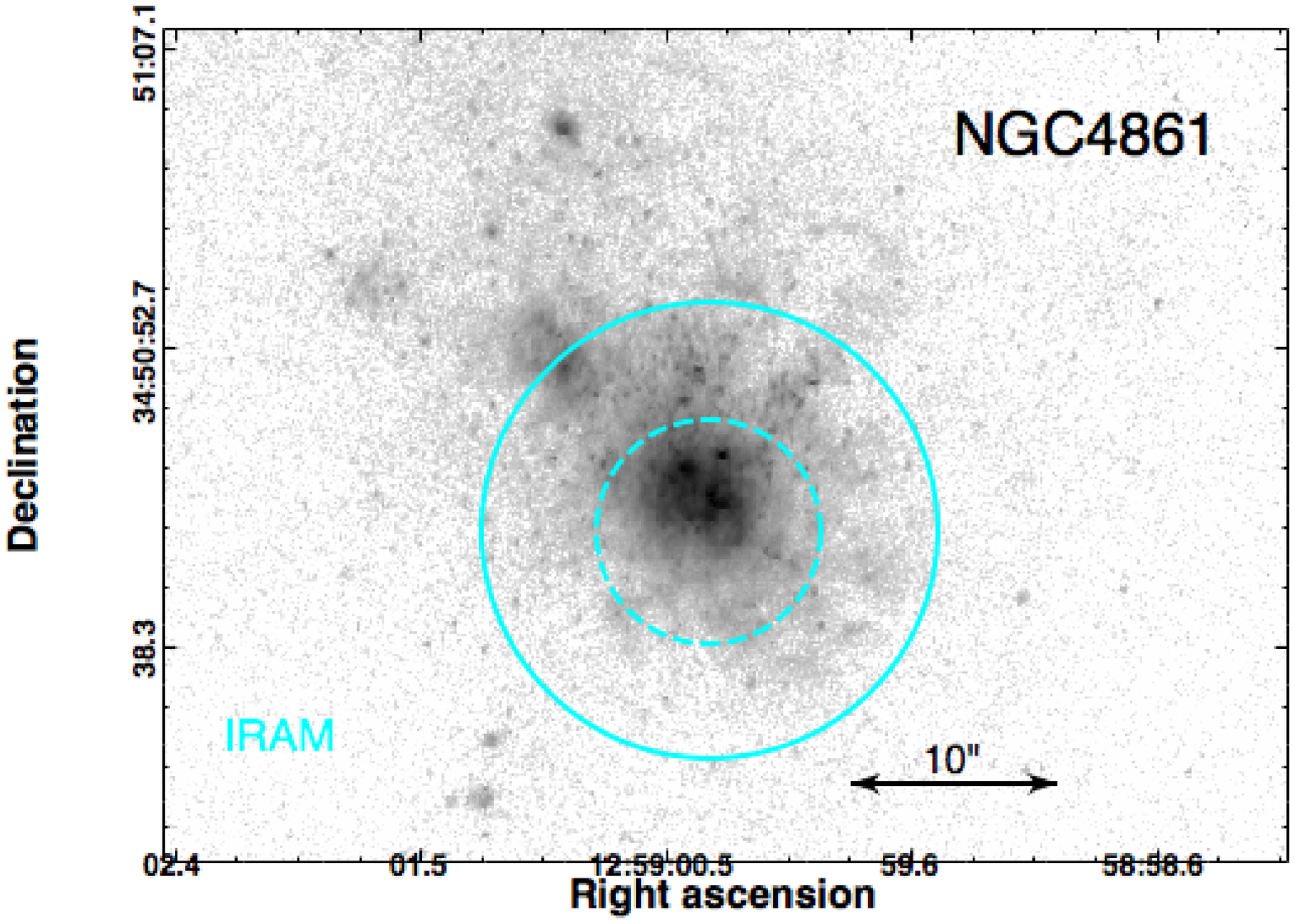}
\includegraphics[clip,trim=0 .4cm .3cm 0, width=6cm]{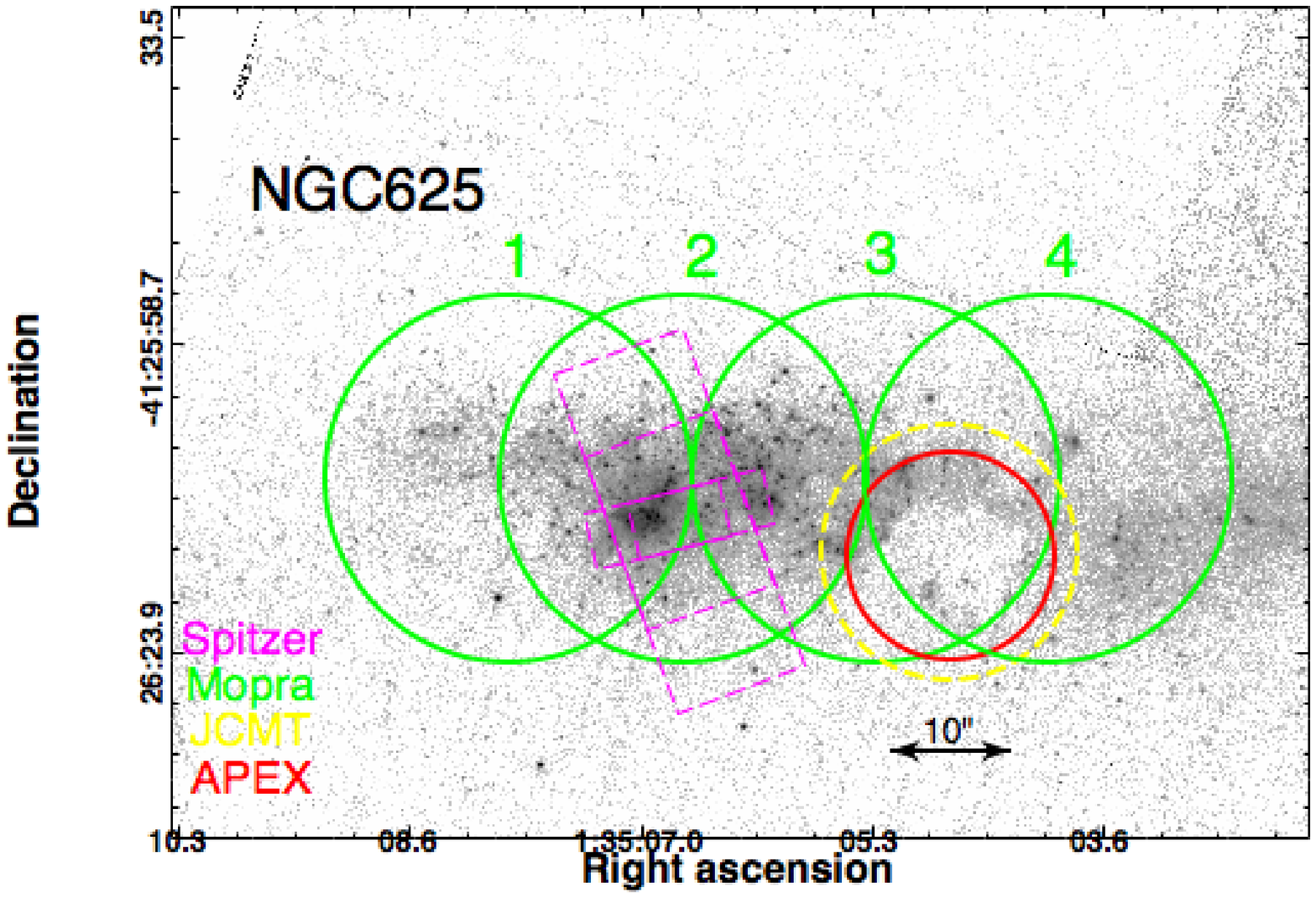}
\includegraphics[clip,trim=.2cm .4cm .3cm 0, width=6cm]{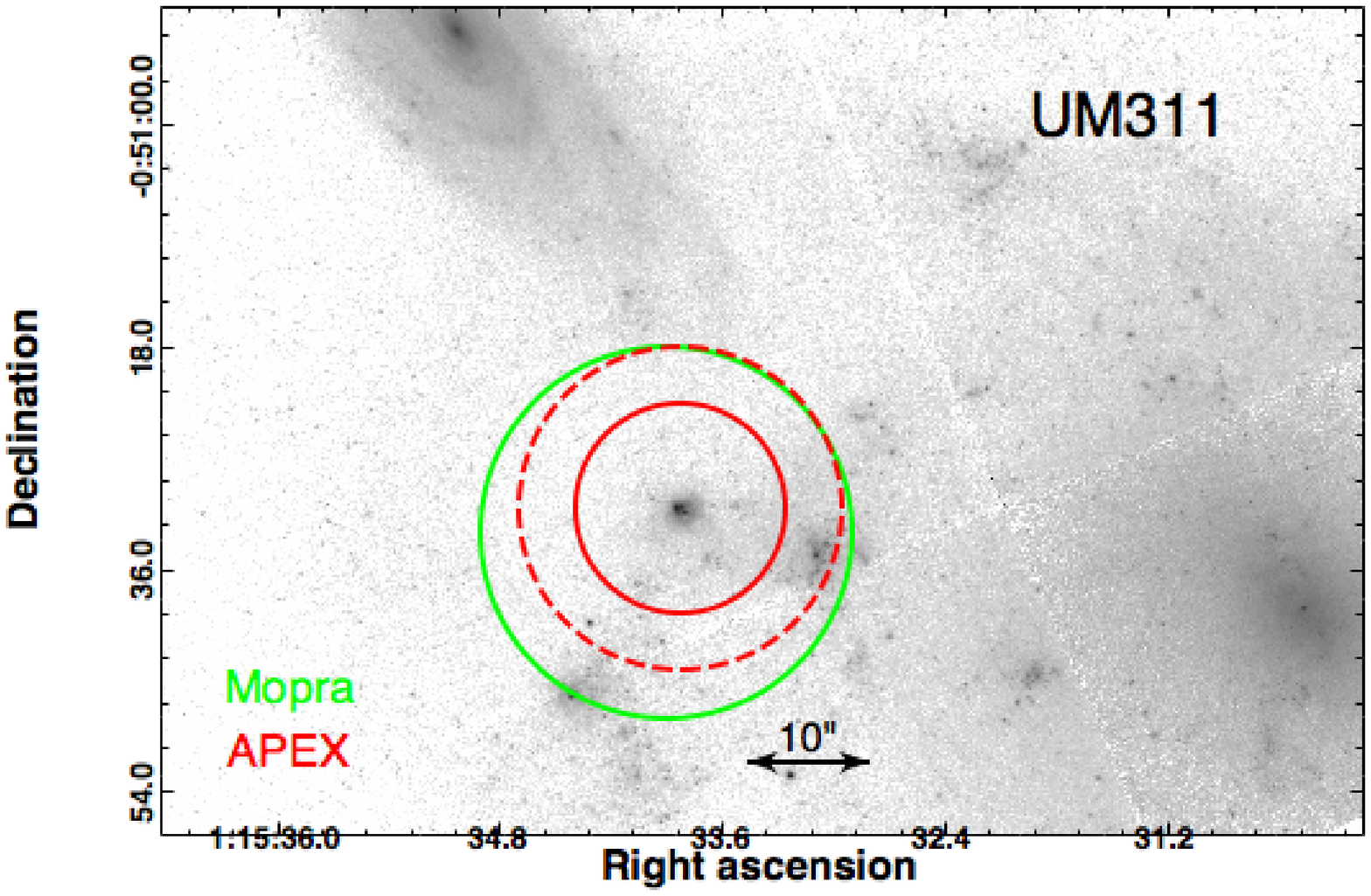}
\caption{
{\it HST} broad band images of the 6 galaxies surveyed in CO 
{(WFPC2 or ACS: F300 for NGC\,625, F814 for UM\,311, 
and F606 for the others)}, 
downloaded from the Hubble Legacy Archive (\protect\url{http://hla.stsci.edu/}). 
The Mopra beam is shown in green (CO(1-0):~30$^{\prime\prime}$), 
APEX in red (CO(2-1):~26$^{\prime\prime}$, 
CO(3-2):~17$^{\prime\prime}$), 
IRAM in blue (CO(1-0):~21$^{\prime\prime}$, 
CO(2-1), dashed:~11$^{\prime\prime}$), and the 
{\it JCMT} in yellow (CO(2-1), dashed:~22$^{\prime\prime}$) 
for NGC\,625. 
The {\it Spitzer}/IRS footprints are also shown in magenta 
for Haro\,11 and NGC\,625. 
}
\label{fig:cobeam}
\end{minipage}
\end{figure*}

To complement the existing CO data for the galaxies in the DGS, 
we have observed CO in 6 additional low-metallicity galaxies of the DGS: 
Haro\,11, Mrk\,930, Mrk\,1089, NGC\,625, NGC\,4861, and UM\,311. 
Global properties of these galaxies are summarized in Table~\ref{table:subgeneral}. 
{They are all sub-solar metallicity and larger 
than a kpc in size. In particular, NGC\,625 and NGC\,4861 are classified 
as SBm galaxies and have lower (optical and IR) luminosities than the other galaxies}. 
This target selection is essentially limited by instrument sensitivity: our sample 
consists of the brightest -- in the infrared -- galaxies of the DGS that lack CO data. 
Mrk\,930, NGC\,4861, and UM\,311 have no CO observations published at all. 
CO(1-0) observations were attempted in Haro\,11 by \cite{bergvall-2000} 
without success. We observed Haro\,11 and UM\,311 in the 3 transitions 
CO(1-0), CO(2-1), and CO(3-2); and Mrk\,930 and 
{the giant H~{\sc ii} region I\,Zw\,49 in NGC\,4861} in CO(1-0) and CO(2-1). 
Data of CO(1-0) already existed for Mrk\,1089 \citep{leon-1998}, 
and of CO(2-1) and CO(3-2) for NGC\,625 (partial coverage, P.I. Cannon). 
Thus for these two galaxies we observed the CO transitions that were lacking. 
NGC\,625 was observed in CO(1-0) with 4x1 pointings, while 
the other galaxies are more compact and required only single pointings. 
The CO beams are overlaid on {\it HST} images in Figure~\ref{fig:cobeam}.

\textit{Haro\,11 --} also known as ESO\,350-IG\,038, 
is a blue compact galaxy (BCG) located at 92~Mpc, 
with metallicity Z$\sim$1/3~Z$_{\odot}$ \citep{james-2013}. 
Haro\,11 hosts extreme starburst conditions with 
star formation rate $\sim$25~$\rm{M_{\odot}~yr^{-1}}$ \citep{grimes-2007}. 
It was previously observed in H~{\sc i} and CO but not detected \citep{bergvall-2000}. 
Interestingly, H~{\sc i} was recently detected in absorption by \cite{machattie-2013}, 
yielding a mass of M(H{\sc i}) between 3$\times$10$^8$ and 10$^9$~M$_{\odot}$, 
{and in emission with the GBT, yielding an H~{\sc i} mass of 
$4.7 \times 10^8$~M$_{\odot}$ (S.~Pardy priv. comm.). 
Both H~{\sc i} masses agree with each other and are higher than the upper limit 
from \cite{bergvall-2000}, thus we use the average value of $5 \times 10^8$~M$_{\odot}$}.

\textit{Mrk\,1089 --} is part of the Hickson Group~31 \citep{hickson-1982}. 
It is an irregular starburst (HCG31~C) at 57~Mpc, with metallicity 
$\sim$1/3~Z$_{\odot}$, SFR$\simeq$16~M$_{\odot}$~yr$^{-1}$ \citep{hopkins-2002}, 
and in interaction with the galaxy NGC\,1741 (HCG31~A). 
Since the two galaxies are not clearly separated in the mid-infrared (MIR)-FIR 
bands, we refer to the group as Mrk\,1089, as in \cite{galametz-2009}. 
HCG31~C dominates the emission in these bands and is the only 
one detected in CO \citep{leon-1998}.

\textit{Mrk\,930 --} is a blue compact galaxy located at 78~Mpc with 
metallicity $\sim$1/5~Z$_{\odot}$. It is also classified as a Wolf-Rayet (WR) 
galaxy \citep{izotov-1998}, with WR signatures indicating that it is 
undergoing a young burst episode. It exhibits an intricate morphology 
with several sites of active star formation, with a star cluster age peaking 
at 4~Myr \citep{adamo-2011} and a global SFR of $\sim$6~M$_{\odot}$~yr$^{-1}$.

\textit{NGC\,4861 --} is an irregular galaxy with a cometary shape, 
located at 7.5~Mpc, and with metallicity Z$\sim$1/6~Z$_{\odot}$. 
It is classified as a SB(s)m galaxy {in the \cite{rc-1991} catalogue}, 
but shows no evidence of spiral structure \citep{wilcots-1996}. 
It hosts several knots of star formation, dominated by 
the south-west region I\,Zw\,49 (where our CO observations point), as seen 
in H$\alpha$ \citep{van-eymeren-2009}. The H$\alpha$ and H~{\sc i} 
distributions coincide in the center, but the H~{\sc i} extends to form a larger envelope. 
It SFR is $\sim$0.05~M$_{\odot}$~yr$^{-1}$.

\textit{NGC\,625 --} is an irregular elongated dwarf galaxy in the Sculptor Group 
located at 4~Mpc and with metallicity $\sim$1/3~Z$_{\odot}$. 
It is undergoing an extended starburst $\ge$50~Myr, with 
SFR$\simeq$0.05~M$_{\odot}$~yr$^{-1}$ \citep{skillman-2003a}, which 
resulted in an outflow visible in H~{\sc i} \citep{cannon-2004,cannon-2005}. 
It is detected in CO(2-1) and CO(3-2) toward 
the main dust concentration, and shows MIR H$_{\rm 2}$ lines 
in its {\it Spitzer}/IRS spectrum (section~\ref{sect:spitzer}).

\textit{UM\,311 --} is one of a group of 3 compact H~{\sc ii} galaxies at 
a distance of 23~Mpc and with metallicity $\sim$1/2~Z$_{\odot}$, located 
between the pair of spiral galaxies NGC\,450 and UGC\,807. 
It is bright in H$\alpha$ \citep{guseva-1998} and H~{\sc i} \citep{smoker-2000}, 
with SFR$\simeq$1~M$_{\odot}$~yr$^{-1}$ \citep{hopkins-2002}. 
The individual sources of the group are separated by $\sim$15$^{\prime\prime}$ 
and are not resolved in the FIR photometry, although UM\,311 dominates 
the emission. When referring to UM\,311, we encompass the full complex 
(the three H~{\sc ii} galaxies and the two spirals).

%%%
\begin{center}
\begin{table*}[!htp]\tiny
  \caption{General properties of the DGS dwarf subsample.}
  \begin{tabular}{l c c c c c c c c c c}
    \hline\hline
    \vspace{-8pt}\\
    \multicolumn{1}{l}{Galaxy} & 
    \multicolumn{2}{c}{Coordinates} &
    \multicolumn{1}{c}{Dist.} & 
    \multicolumn{1}{c}{Optical size} & 
    \multicolumn{1}{c}{Metallicity} & 
    \multicolumn{1}{c}{$M_{\rm B}$} & 
    \multicolumn{1}{c}{$M_{\rm H~I}$} & 
    \multicolumn{1}{c}{$L_{\rm TIR}$} & 
    \multicolumn{1}{c}{$L_{\rm H\alpha}$} & 
    \multicolumn{1}{c}{$L_{\rm FUV}$} \\ \cline{2-3}
    
    \multicolumn{1}{l}{} & 
    \multicolumn{2}{c}{(J2000)} &
    \multicolumn{1}{c}{(Mpc)} & 
    \multicolumn{1}{c}{} & 
    \multicolumn{1}{c}{O/H} & 
    \multicolumn{1}{c}{(mag)} & 
    \multicolumn{1}{c}{($\rm{10^{9}~M_{\odot}}$)} & 
    \multicolumn{1}{c}{($\rm{10^{9}~L_{\odot}}$)} & 
    \multicolumn{1}{c}{($\rm{10^{7}~L_{\odot}}$)} & 
    \multicolumn{1}{c}{($\rm{10^{9}~L_{\odot}}$)} \\
    \hline
    \vspace{-8pt}\\
	Haro\,11		& 00h36$^{\prime}$52.5$^{\prime\prime}$	& -33$^{\circ}$33$^{\prime}$19$^{\prime\prime}$ 	& 
				92 	& 0.4$^{\prime}$$\times$0.5$^{\prime}$ 	& 8.20$^{(a)}$ 	& -20.6 	& 0.5$^{(b)}$ 	& 185	& 92 		& 22	 	 \\
				&	&	&	& (11\,kpc $\times$ 13\,kpc) 	&	&	&	&	&	& \\
	Mrk\,1089		& 05h01$^{\prime}$37.8$^{\prime\prime}$	& -04$^{\circ}$15$^{\prime}$28$^{\prime\prime}$ 	& 
				57 	& 0.6$^{\prime}$$\times$0.2$^{\prime}$ 	& 8.10 	& -20.5 	& 24 	 	& 34.3	& 23		& - 	 	 \\
				&	&	&	& (10\,kpc $\times$ 3\,kpc) 	&	&	&	&	&	& \\
	Mrk\,930		& 23h31$^{\prime}$58.3$^{\prime\prime}$	& +28$^{\circ}$56$^{\prime}$50$^{\prime\prime}$ 	& 
				78 	& 0.4$^{\prime}$$\times$0.4$^{\prime}$ 	& 8.03 	& -19.5	& 3.0	 	& 18.8	& 23		& -		 \\
				&	&	&	& (9\,kpc $\times$ 9\,kpc) 	&	&	&	&	&	& \\
	NGC\,4861	& 12h59$^{\prime}$02.3$^{\prime\prime}$	& +34$^{\circ}$51$^{\prime}$34$^{\prime\prime}$ 	& 
				7.5 	& 4.0$^{\prime}$$\times$1.5$^{\prime}$ 	& 7.89 	& -16.8 	& 0.48	 & 0.323	& 0.04 	& 0.46  \\
				&	&	&	& (9\,kpc $\times$ 3\,kpc) 	&	&	&	&	&	& \\
	NGC\,625		& 01h35$^{\prime}$04.6$^{\prime\prime}$	& -41$^{\circ}$26$^{\prime}$10$^{\prime\prime}$ 	& 
				3.9 	& 5.8$^{\prime}$$\times$1.9$^{\prime}$ 	& 8.22 	& -16.2 	& 0.13 	& 0.280	& 0.18 	& 0.11  \\
				&	&	&	& (7\,kpc $\times$ 2\,kpc) 	&	&	&	&	&	& \\
	UM\,311		& 01h15$^{\prime}$34.0$^{\prime\prime}$	& -00$^{\circ}$51$^{\prime}$32$^{\prime\prime}$ 	& 
				24 	& 0.1$^{\prime}$$\times$0.1$^{\prime}$ 	& 8.38 	& -19.2 	& 3.0 	& 5.18 	 & 0.48 	& - 	 	 \\
				&	&	&	& (0.7\,kpc $\times$ 0.7\,kpc) 	&	&	&	&	&	& \\
    \hline \hline
  \end{tabular}
  \begin{flushleft}
  \hfill{}
    \vspace{-1pt}\\
  Coordinates and optical sizes are taken from the NASA/IPAC Extragalactic Database. 
  The size of UM\,311 is that of the individual H~{\sc ii} galaxy, but the complex that we consider is larger ($\sim$3$^{\prime}\times2^{\prime}$). 
  Distances, metallicities, $M_{\rm B}$, and H~{\sc i} masses are from \cite{madden-2013} (see references therein). 
  For Mrk\,930, $M_{\rm B}$ is from \cite{markarian-1989}. 
  Luminosities and masses are scaled to the quoted distances. 
  L$_{\rm TIR}$ is the total infrared (TIR) luminosity measured by integrating SEDs from 8 to 1000$\mu$m, from \cite{remy-2013b}. 
  $(a)$~For Haro\,11, we keep the metallicity used in \cite{cormier-2012}. 
  $(b)$~From \cite{machattie-2013} and {S.~Pardy (priv. comm.), see details in section~\ref{sect:obs}}. 
  {\it References for $L_{\rm H\alpha}$}: Haro\,11: \cite{ostlin-1999}, UM\,311: \cite{terlevich-1991}, 
  Mrk\,930: \cite{adamo-2011}, NGC\,4861: \cite{schmitt-2006}, Mrk\,1089: \cite{iglesias-paramo-1997}, and NGC\,625: \cite{kennicutt-2008}. 
  {\it References for $L_{\rm FUV}$}: Haro\,11: \cite{grimes-2007}, NGC\,4861: \cite{de-paz-2007}, and NGC\,625: \cite{lee-2011}. 
  \end{flushleft}
  \label{table:subgeneral}
\end{table*}
\end{center}

%%%%%
\subsection{Mopra data: CO(1-0) in southern hemisphere targets}
\label{sect:mopra}
%%%%%
We observed the CO(1-0) line in Haro\,11, NGC\,625, and UM\,311 
with the ATNF Mopra 22-m telescope as part of the program M596, from 
September 13 to 18 2011.
Observations consisted of a single pointing for Haro\,11 and UM\,311, 
and of a 4$\times$1 pointed map with a half-beam step size for NGC\,625. 
The size of the beam is $\sim$30$^{\prime\prime}$ at 115~GHz. 
The position switching observing mode consists of a pattern of 4\,minutes: 
OFF1-ON-ON-OFF2, where OFF1 and OFF2 are two different reference positions 
in the sky taken $\ge$1.5$^{\prime}$ away from the source, to remove 
the contribution from the sky. 
We used the 3-mm band receiver and the backend 
MOPS in wideband mode, scanning a total bandwidth of 8.2~GHz 
with spectral resolution 0.25~MHz and central frequency 109.8~GHz 
for Haro\,11, 111.5~GHz for UM\,311, and 112.0~GHz for NGC\,625. 
The above-atmosphere system temperature was estimated every 20\,minutes 
using an ambient (hot) load. Between hot load measurements, fluctuations 
in the system temperature are monitored using a noise diode.
The total (ON+OFF) integration times were 14.1\,h, 10.3\,h, and 3.0\,h 
for Haro\,11, UM\,311, and NGC\,625 respectively.
The average system temperatures were 265, 375, and 370~Kelvin.
We used Orion KL to monitor the absolute flux calibration and the following 
pointing sources: R\,Hor, R\,Dor, S\,Scl, R\,Apr, and o\,Ceti.

%%%
We used our own Interactive Data Language (IDL)-written script for the data reduction.
We first create a quotient spectrum taking the average of the two 
reference position spectra before and after the on-source spectrum. 
The equation for the quotient spectrum $Q$ with continuum removal is: 
$Q = T_{OFF} \times (ON/OFF) - T_{ON}$, where $ON$ and $OFF$ are 
the on-source and off-source spectra. 
We then fit a spline function to remove the large scale variations and 
shift the signal to a median level of zero.
The data are calibrated using a main beam efficiency of 
$\eta_{mb}=0.42 \pm 0.02$ \citep{ladd-2005}. 
The pointing errors are estimated to be $\le$3.5$^{\prime\prime}$. 

\begin{figure}[!t]
\centering
\includegraphics[clip, width=8.8cm,trim=1cm 1.5cm 1cm .4cm]{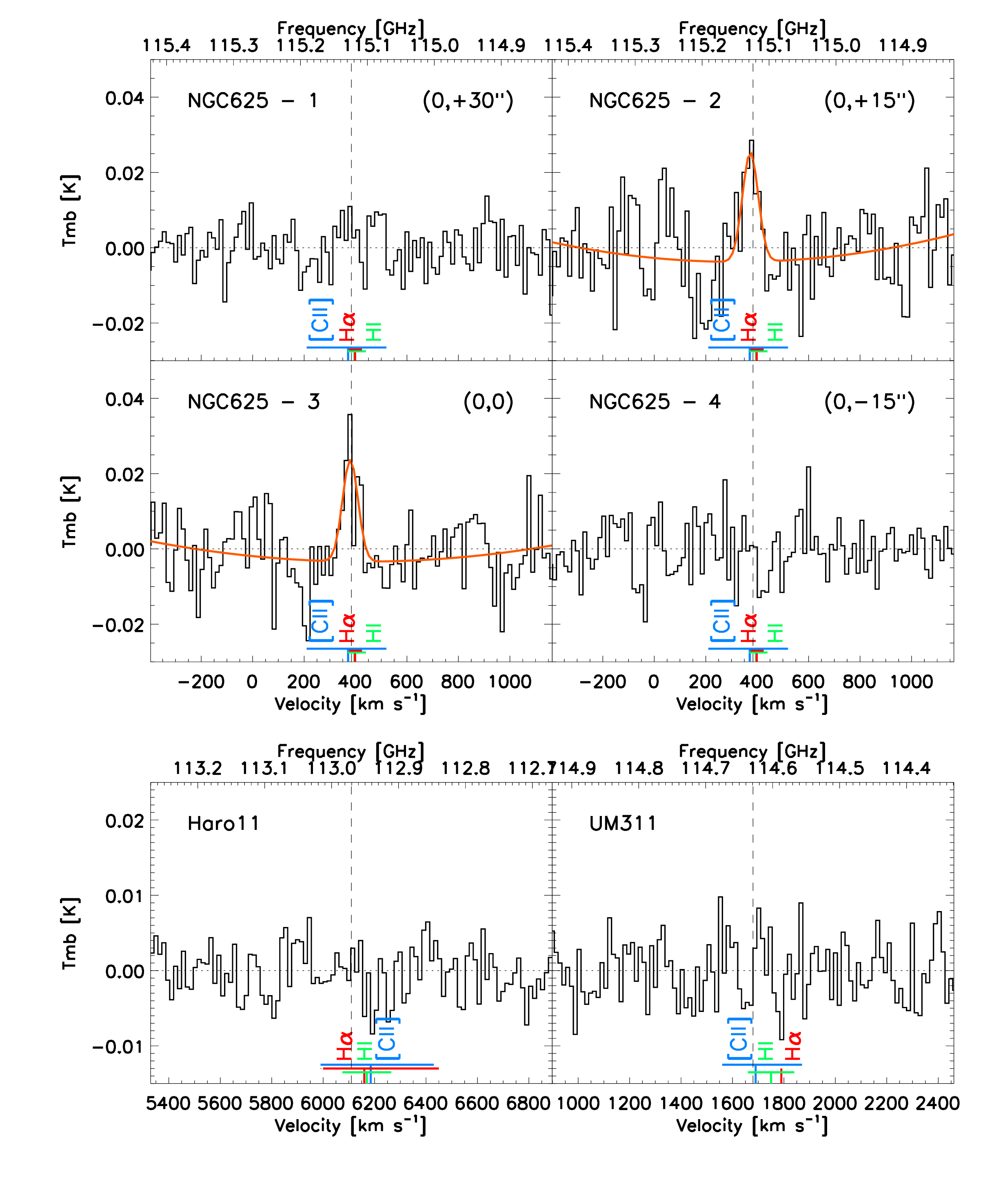}
\caption{
Mopra spectra ($T_{mb}$) of the CO(1-0) line 
at 115~GHz in NGC\,625, Haro\,11, and UM\,311. 
The $(0,0)$ position in NGC\,625 (pointing~3) has coordinates 
$(RA,Dec) = (01h35m05.46s,-41d26m09.6s)$. 
The spectral resolution for the display is 15~km~s$^{-1}$, 
corresponding to $\sim$5.8~MHz. 
The orange curve shows the fit to the line when detected. 
The vertical dashed lines indicate the expected position of the line, 
based on the central velocity of the other CO transitions. 
Central velocities of [C~{\sc ii}], H~{\sc i}, and H$\alpha$ and their 
dispersion are also indicated in blue, green, and red, respectively. 
All velocities are indicated in the local standard-of-rest (LSR) reference frame. 
}
\label{fig:mopra_spec}
\end{figure}

With Mopra, we detect CO(1-0) in NGC\,625, in the beam 
position~3 (counting from east to west, see Fig.~\ref{fig:cobeam}), 
and a possible detection in position~2. 
It is surprising that no line is detected in position~4, since both positions~3 and~4 
cover the area where the ancillary CO(2-1) and 
CO(3-2) observations were taken and the lines detected. 
This could indicate that the CO emission detected in position~3 comes from 
a region not covered by position~4 or the ancillary data, and the emission 
from the ancillary data is not seen by Mopra because of its limited sensitivity. 
Hence we can only compute an upper limit on the flux in the area covered by the 
ancillary data to be able to compare the CO(1-0) to the other transitions. 
The spectra from all positions are displayed in Figure~\ref{fig:mopra_spec} 
and the corresponding line parameters are reported in Table~\ref{table:lines}. 
For the total CO emission, we consider the average flux of all 4 positions and 
an average beam size taken as the geometric area covered by the 4 pointings. 

CO(1-0) is not detected with Mopra in Haro\,11 nor in UM\,311. 
Their spectra are strongly affected by $\sim$30~MHz baseline ripples, 
caused by standing waves in the receiver-subreflector cavity.
To counter this effect, we selected two different reference positions and 
applied a defocusing procedure while observing, but the standing 
waves were still present in our data. 
Since they change with frequency, it is difficult to model or filter them.
As a first step, we identify by eye the spectra affected by the ripples.
10\% show ripples with strong amplitude, and standard deviation $>$110~mK, 
and 20\% with ripples of lower amplitude but clear wave shapes.
Flagging out these spectra, we then rebin the data to a spectral resolution 
of 10~km~s$^{-1}$, and do a sigma-weighted average to obtain a final spectrum, 
with no additional correction. 
The CO(1-0) lines remain undetected in Haro\,11 and UM\,311. 
To further improve our final rms, we use a defringing tool based 
on a routine originally developed for removing fringes from ISO SWS 
spectroscopic data \citep{kester-2003}. 
The routine was modified for our purpose and tuned to specifically search 
for a maximum of ten significant fringes in the frequency domain of the fringes 
seen in the deep Mopra data. It is applied on all individual spectra. 
After subtracting the ripples, the noise level is greatly reduced.
Again, the final spectra, shown in Fig.~\ref{fig:mopra_spec}, 
are obtained with a sigma-weighted average of all spectra. 
\cite{bergvall-2000} report an incredibly low limit 
of $L_{\rm CO(1-0)} < 260$~L$_{\odot}$ in Haro\,11, 
but without giving details on their calculation. 
In order to be conservative, we prefer to use our upper limit throughout this paper.

%%%

\begin{center}
\begin{table*}[!t]
  \caption{CO line parameters and intensities.}
  \hfill{}
  \begin{tabular}{l l l l c c c c c c}
    \hline\hline
    \vspace{-8pt}\\
    \multicolumn{1}{l}{Galaxy} & 
    \multicolumn{1}{l}{Line} & 
    \multicolumn{1}{l}{Center} & 
    \multicolumn{1}{l}{$\theta_{FWHM}$} & 
    \multicolumn{1}{c}{$T_{\rm mb}$} &
    \multicolumn{1}{c}{V} &
    \multicolumn{1}{c}{$\Delta$V} &
    \multicolumn{1}{c}{$I_{\rm CO}$} &
    \multicolumn{1}{c}{$L_{\rm CO}$} &
    \multicolumn{1}{c}{Reference} \\
    \multicolumn{1}{l}{} & 
    \multicolumn{1}{l}{} & 
    \multicolumn{1}{l}{} & 
    \multicolumn{1}{l}{($^{\prime\prime}$)} & 
    \multicolumn{1}{c}{(mK)} &
    \multicolumn{1}{c}{(km~s$^{-1}$)} &
    \multicolumn{1}{c}{(km~s$^{-1}$)} &
    \multicolumn{1}{c}{(K~km~s$^{-1}$)} &
    \multicolumn{1}{c}{($L_{\odot}$)} &
    \multicolumn{1}{c}{} \\
    \hline 
    \vspace{-8pt}\\
	Haro\,11 		& {\sc CO(1-0)}	&	& 30 $^{(1)}$ 	& $<$~10.7 $^{(a)}$	& - 	& - 	& $<$~0.69 		& $<$~6.6$\times10^{3}$	& this work 	\\
				& {\sc CO(2-1)}	&	& 26 $^{(2)}$	& 8.63	& 6113	& 59 		& 0.54 $\pm$ 0.19 	& 3.2$\times10^{4}$		& this work 	\\
				& {\sc CO(3-2)}	&	& 17 $^{(2)}$	& 7.20	& 6103	& 63		& 0.48 $\pm$ 0.17	& 4.0$\times10^{4}$		& this work 	\\
	Mrk\,1089 	& {\sc CO(1-0)}	&	& 21 $^{(3)}$	& 26.3	& 3987	& 148	& 2.56 $\pm$ 0.50 	& 4.7$\times10^{3}$		& \cite{leon-1998}	\\
				& {\sc CO(2-1)}	&	& 26			& 14.1	& 3990	& 111	& 1.67 $\pm$ 0.40	& 3.8$\times10^{4}$		& this work 	\\
				& {\sc CO(3-2)}	&	& 17			& 11.5	& 3976	& 95		& 1.16 $\pm$ 0.32	& 3.8$\times10^{4}$		& this work 	\\
	Mrk\,930	 	& {\sc CO(1-0)}	&	& 21 			& 2.17	& 5491	& 62		& 0.14 $\pm$ 0.07 	& 4.7$\times10^{2}$		& this work	\\
				& {\sc CO(2-1)}	&	& 11	$^{(3)}$	& $<$~6.44	& -	& -		& $<$~0.34		& $<$~2.6$\times10^{3}$		& this work 	\\
	NGC\,4861 	& {\sc CO(1-0)}	&	& 21 			& $<$~6.51	& -	& -		& $<$~0.35		& $<$~1.1$\times10^{1}$		& this work	\\
			 	& {\sc CO(1-0)}	&	& $total ^{(b)}$	& -	& -	& -				& $<$~1.23		& $<$~3.9$\times10^{1}$		& this work	\\
				& {\sc CO(2-1)}	&	& 11			& $<$~8.74	& -	& -		& $<$~0.47		& $<$~3.3$\times10^{1}$		& this work 	\\
	NGC\,625 	& {\sc CO(1-0)}	& $(0,+30^{\prime\prime})$	& 30 			& $<$~24.9	& - 	& - 		& $<$~1.86 		& $<$~3.3$\times10^{1}$	& this work 	\\
				& {\sc CO(1-0)}	& $(0,+15^{\prime\prime})$	& 30 			& 29.0 	& 373 	& 72 		& 2.23 $\pm$ 0.72 	& 4.0$\times10^{1}$		& this work 	\\
			 	& {\sc CO(1-0)}	& $(0,0)$					& 30 			& 27.1 	& 379 	& 71 		& 2.05 $\pm$ 0.63 	& 3.6$\times10^{1}$		& this work 	\\
				& {\sc CO(1-0)}	& $(0,-15^{\prime\prime})$	& 30 			& $<$~21.6 	& - 	& - 		& $<$~1.61 		& $<$~2.9$\times10^{1}$	& this work 	\\
			 	& {\sc CO(1-0)}	&						& $total ^{(b)}$ 	& 57.7 	& 374 	& 70 		& 4.26 $\pm$ 0.62	& 7.5$\times10^{1}$		& this work 	\\
			 	& {\sc CO(1-0)}	& $ref ^{(c)}$				& 30 			& - 		& - 	& - 			& $<$~1.61$^{(c)}$ 	& $<$~2.9$\times10^{1}$	& this work 	\\
				& {\sc CO(2-1)}	&	& 22	$^{(4)}$	& 58.0	& 390	& 27		& 1.64 $\pm$ 0.40	& 1.0$\times10^{2}$		& \cite{cannon-phd} 	\\
				& {\sc CO(3-2)}	&	& 17			& 59.0	& 381	& 23		& 1.44 $\pm$ 0.38	& 2.2$\times10^{2}$		& this work \\ %ESO Archive$^{(d)}$ 	\\
	UM\,311 		& {\sc CO(1-0)}	&	& 30 			& $<$~11.4	& - 	& - 		& $<$~0.61 		& $<$~4.1$\times10^{2}$	& this work 	\\
		 		& {\sc CO(1-0)}	&	& $total ^{(b)}$ 		& -	& - 	& - 			& $<$~3.05 		& $<$~2.0$\times10^{3}$		& this work 	\\
				& {\sc CO(2-1)}	&	& 26			& 10.9	& 1678	& 41		& 0.47 $\pm$ 0.15	& 1.9$\times10^{3}$		& this work 	\\
				& {\sc CO(3-2)}	&	& 17			& 8.62	& 1686	& 44 		& 0.40 $\pm$ 0.14	& 2.3$\times10^{3}$		& this work 	\\
    \hline \hline
  \end{tabular}
  \hfill{}
  \newline
    \vspace{-1pt}\\
    $(1)$~ATNF/Mopra, $(2)$:~APEX/SHeFI, $(3)$:~IRAM/30-m, $(4)$:~{\it JCMT}/RxA. \\
     ${(a)}$ The upper limits are 3-$\sigma$. 
     ${(b)}$ Total emission for extended galaxies, estimated from the other Mopra pointings for NGC\,625 
     	and extrapolated from the 100$\mu$m continuum emission for NGC\,4861 and UM\,311 (see section~\ref{sect:comiss}). 
     ${(c)}$ Estimated CO(1-0) line intensity at the position covered by the CO(2-1) and CO(3-2) lines. 
  \label{table:lines}
\end{table*}
\end{center}

%%%%%
\subsection{APEX data: CO(2-1) and CO(3-2) in southern hemisphere targets}
 \label{sect:apex}
%%%%%
We observed the CO(2-1) and CO(3-2) lines 
in Haro\,11, Mrk\,1089, and UM\,311 with the Swedish Heterodyne Facility 
Instrument (SHeFI) on the Atacama Pathfinder EXperiment (APEX). 
These observations are part of the program, ID~088.F-9316A, and were 
performed from September 3 to 11 2011. 
The beam size is $\sim$26$^{\prime\prime}$ at 230~GHz and $\sim$17$^{\prime\prime}$ 
at 345~GHz. We made single pointings in ON/OFF mode with the receivers APEX~1 
and APEX~2 and backend XFFTS. 
The bandwidth covered is 2.5~GHz, with initial spectral resolution 88.5~kHz. 
Total integrations times for each source (ON source + OFF source) were
82 and 120\,minutes for Haro\,11, 79 and 140\,minutes for UM\,311, 
135 and 110\,minutes for Mrk\,1089, for the CO(2-1) and CO(3-2) lines 
respectively. Calibration was done every 10\,minutes. 
The average system temperatures were 150, 248, 156, 214, 153, 208\,K.
The following pointing and focus sources were used: 
PI1-GRU, WX-PSC, O-CETI, R-LEP, TX-PSC, RAFGL3068, IK-TAU, R-SCL, Uranus. 
The pointing accuracy is typically 2$^{\prime\prime}$.

We also present CO(3-2) observations of NGC\,625, 
obtained between August 17 and 26 2005 as part of the program 
00.F-0007-2005 (P.I.~Cannon) and retrieved from the ESO archive. 
Observations consisted of a single pointing (Fig.~\ref{fig:cobeam}) with 
the receiver APEX~2 and backend FFTS (decommissioned), which 
is composed of two 1~GHz units achieving a spectral resolution of 122~kHz. 
The total integration time was 155\,minutes and the average system temperature was 150\,K. 
The sources R-FOR and R-SCL were used for pointing, and Jupiter for focus.

\begin{figure}[!ht]
\begin{minipage}{8.8cm}
\centering
\includegraphics[height=4.35cm,angle=-90]{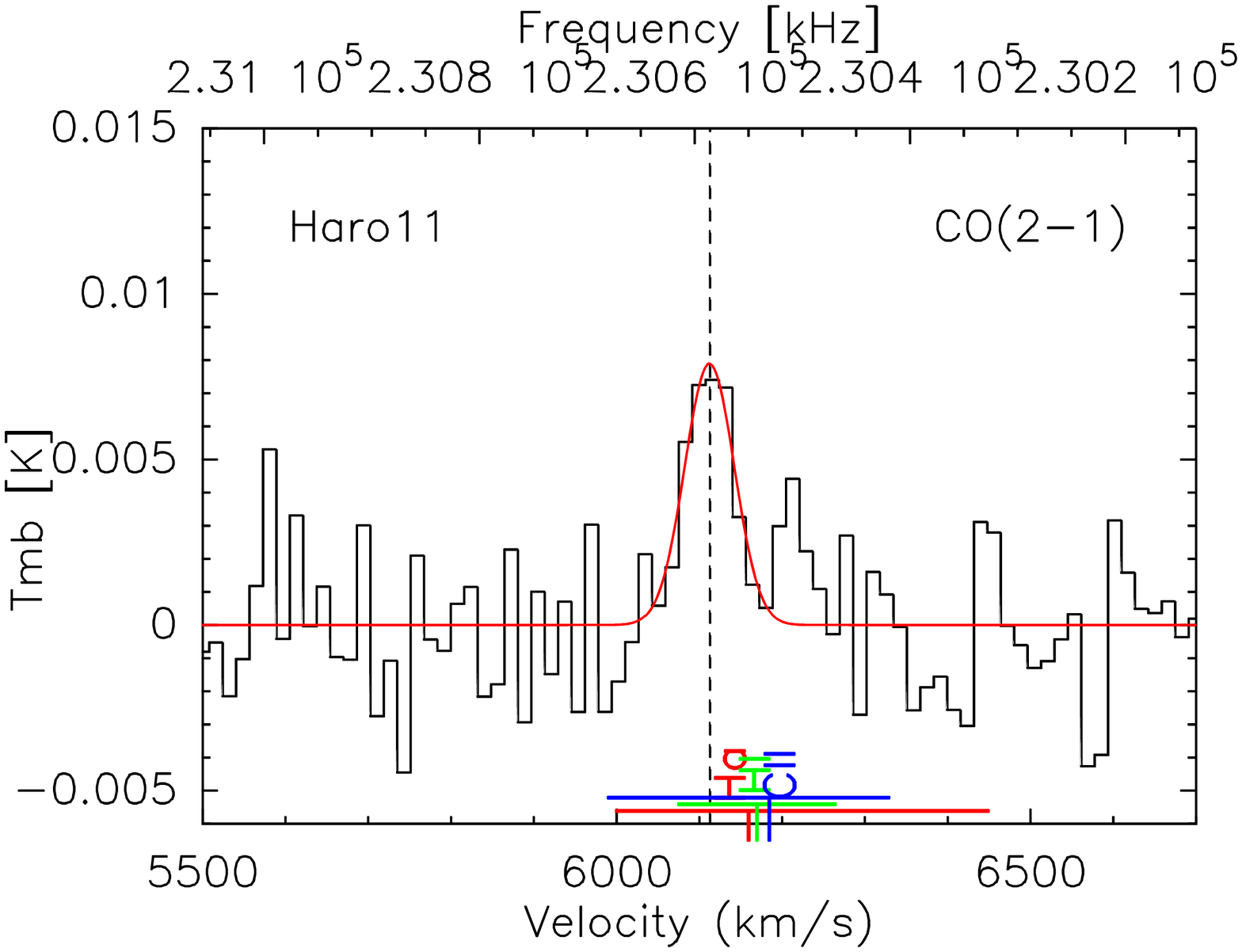}
\includegraphics[height=4.35cm,angle=-90]{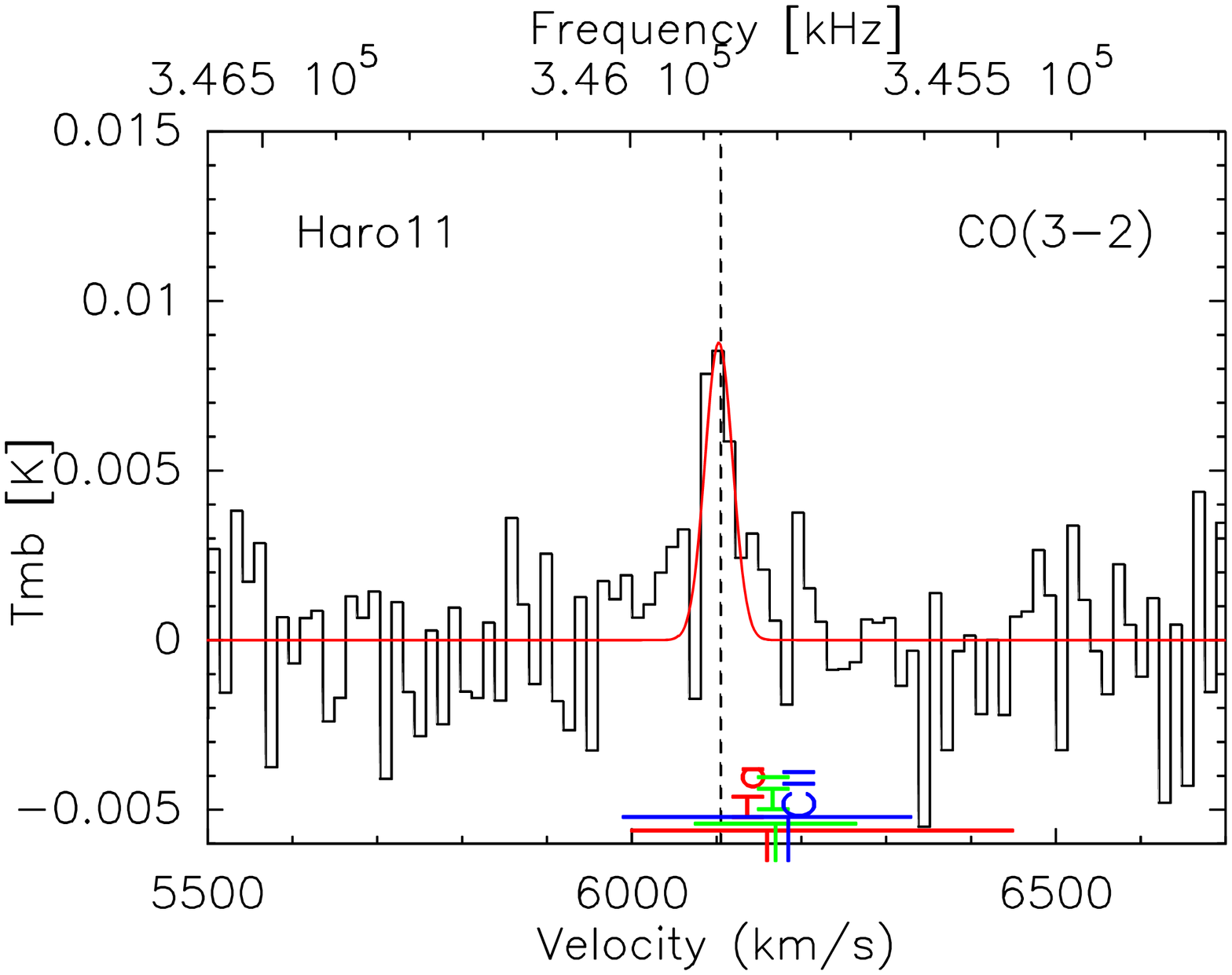}
\includegraphics[height=4.35cm,angle=-90]{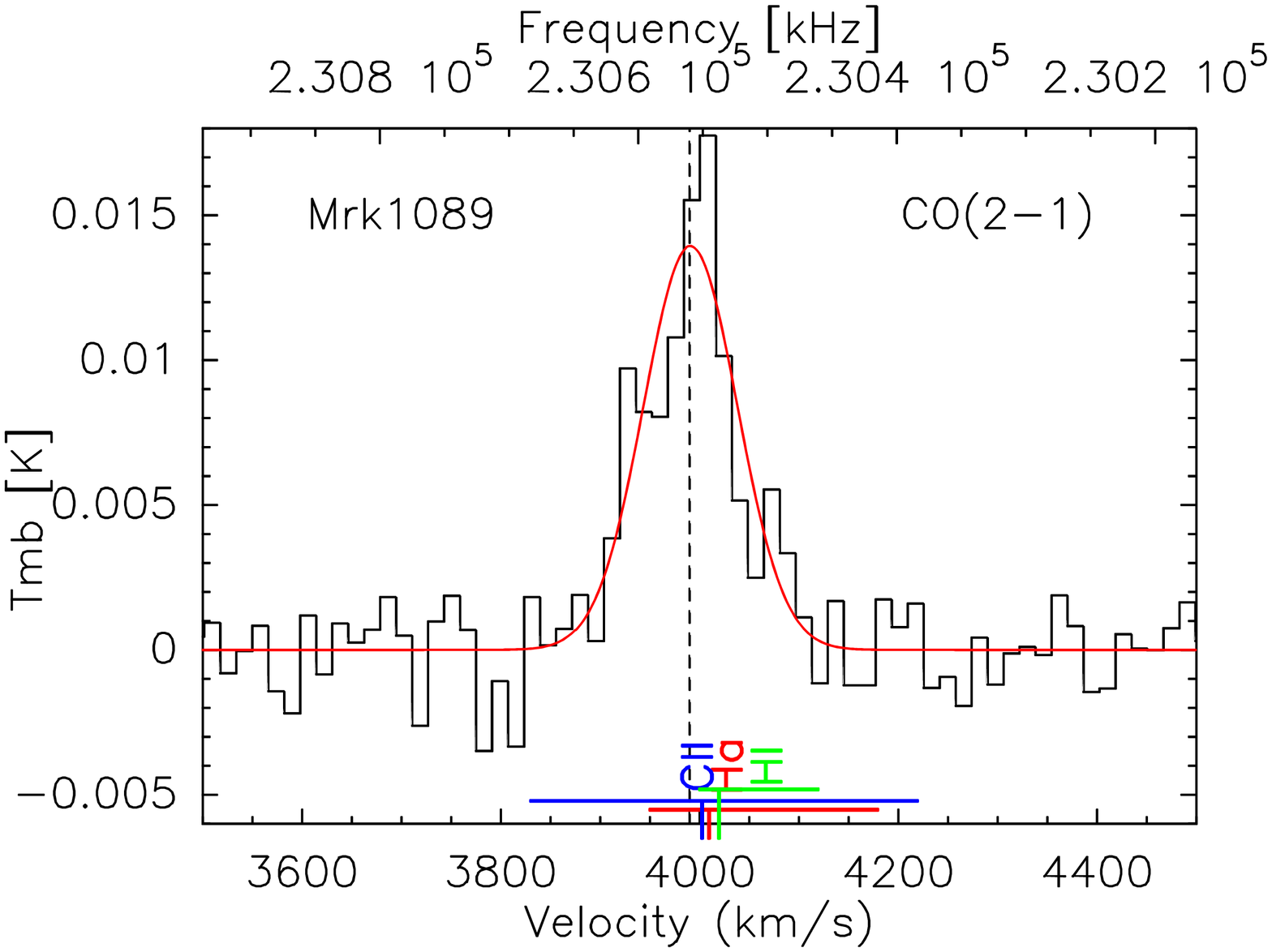}
\includegraphics[height=4.35cm,angle=-90]{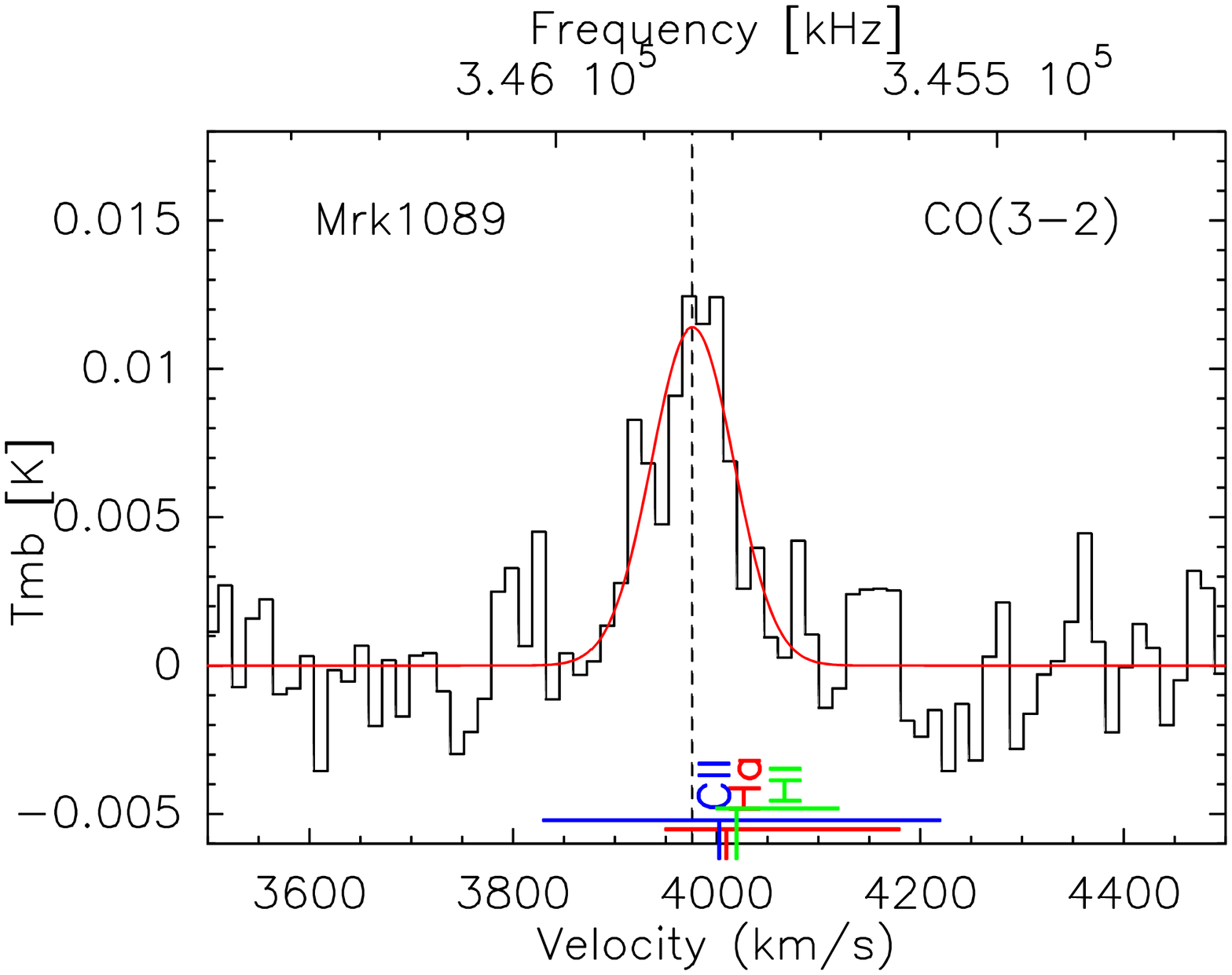}
\includegraphics[height=4.35cm,angle=-90]{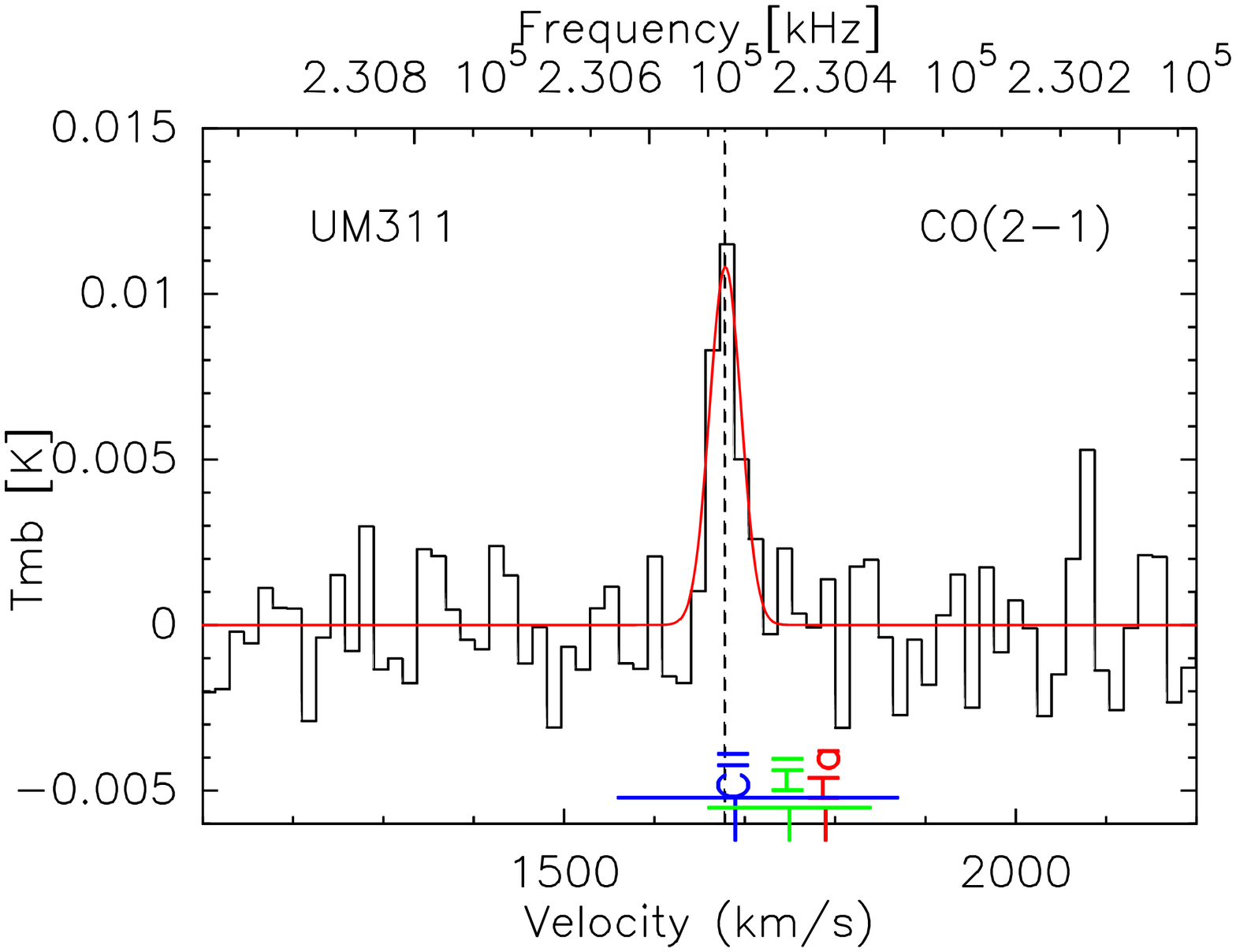}
\includegraphics[height=4.35cm,angle=-90]{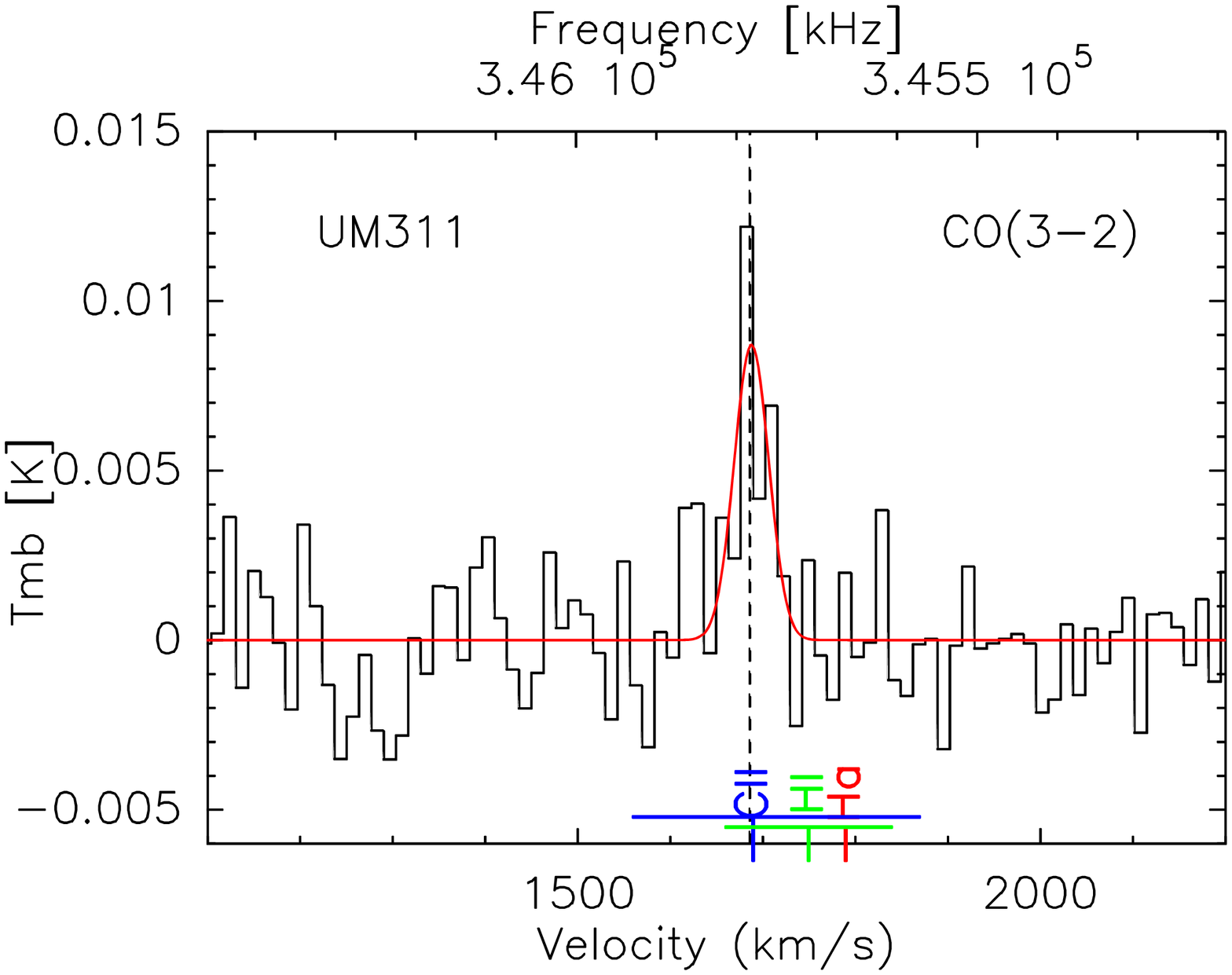}
\includegraphics[height=4.35cm,angle=-90]{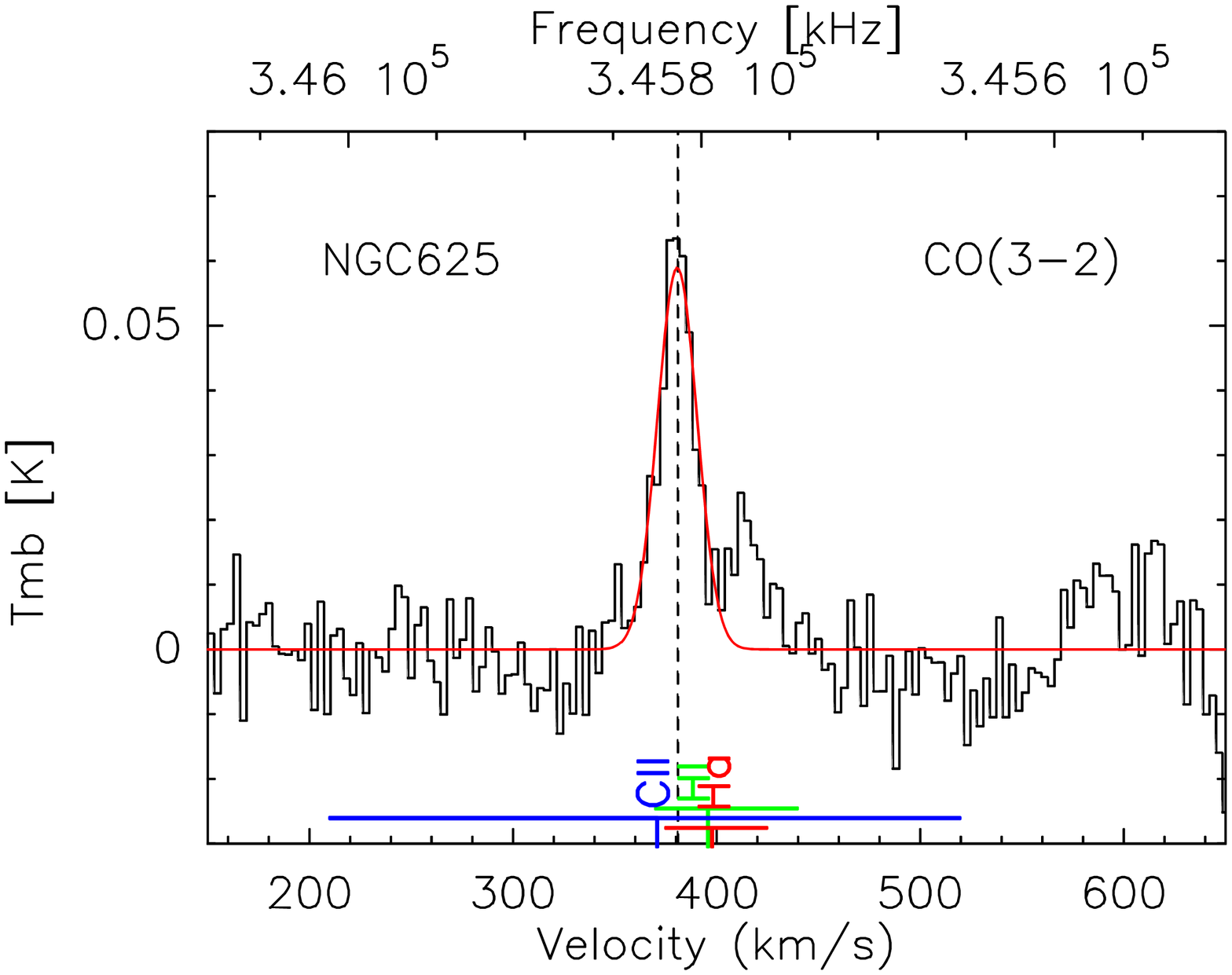}
\caption{
APEX/SHeFI spectra ($T_{mb}$) of the CO(2-1) ({\it left}) and 
CO(3-2) ({\it right}) lines at 230~GHz and 345~GHz, 
in Haro\,11, Mrk\,1089, UM\,311, and NGC\,625 (from {\it top} to {\it bottom}). 
The spectra are rebinned to a common resolution of $\sim$10-15~km~s$^{-1}$ 
(and 3~km~s$^{-1}$ for NGC\,625). 
Central velocities of [C~{\sc ii}], H~{\sc i}, and H$\alpha$ and their 
dispersion are also indicated in blue, green, and red, respectively. 
}
\label{fig:apex_lines}
\end{minipage}
\end{figure}

%%%
The data reduction was performed with the 
\textsc{GILDAS}\footnote{\url{http://www.iram.fr/IRAMFR/GILDAS}} 
software \textsc{CLASS}. 
The spectra are all averaged together and smoothed to a resolution of 
$\rm{\sim10-15~km~s^{-1}}$ (3~km~s$^{-1}$ for NGC\,625). They are converted 
to main beam brightness temperature, $T_{mb} = T_a^* / \eta_{mb}$, 
with main beam efficiencies\footnote{\url{http://www.apex-telescope.org/telescope/efficiency/}} 
of 0.75 and 0.73 at 230 and 345~GHz (with typical uncertainties of 5-10\%). 
We apply a third order polynomial fit to the baseline and 
a gaussian fit to the line to derive the line intensity. 
A sinusoid baseline subtraction is also applied to NGC\,625. 
For total uncertainties, we add 20\% to the uncertainties on the line fits. 
We detect the CO(2-1) and CO(3-2) lines in all observations 
(Figure~\ref{fig:apex_lines} and Table~\ref{table:lines}).

%%%

%%%%%
\subsection{IRAM observations: CO(1-0) and CO(2-1) in northern hemisphere targets}
%%%%%
We observed the CO(1-0) and CO(2-1) lines in 
Mrk\,930 and NGC\,4861 with the IRAM 30-m telescope from July 19 to 24 2012.
Observations were done in wobbler switching mode, with a throw of 4$^{\prime}$ 
for NGC\,4861 and 1.5$^{\prime}$ for Mrk\,930, and consisted of single pointings 
toward the peak of the [C~{\sc ii}] emission. 
The beam sizes are $\sim$21$^{\prime\prime}$ at 115~GHz and 
$\sim$11$^{\prime\prime}$ at 230~GHz. 
We used the EMIR receivers E090 and E230 simultaneously, with 
bandwidth coverage 8~GHz each, and with the backends WILMA (2~MHz resolution) 
for CO(1-0) and FTS (200~kHz resolution) for CO(2-1).  
The total integration times (ON+OFF) were 455 and 407\,min for NGC\,4861 
and 538 and 548\,minutes for Mrk\,930, for CO(1-0) and CO(2-1) 
respectively. The average system temperatures were 362 and 408\,K 
for NGC\,4861 and 190 and 355\,K for Mrk\,930. 
The temperature of the system was calculated every 15\,minutes, pointings were 
done every 1\,h and focus every 2\,h on average.
For pointing and focus, we use the sources: Mercury, Mars, Uranus
1039+811, 2251+158, 2234+282, 2200+420, NGC7027, 
and for line calibrators: IRC+10216 and CRL2688.
The pointing accuracy is estimated to be $\sim$3$^{\prime\prime}$.

\begin{figure}[!ht]
\begin{minipage}{8.8cm}
\centering
\includegraphics[height=4.35cm,angle=-90]{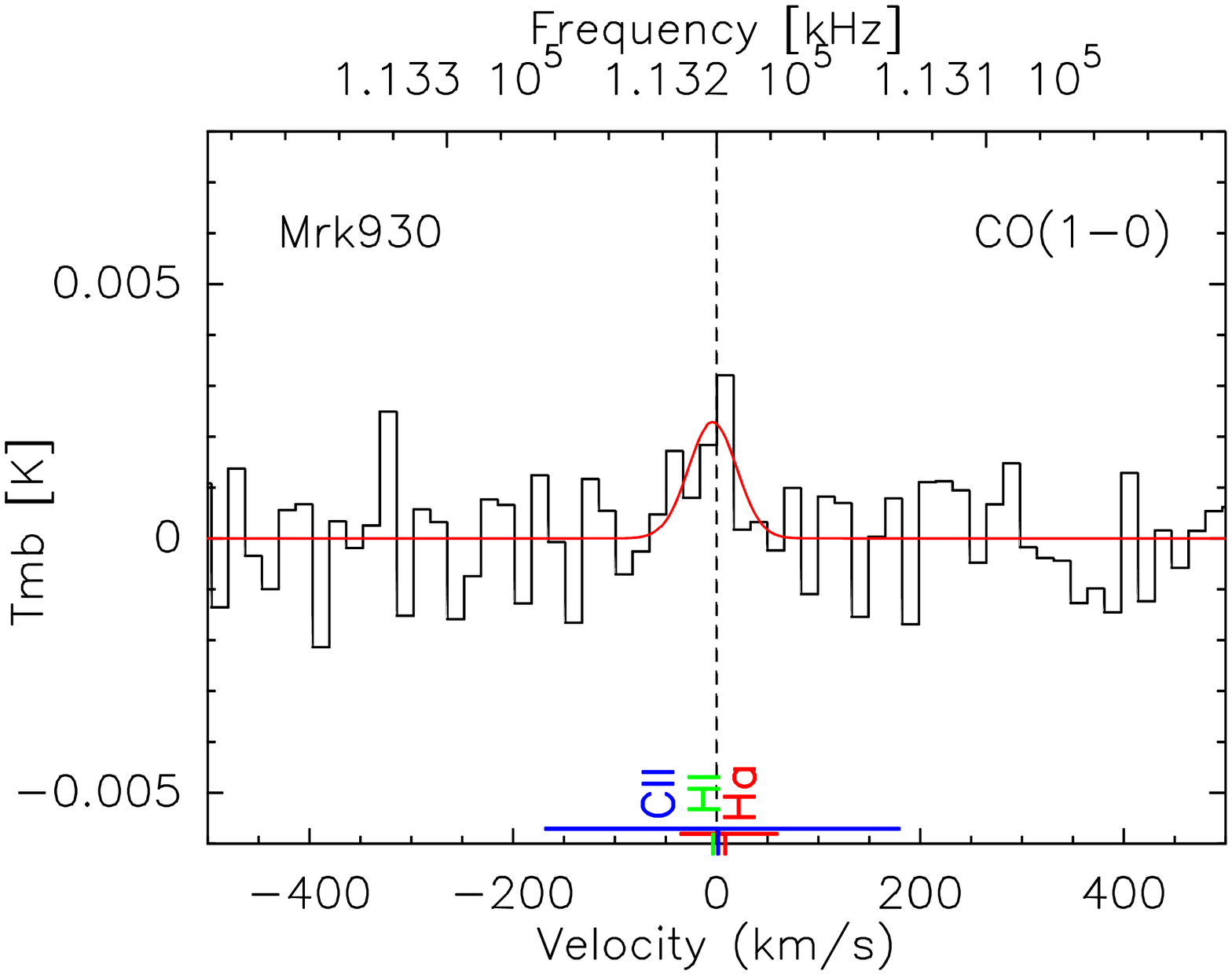}
\includegraphics[height=4.35cm,angle=-90]{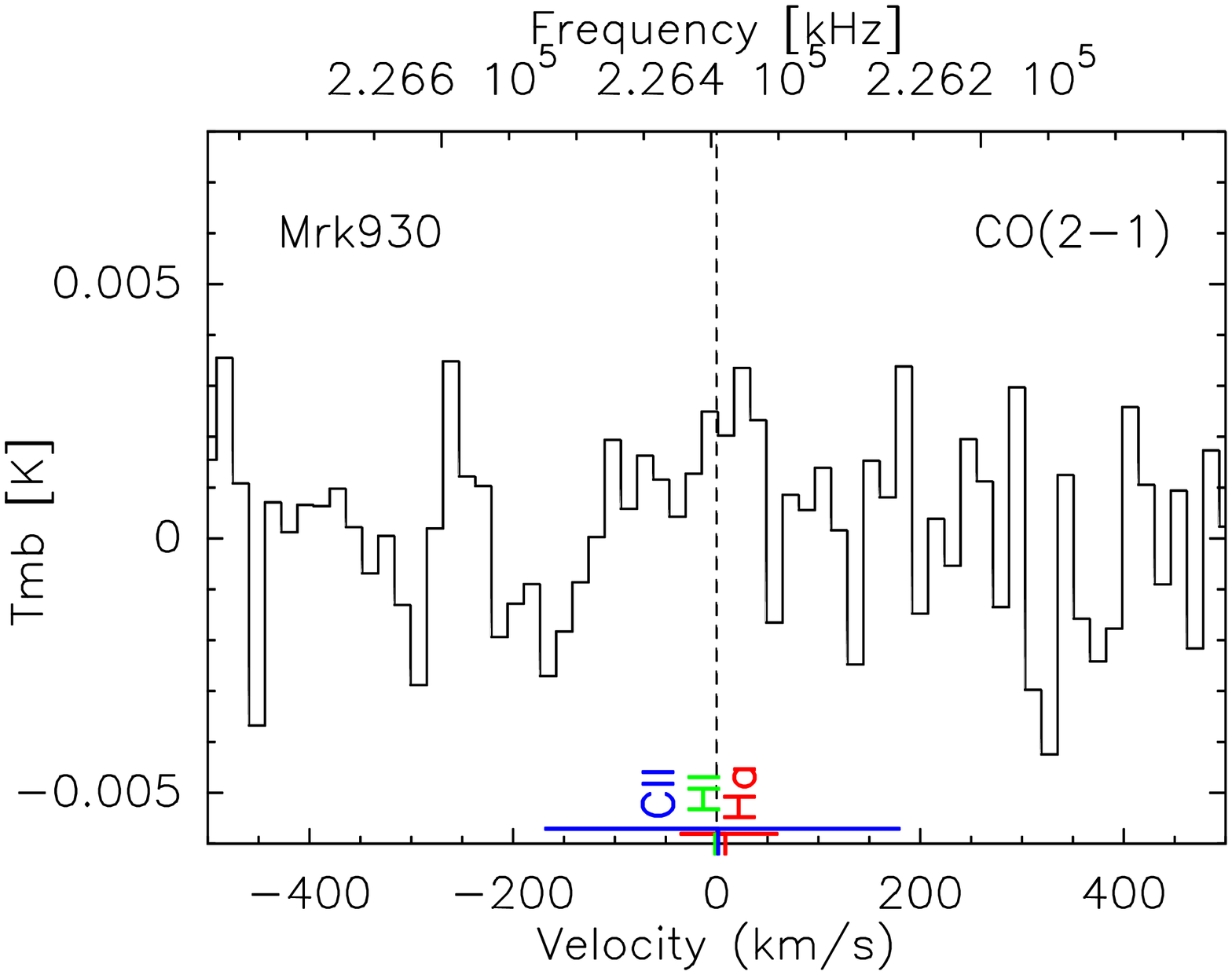}
\includegraphics[height=4.35cm,angle=-90]{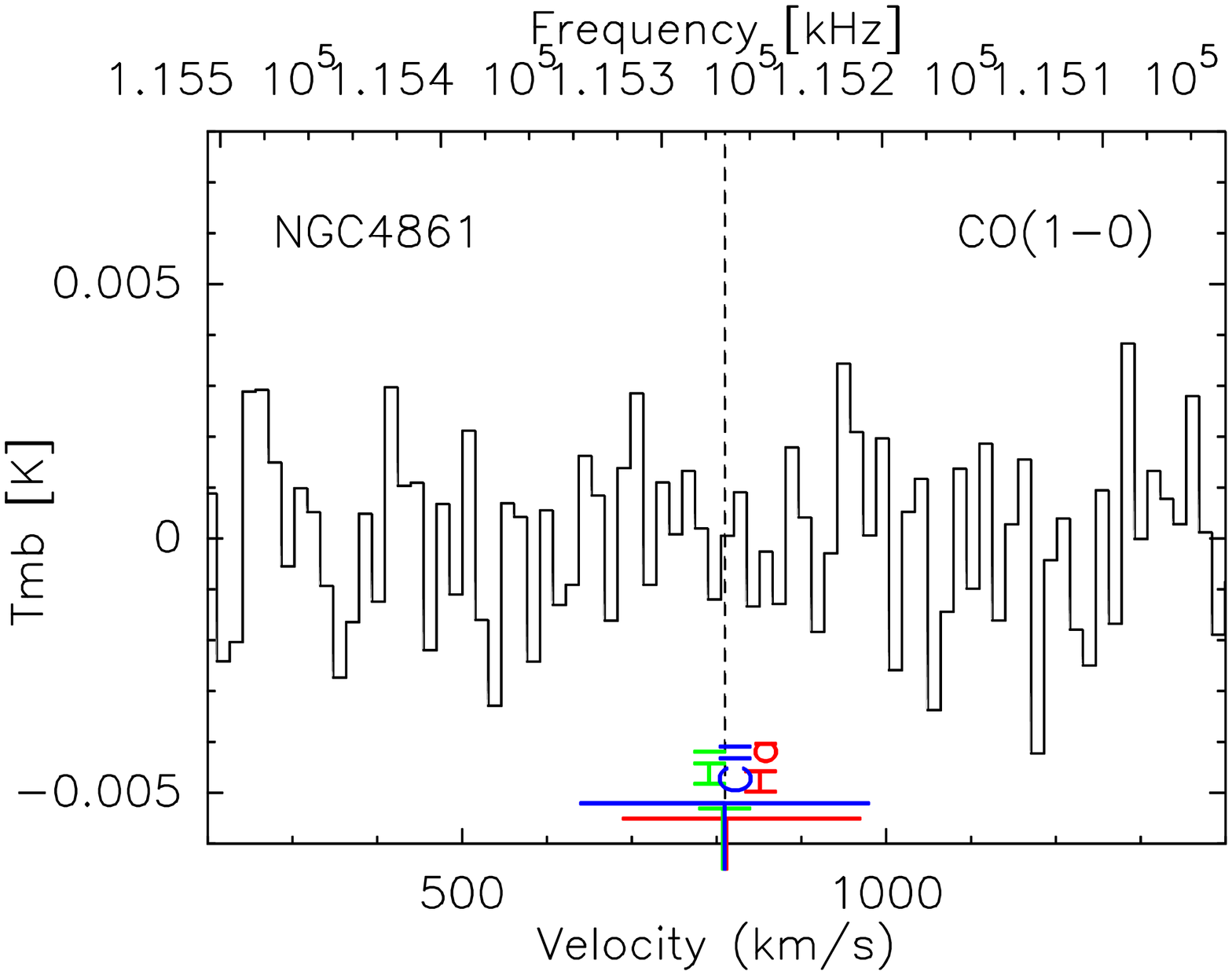}
\includegraphics[height=4.35cm,angle=-90]{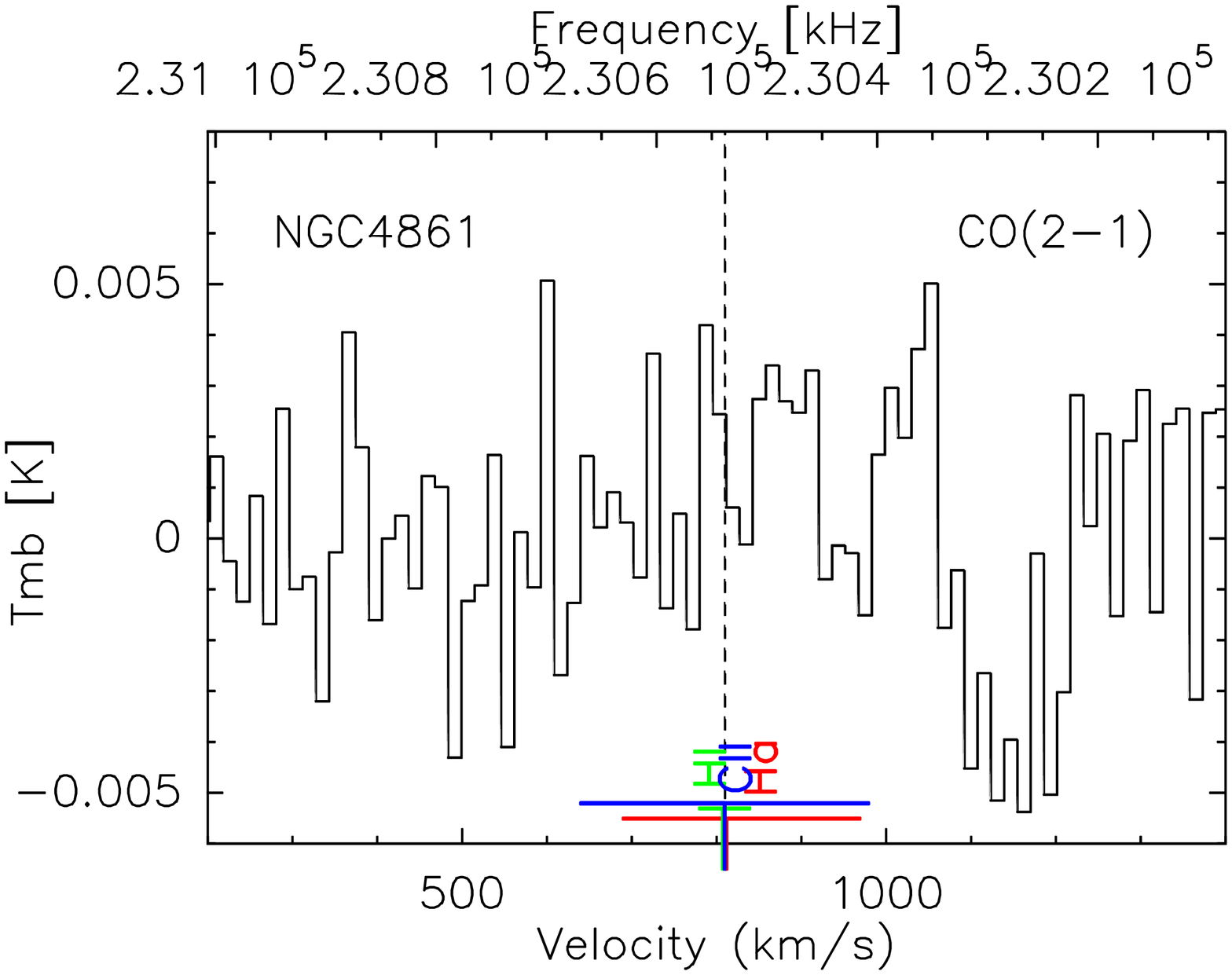}
\caption{
IRAM-30m spectra ($T_{mb}$) of the CO(1-0) ({\it left}) and 
CO(2-1) ({\it right}) lines at 115~GHz and 230~GHz, 
in Mrk\,930 ({\it top}) and NGC\,4861 ({\it bottom}). 
The spectra are rebinned to a common resolution of $\sim$15~km~s$^{-1}$. 
Central velocities of [C~{\sc ii}], H~{\sc i}, and H$\alpha$ and their 
dispersion are also indicated in blue, green, and red, respectively. 
}
\label{fig:iram_lines}
\end{minipage}
\end{figure}

%%%
The 30-m data are also reduced with \textsc{CLASS}. 
Antenna temperatures are corrected for main beam efficiencies 
$B_{eff}$\footnote{\url{http://www.iram.es/IRAMES/mainWiki/Iram30mEfficiencies}} 
of 0.78 and 0.60 at 115 and 230~GHz, using $T_{mb} = T_a^* \times F_{eff} / B_{eff}$. 
We have a marginal detection of the CO(1-0) line in Mrk\,930, 
while CO is not detected in the other observations (Figure~\ref{fig:iram_lines}), 
with rms values $\sim$2~mK. 
For total uncertainties, we add 20\% to the uncertainties on the line fits. 
The line parameters, intensities, and upper limits can be found in Table~\ref{table:lines}.

%%%

%%%%%
\subsection{{\it Herschel} data: [C~{\sc ii}] and [O~{\sc i}]}
\label{sect:herschel}
%%%%%
The [C~{\sc ii}]~157$\mu$m and [O~{\sc i}]~63$\mu$m fine-structure cooling lines 
were observed in the six galaxies with the {PACS} spectrometer \citep{poglitsch-2010} 
onboard the {\it Herschel} Space Observatory \citep{pilbratt-2010} as part of 
two Guaranteed Time Key Programs, the DGS \citep{madden-2013} 
and SHINING (P.I. Sturm). 
The results of the full {\it Herschel} spectroscopic survey in dwarf galaxies 
is presented in \cite{cormier-2014}.
The PACS array is composed of 5x5 spatial pixels 
of size 9.4$^{\prime\prime}$ each, covering a field-of-view of 
47$^{\prime\prime}$x47$^{\prime\prime}$. 
The [C~{\sc ii}] observations consist of a 5x2 raster map for NGC\,4861, and 
of 2x2 raster maps for the other galaxies. The [O~{\sc i}] observations consist 
of a 3x3 raster map for Mrk\,1089, 2x2 raster maps for Haro\,11 and UM\,311, 
and single pointings for Mrk\,930, NGC\,625, NGC\,4861. 
All observations were done in chop-nod mode
with a chop throw of 6$^{\prime}$ off the source (free of emission). 
The beam sizes are $\sim$9.5$^{\prime\prime}$ and 11.5$^{\prime\prime}$, and 
the spectral resolution $\sim$90~km~s$^{-1}$ and 240~km~s$^{-1}$ at 60$\mu$m 
and 160$\mu$m respectively (PACS Observer's Manual 2011).

%%%
The data were reduced with the {\it Herschel} Interactive Processing Environment 
User Release v9.1.0 \citep{ott-2010}, using standard scripts of the {PACS} 
spectrometer pipeline, and analyzed with the in-house 
software {\sc PACSman} v3.52 \citep{lebouteiller-2012}. 
For the line fitting, the signal from each spatial position of the PACS array is 
fit with a second order polynomial plus Gaussian for the baseline and line. 
The rasters are combined by drizzling to produce final maps of 3$^{\prime\prime}$ 
pixel size.

\begin{figure*}[!thp]
\begin{minipage}{18cm}
\centering
\includegraphics[clip,width=5.9cm,height=5.6cm]{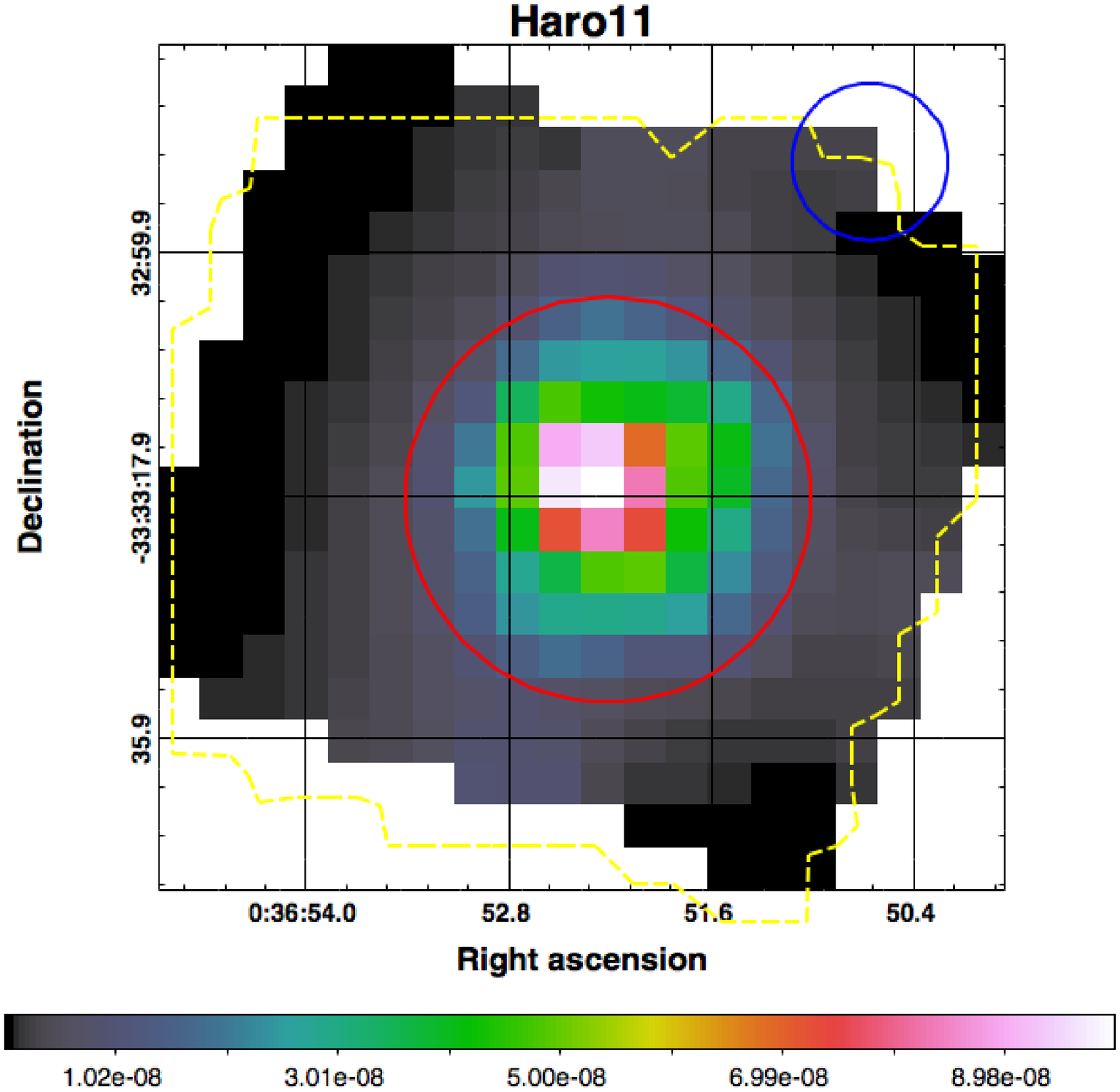}
\includegraphics[clip,width=5.9cm,height=5.6cm]{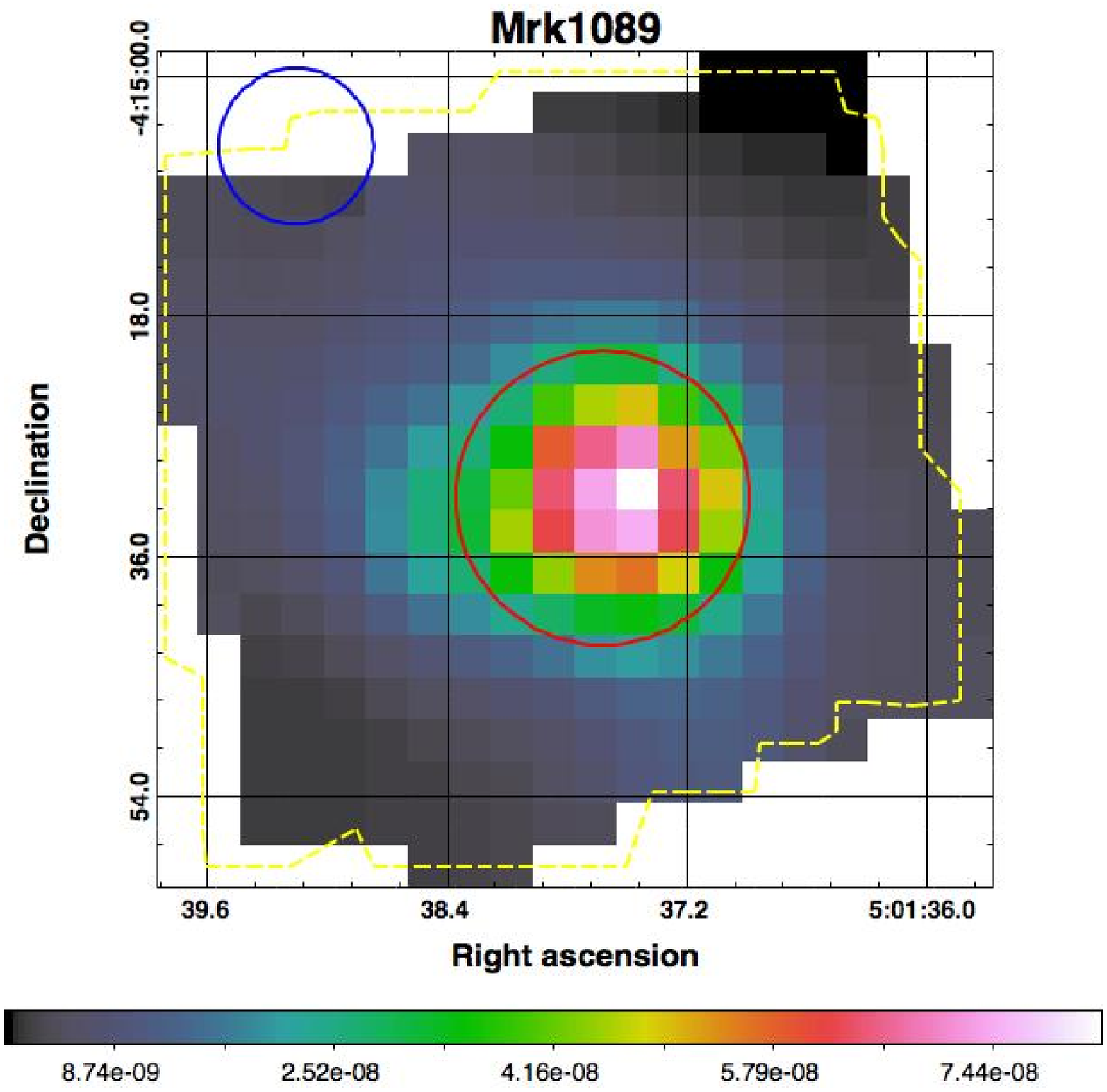}
\includegraphics[clip,width=5.9cm,height=5.6cm]{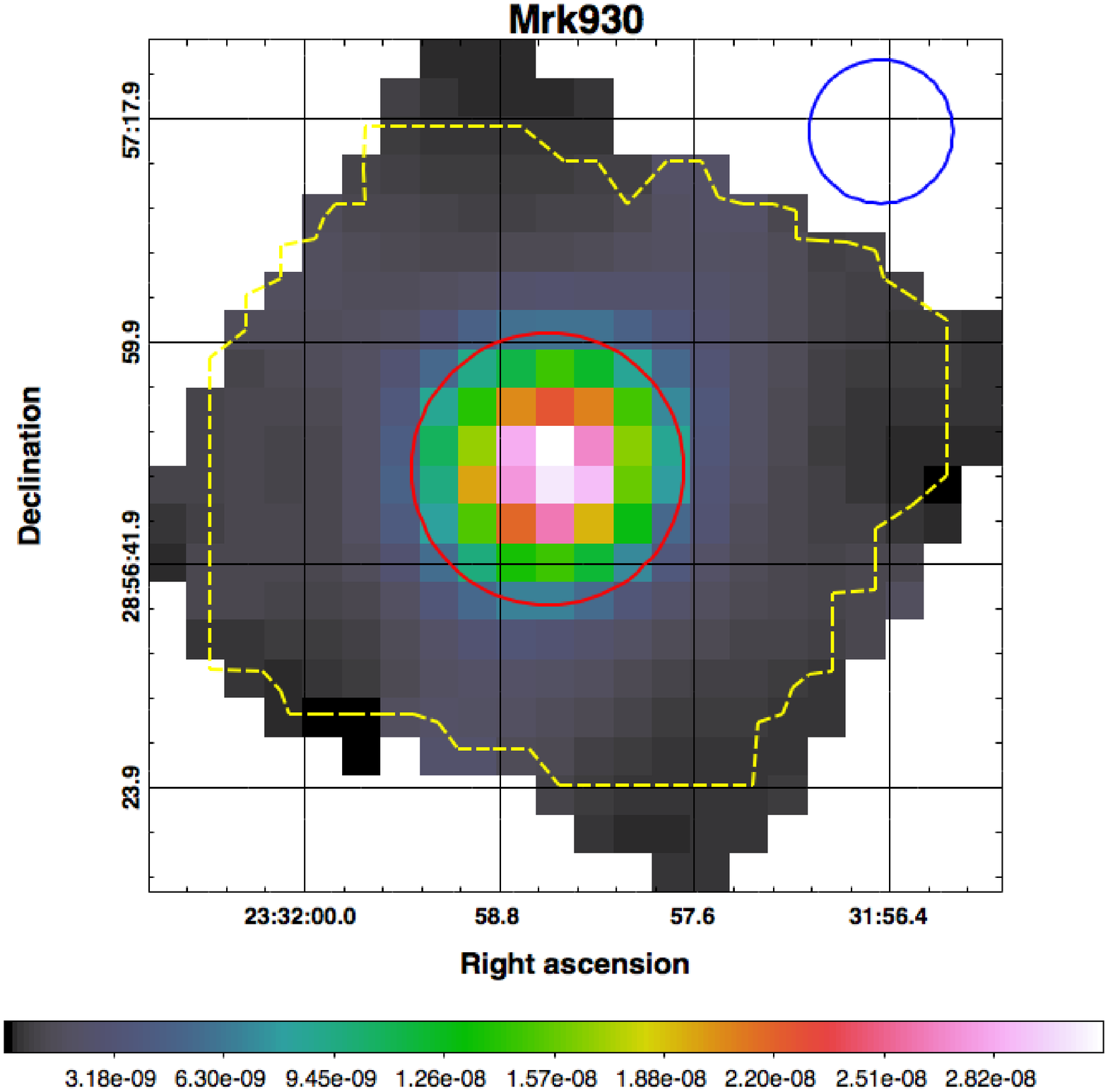}
\includegraphics[clip,width=5.9cm,height=5.6cm]{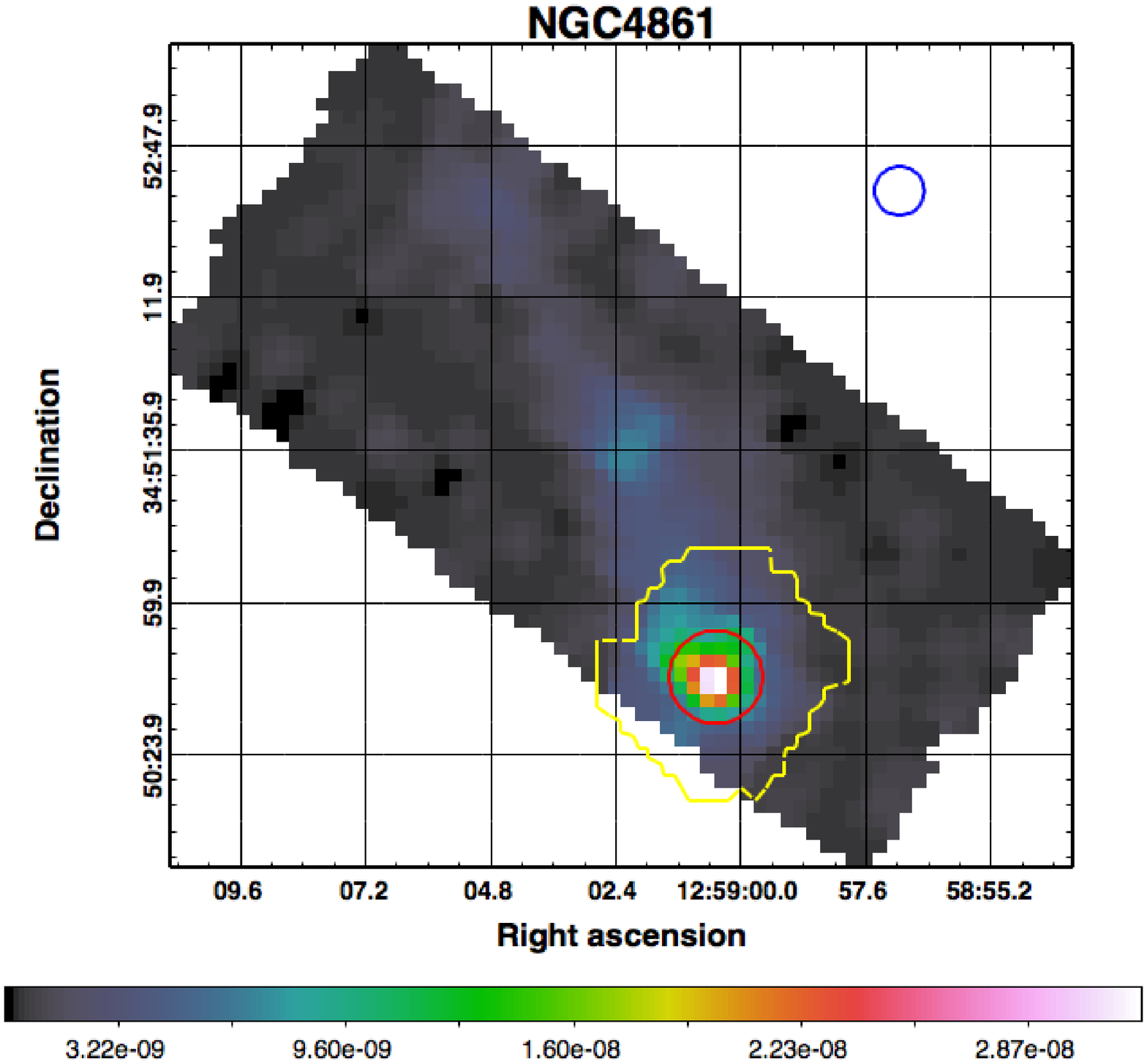}
\includegraphics[clip,width=5.9cm,height=5.6cm]{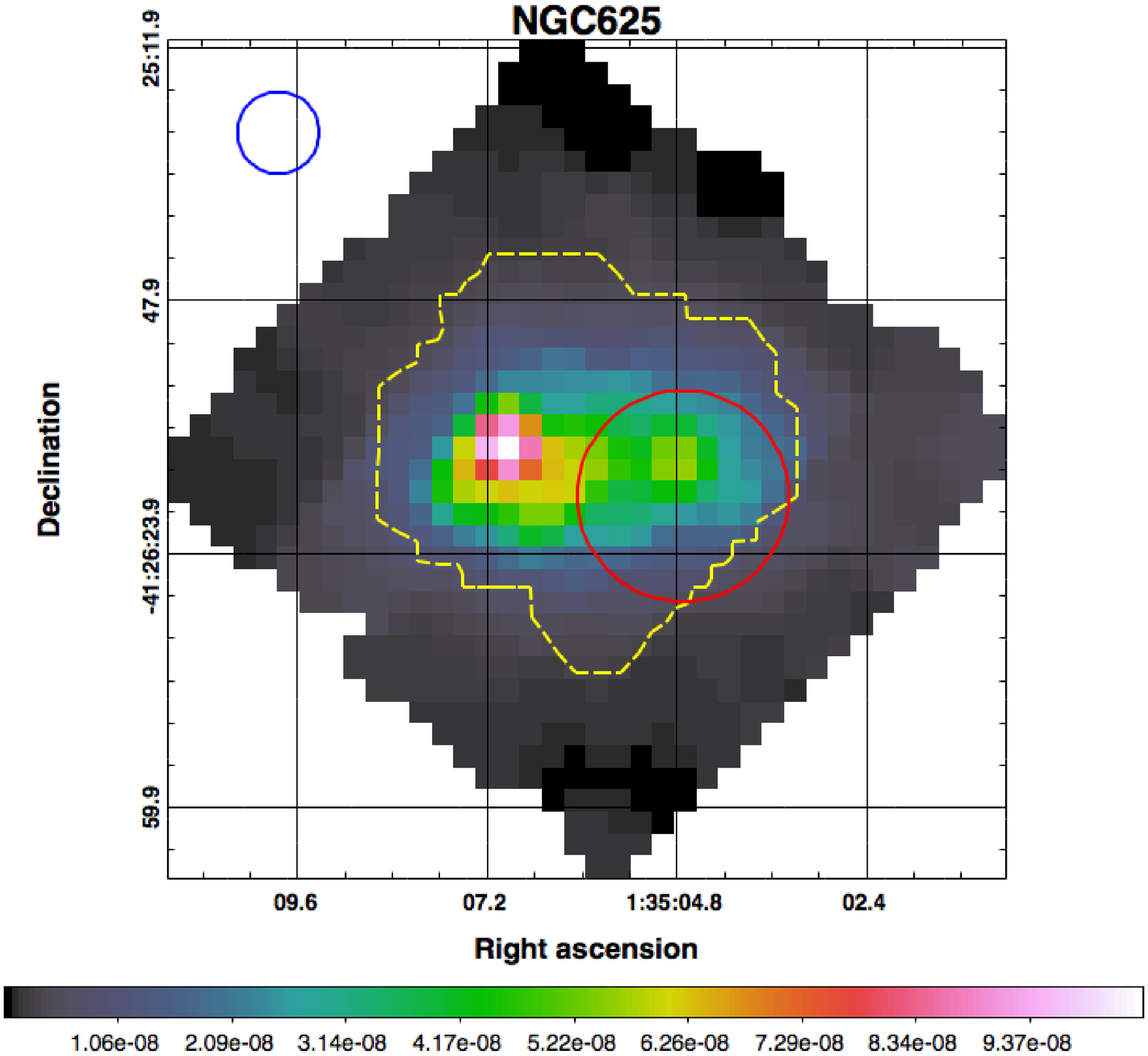}
\includegraphics[clip,width=5.9cm,height=5.6cm]{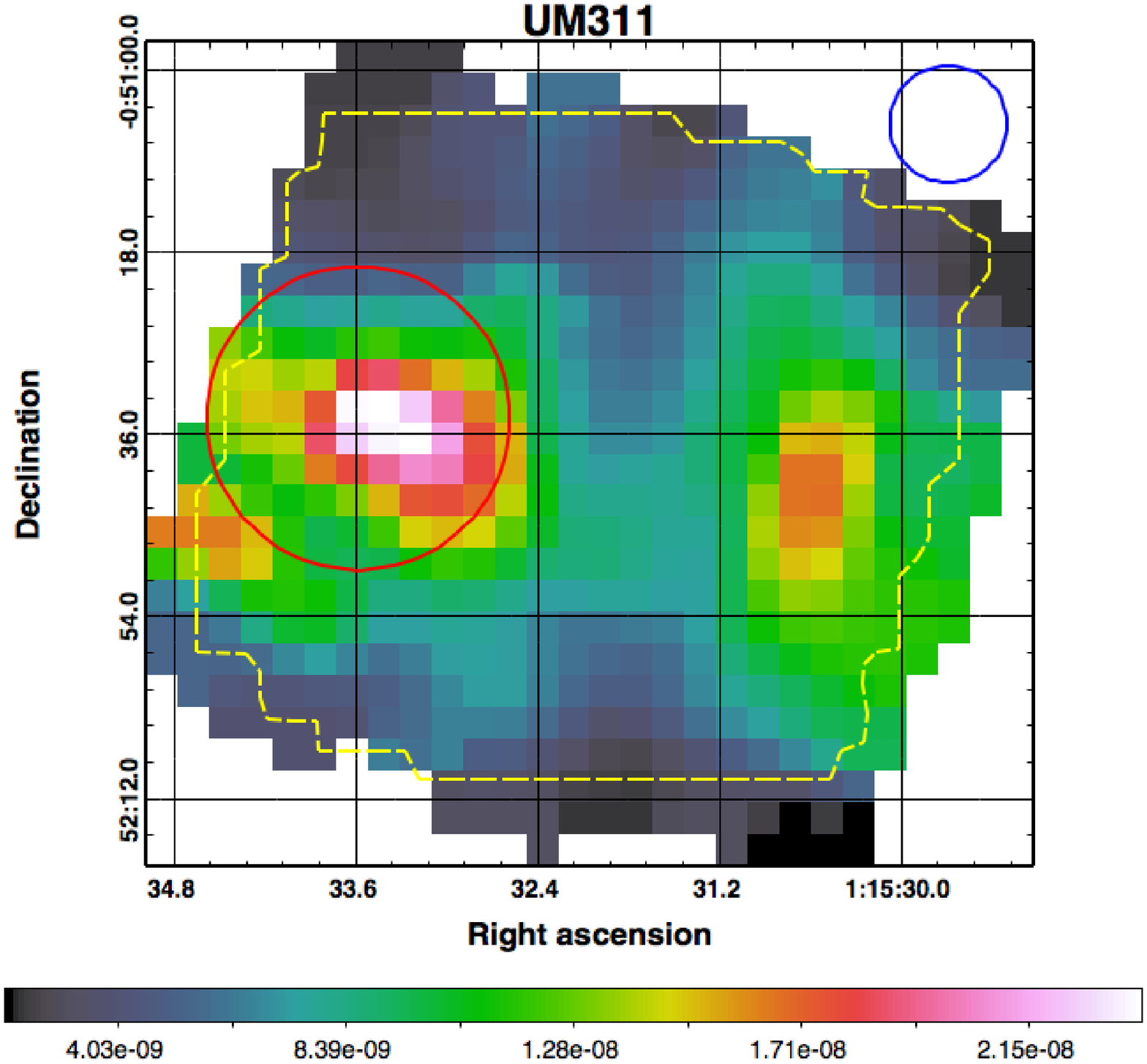}
\caption{
{\it Herschel}/PACS [C~{\sc ii}]~157$\mu$m intensity maps 
(units of W~m$^{-2}$~sr$^{-1}$). 
The PACS beam is indicated in blue ($\sim$11.5$^{\prime\prime}$) 
and the CO(1-0) aperture used to integrate the flux in red. 
For NGC\,625, the aperture is not centered on the [C~{\sc ii}] peak but 
on the same region covered by the CO(2-1) and CO(3-2) observations. 
The yellow contours show the area mapped in the [O~{\sc i}]~63$\mu$m 
line by PACS. 
}
\label{fig:pacsmap}
\end{minipage}
\end{figure*}

The line intensity maps of [C~{\sc ii}]~157$\mu$m are displayed 
in Figure~\ref{fig:pacsmap}. 
The [C~{\sc ii}] emission is extended in NGC\,625, NGC\,4861, 
and UM\,311, and marginally in Haro\,11, Mrk\,1089, and Mrk\,930. 
The [O~{\sc i}] emission is relatively more compact. 
The signal-to-noise is $\ge$15-$\sigma$ on the peak of emission 
for both [C~{\sc ii}] and [O~{\sc i}]. 
We measure line fluxes both from the area covered by the maps 
and in a circular aperture of size that of the CO(1-0) beam (Table~\ref{table:pacs}).
The uncertainties come from the noise and from fitting the spectra. 
Calibration errors add an additional $\sim$30\% systematic uncertainty 
\citep{poglitsch-2010} and thus dominate the total uncertainties.

%%%

%%%
\begin{center}
\begin{table*}[ht]
  \caption{{\it Herschel}/PACS [C~{\sc ii}]~157$\mu$m line observations and fluxes.}
  \hfill{}
  \begin{tabular}{l l l c c c c c c c}
    \hline\hline
    \vspace{-8pt}\\
    \multicolumn{1}{l}{Galaxy} & 
    \multicolumn{1}{l}{Program} & 
    \multicolumn{1}{l}{Line} & 
    \multicolumn{1}{c}{OD$^{(a)}$} & 
    \multicolumn{1}{c}{OBSID} & 
    \multicolumn{1}{c}{Map area} &
    \multicolumn{1}{c}{V} &
    \multicolumn{1}{c}{$\Delta$V$^{(b)}$} &
    \multicolumn{1}{c}{Total Flux} &
    \multicolumn{1}{c}{Flux in CO(1-0) area$^{(c)}$} \\ 
    \hline 
    \vspace{-8pt}\\
	Haro\,11 	& DGS		& [C~{\sc ii}] 	& 409	& 1342199236		& 51$^{\prime\prime}$x51$^{\prime\prime}$	& 6193	& 160	& 6.48 $\pm$ 0.02 	& 5.28 $\pm$ 0.01	\\
		 	& 			& [O~{\sc i}] 	& 409	& 1342199237		& 51$^{\prime\prime}$x51$^{\prime\prime}$	& 6184	& 140	& 6.45 $\pm$ 0.10 	& 5.34 $\pm$ 0.02	\\
	Mrk\,1089 & DGS		& [C~{\sc ii}] 	& 690	& 1342217859		& 51$^{\prime\prime}$x51$^{\prime\prime}$	& 4020	& -		& 8.84 $\pm$ 0.14	& 4.08 $\pm$ 0.01	\\
			 & 			& [O~{\sc i}] 	& 690	& 1342217861		& 53$^{\prime\prime}$x53$^{\prime\prime}$	& 4022	& 80		& 2.83 $\pm$ 0.06	& 1.48 $\pm$ 0.01	\\
	Mrk\,930 	& DGS		& [C~{\sc ii}] 	& 607 	& 1342212520		& 51$^{\prime\prime}$x51$^{\prime\prime}$	& 5486	& 100	& 2.31 $\pm$ 0.55	& 1.40 $\pm$ 0.01	\\
		 	& 			& [O~{\sc i}] 	& 607 	& 1342212518		& 47$^{\prime\prime}$x47$^{\prime\prime}$	& 5488	& 120	& 1.58 $\pm$ 0.10	& 0.89 $\pm$ 0.01	\\
	NGC\,4861 & SHINING	& [C~{\sc ii}] 	& 549	& 1342208902		& 1.4$^{\prime}$x3.6$^{\prime}$			& 801	& -		& 6.96 $\pm$ 0.26	& 1.39 $\pm$ 0.01	\\
			 & 			& [O~{\sc i}] 	& 745	& 1342221887		& 47$^{\prime\prime}$x47$^{\prime\prime}$	& 799	& 40		& 1.95 $\pm$ 0.07	& 1.23 $\pm$ 0.01	\\
	NGC\,625 & SHINING	& [C~{\sc ii}] 	& 754	& 1342222218		& 1.4$^{\prime}$x1.4$^{\prime}$			& 383	& -		& 16.9 $\pm$ 0.15	& 4.07 $\pm$ 0.01	\\
			 & 			& [O~{\sc i}] 	& 754	& 1342222217		& 47$^{\prime\prime}$x47$^{\prime\prime}$	& 388	& 100	& 4.85 $\pm$ 0.07	& 1.25 $\pm$ 0.10	\\
	UM\,311 	& DGS		& [C~{\sc ii}] 	& 621	& 1342213288		& 1.2$^{\prime}$x1.2$^{\prime}$			& 1698	& -		& 9.98 $\pm$ 0.17	& 2.23 $\pm$ 0.05	\\
		 	& 			& [O~{\sc i}] 	& 621	& 1342213291		& 1.0$^{\prime}$x1.0$^{\prime}$			& 1683	& 75		& 3.28 $\pm$ 0.14	& 1.28 $\pm$ 0.28	\\
    \hline \hline
  \end{tabular}
  \hfill{}
  \newline
    \vspace{-1pt}\\
    Fluxes are given in 10$^{-16}$~W~m$^{-2}$ and 
    velocities in km\,s$^{-1}$ in the LSR reference frame. 
    The uncertainties on the flux quoted here are those associated 
    with the line fitting and do not take into account the 30\% calibration errors. 
    $(a)$~Observation Day of the {\it Herschel} mission. 
    $(b)$~Intrinsic broadening of the line in km\,s$^{-1}$. 
    $(c)$~The aperture over which the fluxes 
    are integrated are shown in Figure~\ref{fig:pacsmap}. 
  \label{table:pacs}
\end{table*}
\end{center}

%%%%%
\subsection{{\it Spitzer}/IRS data: warm H$_{\rm 2}$}
\label{sect:spitzer}
%%%%%
All galaxies were observed with the Infrared Spectrograph \citep[IRS;][]{houck-2004} 
onboard the {\it Spitzer} Space Telescope \citep{werner-2004}, but 
H$_{\rm 2}$ rotational lines are clearly seen in the spectra of Haro\,11 and 
NGC\,625 only, hence we present the IRS data of these two galaxies. 
NGC\,625 was observed on the 11th of July 2005 (P.I. Gehrz; AORKey 5051904) 
in staring mode with the high-resolution modules SH ($\approx10-20~\mu$m; slit size 
$4.7^{\prime\prime} \times 11.3^{\prime\prime}$) and LH 
($\approx19-37~\mu$m; slit size $11.1^{\prime\prime} \times 22.3^{\prime\prime}$). 
The observation points toward the coordinates (RA, Dec)=(1h35m06.8s, -41d26m13s), 
where the [C~{\sc ii}] emission peaks. 
The processing of the Haro\,11 observations is described in \cite{cormier-2012}.

%%%
The data reduction and analysis was performed with SMART v8.2.2
\citep{lebouteiller-2010,higdon-2004}. 
IRS high-resolution observations are strongly affected 
by bad pixels in the detectors. Lacking the observation of an offset field 
for NGC\,625 (and Haro\,11), the quality of the spectra 
mainly relies on the pixel cleaning step (performed with 
IRSCLEAN\footnote{\url{http://irsa.ipac.caltech.edu/data/SPITZER/docs/dataanalysistools/tools/irsclean/}}) 
and on the comparison of the two spectra obtained at the two nod positions. 
The high-resolution spectra were obtained from the full-slit extraction, 
assuming both point-like and extended source calibrations, as NGC\,625 
is extended in the MIR bands while the H$_{\rm 2}$ emission seems to 
be dominated by one knot. 

\begin{figure*}[!thp]
\begin{minipage}{18cm}
\centering
\hspace{4.4cm}
\includegraphics[clip,trim=.7cm .3cm .3cm .5cm,width=4.4cm]{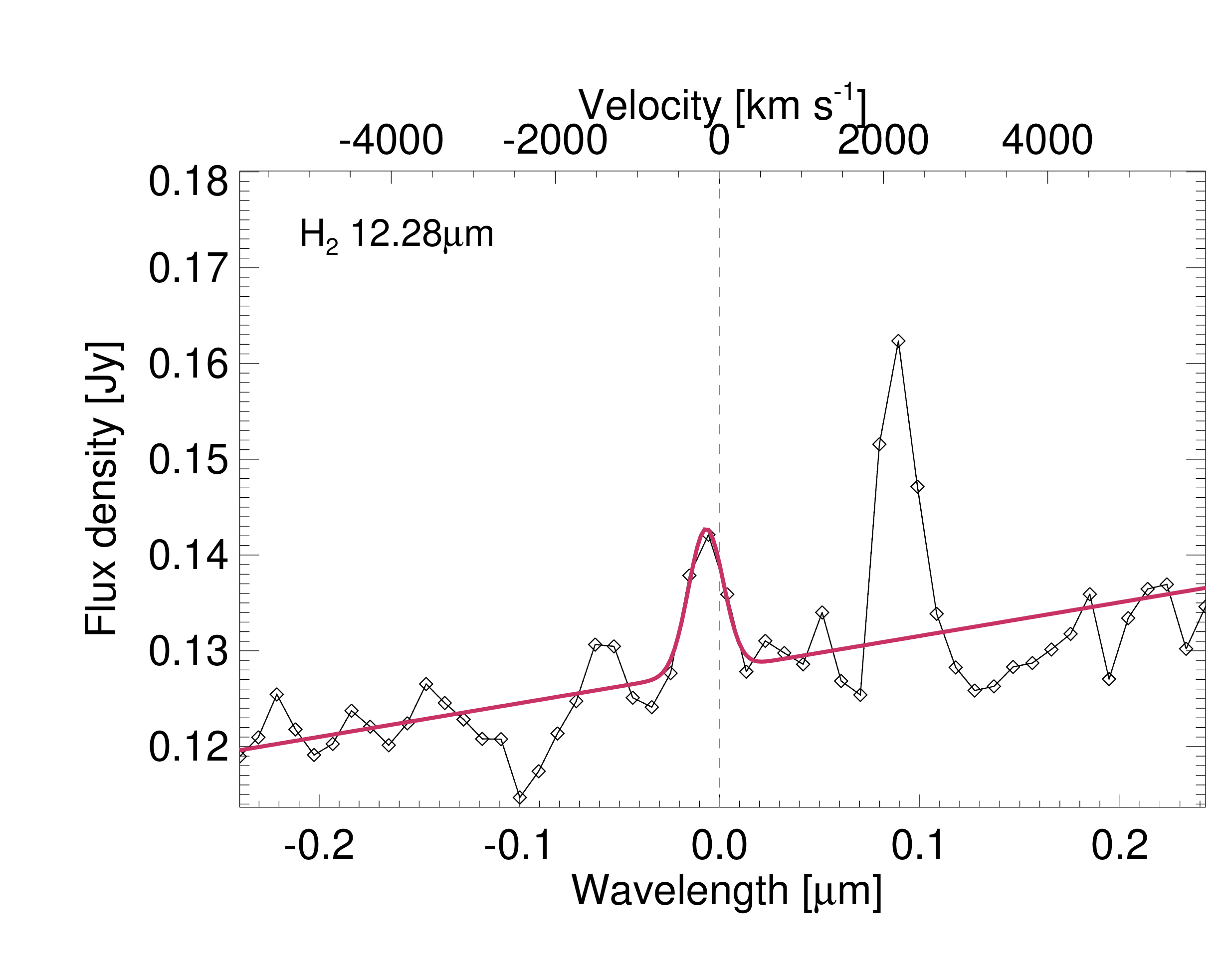}
\includegraphics[clip,trim=.7cm .3cm .3cm .5cm,width=4.4cm]{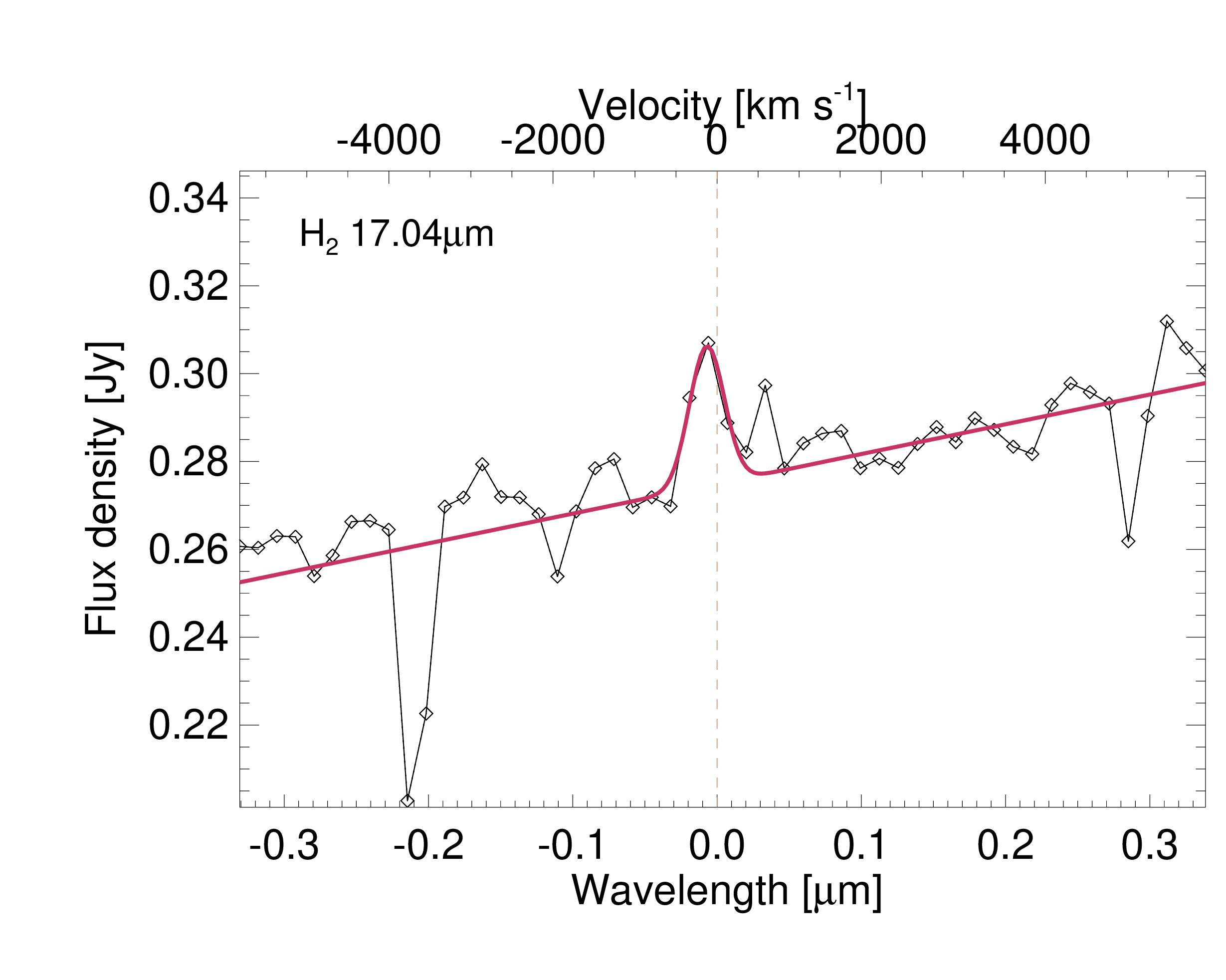}
\includegraphics[clip,trim=.7cm .3cm .3cm .5cm,width=4.4cm]{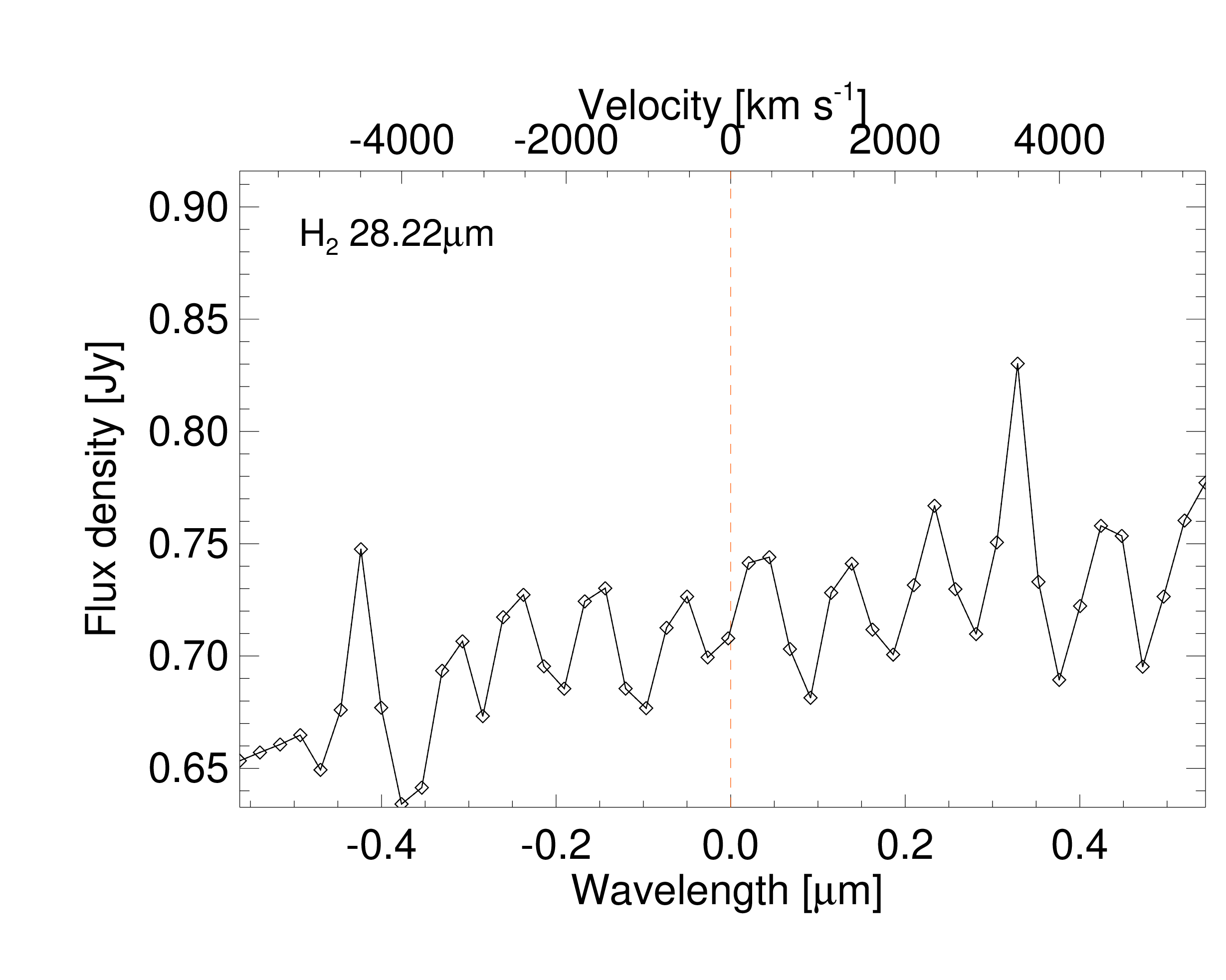}
\includegraphics[clip,trim=1cm .3cm 1cm .5cm,width=4.4cm]{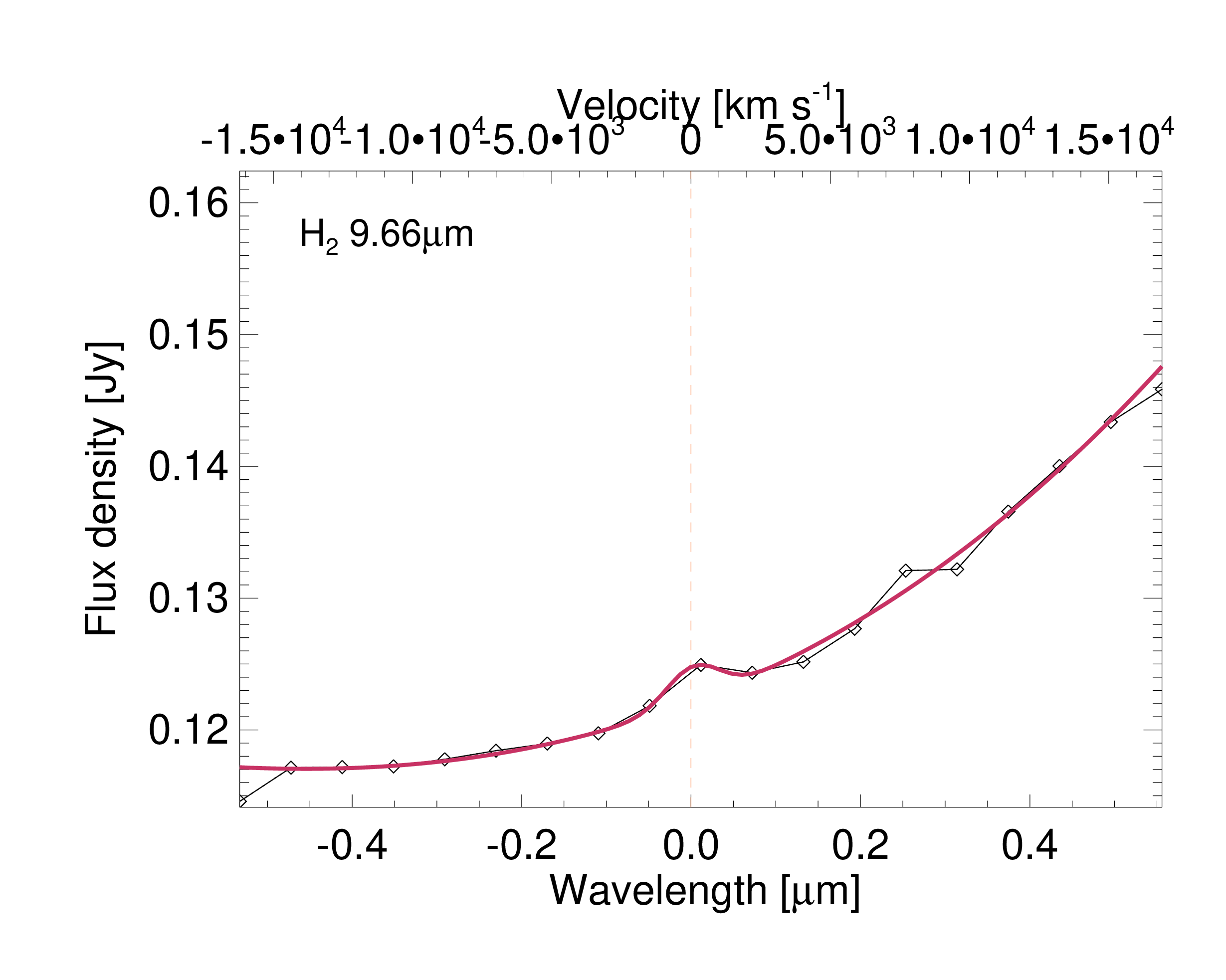}
\includegraphics[clip,trim=1cm .3cm 1cm .5cm,width=4.4cm]{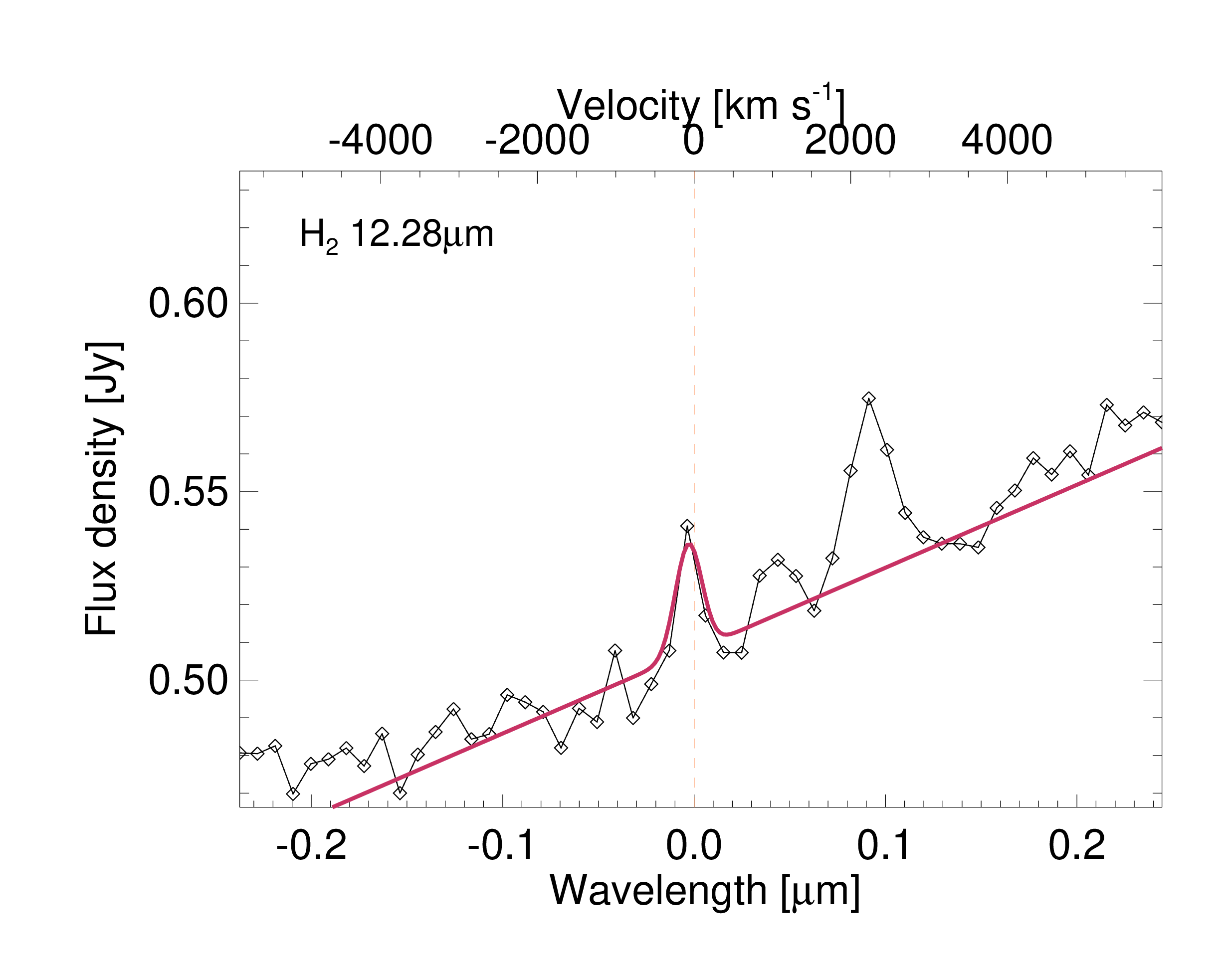}
\includegraphics[clip,trim=1cm .3cm 1cm .5cm,width=4.4cm]{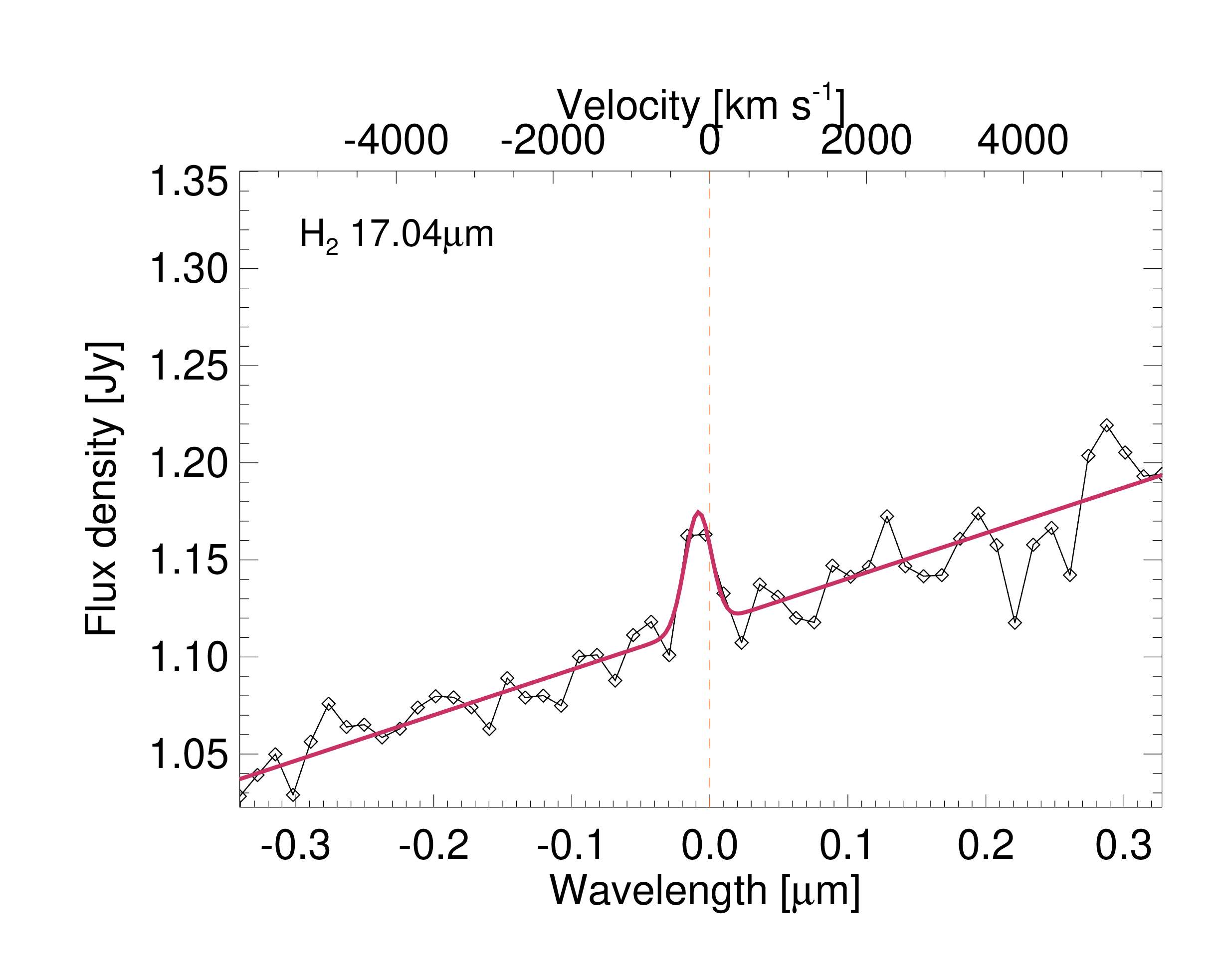}
\includegraphics[clip,trim=1cm .3cm 1cm .5cm,width=4.4cm]{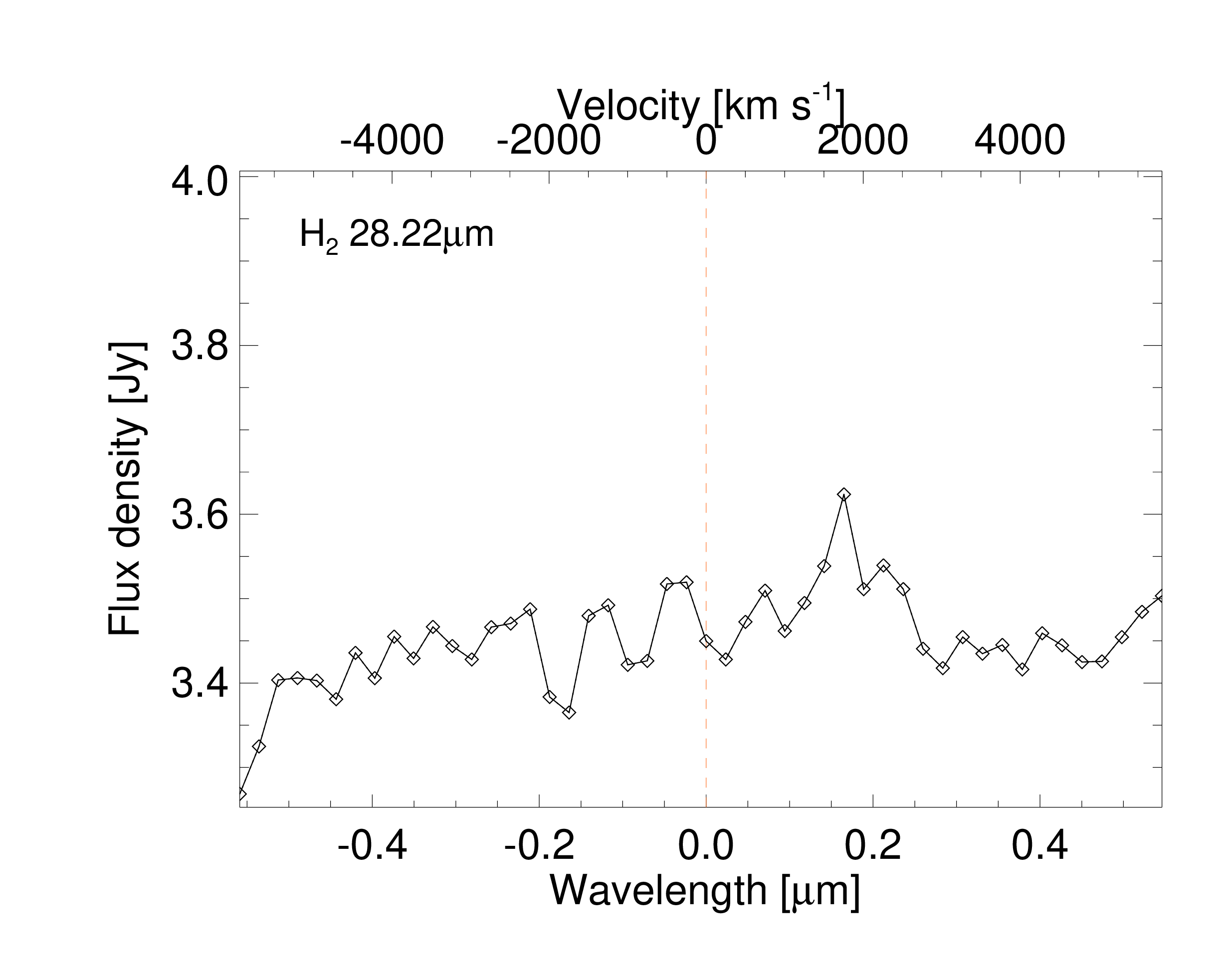}
\centering
\caption{
{\it Spitzer}/IRS spectra of the H$_{\rm 2}$ lines at 9.66, 12.28, 17.04, and 
28.22$\mu$m in NGC\,625 ({\it top} row) and Haro\,11 ({\it bottom} row). 
The red curve indicates the fit to the line when detected. 
}
\label{fig:h2irs}
\end{minipage}
\end{figure*}

We applied a first order polynomial function fit 
to the baseline and a Gaussian profile to fit the line. We also added 10\% of the 
line flux to the total uncertainties in order to take calibration uncertainties 
into account. 
The SH spectrum is multiplied by a factor 1.4 and 2 for the point-like and 
extended calibrations respectively to match the continuum level of the LH 
spectrum (larger aperture). 
Final fluxes are taken as the average of the fluxes obtained with the two 
nods and two different calibrations. 
The IRS observation of NGC\,625 points toward the main starburst region 
where the [C~{\sc ii}] and IR peaks are located, east of the location of the 
CO(2-1) and CO(3-2) pointings (see the IRS footprint on Figure~\ref{fig:cobeam}). 
The IRS slit captures most, but probably not all, of the H$_{\rm 2}$ emission  
since NGC\,625 is extended in the {\it Spitzer} and {\it Herschel} broad bands. 
To estimate the total flux (over the entire galaxy), we scale the H$_{\rm 2}$ emission 
to that of the total 24$\mu$m {\it Spitzer}/MIPS emission (\citealt{bendo-2012}), 
multiplying all line fluxes by another factor 1.5. 
Individual spectra of the H$_{\rm 2}$ lines are displayed in Figure~\ref{fig:h2irs}, 
and the line fluxes are given in Table~\ref{table:h2flux}. 
For Haro\,11, we use the fluxes from \cite{cormier-2012} but re-estimate a more 
robust upper limit on the S(0) line by considering $3\times rms \times FWHM_{inst} / sqrt(nres)$, 
where $nres$ is the number of resolution elements in the instrumental FWHM.

%%%

The H$\rm{_{2}}$ S(1) 17.03$\mu$m and S(2) 12.28$\mu$m lines are detected 
in NGC\,625 (Table~\ref{table:h2flux}), while we measure only an upper limit for the 
S(0) 28.22$\mu$m line, and the S(3) 9.66$\mu$m line falls outside of the observed 
wavelength range. The total luminosity of the two detected H$\rm{_{2}}$ lines is 
$\rm{L_{H_2} = 1.2 \times 10^{4}~L_{\odot}}$ for NGC\,625. 
In the IRS spectrum of Haro\,11, the S(1), S(2), and S(3) lines are detected, 
while the S(0) line is undetected. The total luminosity of the three detected lines 
is $\rm{L_{H_2} = 8.6 \times 10^{6}~L_{\odot}}$.

%%%
\begin{center}
\begin{table}[t!]
  \caption{{\it Spitzer}/IRS H$_{\rm 2}$ observations and fluxes.}
  \hfill{}
  \begin{tabular}{l l c c c}
    \hline\hline
    \vspace{-8pt}\\
    \multicolumn{2}{l}{Galaxy} & 
    \multicolumn{1}{c}{NGC\,625} & 
    \multicolumn{1}{c}{NGC\,625} & 
    \multicolumn{1}{c}{Haro\,11$^{(a)}$} \\ 
    \multicolumn{2}{l}{Observing mode} & 
    \multicolumn{1}{c}{staring} &    
    \multicolumn{1}{c}{{\it total}~$^{(b)}$} &    
    \multicolumn{1}{c}{staring} \\
    \multicolumn{2}{l}{AORKey} & 
    \multicolumn{1}{c}{5051904} & 
    \multicolumn{1}{c}{} &    
    \multicolumn{1}{c}{9007104} \\ \cline{1-5}
    \vspace{-8pt}\\
    \multicolumn{2}{l}{Line flux:} &
    \multicolumn{1}{c}{} &
    \multicolumn{1}{c}{} &
    \multicolumn{1}{c}{} \\
	\hspace{.2cm} & \multicolumn{1}{r}{H{\scriptsize 2} S(0) 28.22$\mu$m}	& $<$ 18.4 & $<$ 27.6 & $<$ 24.9 \\ %61.8 \\
	 & \multicolumn{1}{r}{H{\scriptsize 2} S(1) 17.03$\mu$m}	& 10.8$\pm$2.6 & 16.2$\pm$3.8 	& 16.8$\pm$5.5 \\
	 & \multicolumn{1}{r}{H{\scriptsize 2} S(2) 12.28$\mu$m}	& 6.39$\pm$1.40 & 9.58$\pm$2.50	& 10.1$\pm$1.3 \\
	 & \multicolumn{1}{r}{H{\scriptsize 2} S(3) 9.66$\mu$m~~~}	& - & - & 12.1$\pm$1.8  \\
    \hline \hline
  \end{tabular}
  \hfill{}
  \newline
    \vspace{-1pt}\\
    Fluxes are in 10$^{-18}$~W~m$^{-2}$ and upper limits are given at a 3-$\sigma$ level. 
    $(a)$~estimated total H$_{\rm 2}$ emission from the galaxy by scaling each 
    flux measured in the IRS slit to the 24$\mu$m emission of the galaxy. 
  \label{table:h2flux}
\end{table}
\end{center}

%%%%%
\section{Analysis of the CO observations}
%%%%%
%%%
\subsection{Line profiles and velocities}
\label{sect:lineprofil}
%%%
We show the observed CO line profiles in Figures~\ref{fig:mopra_spec} to \ref{fig:iram_lines}. 
For comparison, we indicate below each CO spectrum the line centers of other relevant tracers: 
H$\alpha$, H~{\sc i}, and [C~{\sc ii}], 
as well as the velocity dispersion, when known. 
This dispersion measure includes broadening due to line fitting uncertainties 
(typically a few \kms) and the instrumental line width (240~\kms for [C~{\sc ii}] 
with {\it Herschel}/PACS), which could blend several emission components. 
In those cases where mapped observations are available from the literature, 
we include also the range of observed peak positions across 
the source in the dispersion measure. 
Typical errors on the CO line centers are of 5-10~\kms. 
{
H$\alpha$ velocity information is taken from the following references: 
\cite{james-2013} for Haro\,11, \cite{rubin-1990} for Mrk\,1089, 
\cite{perez-montero-2011} for Mrk\,930, \cite{van-eymeren-2009} for NGC\,4861, 
\cite{marlowe-1997} for NGC\,625, and \cite{terlevich-1991} for UM\,311. 
H~{\sc i} velocity information is taken from: 
\cite{machattie-2013} for Haro\,11, \cite{williams-1991} for Mrk\,1089, 
\cite{thuan-1999} for Mrk\,930, \cite{van-eymeren-2009} for NGC\,4861, 
\cite{cannon-2004} for NGC\,625, and \cite{smoker-2000} for UM\,311.
}

For Haro\,11, we observe shifts in the velocity centers of the different tracers 
which could be due to the fact that the peak emission regions for atoms/ions 
and molecules are not co-spatial. 
\cite{sandberg-2013} find a blueshift of $\sim 44$~km\,s$^{-1}$ of the 
neutral gas relative to the ionized gas toward knot~B, and a redshift of 
$\sim 32$~km\,s$^{-1}$ toward knot~C. This indicates that the bulk of 
the cold gas emission could be associated with knot~B. 
Interestingly, the PACS lines display rotation along the north-south axis 
and have broad profiles. 
The full width at half maximum (FWHM) of the FIR lines 
are larger than the PACS instrumental line widths. 
Subtracting {quadratically} the instrumental line width 
(240~km~s$^{-1}$ for [C~{\sc ii}] and 90~km~s$^{-1}$ for [O~{\sc i}]~63$\mu$m), 
the intrinsic line width of [C~{\sc ii}] is $\sim$160~km~s$^{-1}$, 
while those of [O~{\sc iii}] and [O~{\sc i}] are 250 and 140~km~s$^{-1}$ 
respectively, which is much larger than the CO line width 
of $\sim$60~km~s$^{-1}$. The [O~{\sc iii}] broadening agrees with that 
of H$\alpha$, found as large as $\sim$280~km~s$^{-1}$ by \cite{james-2013}. 
This confirms the fact that the emission of the FIR lines arises 
from a larger region than CO. The FIR lines are also too broad to be 
solely due to rotation and probably trace the presence of outflows.

For UM\,311, important velocity shifts are observed for the different tracers, 
which are probably due to the crowding of the region, while for the other galaxies, 
CO and the other tracers peak at the same velocity overall. 
Mrk\,1089 has broad CO profiles and the signal-to-noise is high enough 
in the APEX CO(2-1) and CO(3-2) spectra to clearly distinguish 
two velocity components, also present in the data from \cite{leon-1998}. 
The spectral resolution of PACS does not allow us to resolve the separate 
narrow components seen in the CO profiles.  
Our [C~{\sc ii}] velocity map displays rotation, with red-shifted emission 
east of the nucleus and blue-shifted emission west of the nucleus. 
This is in agreement with the velocity analysis of H~{\sc i} and H$\alpha$, although 
the velocity fields are more irregular in the HCG\,31 complex as a whole. 
In NGC\,625, the CO(1-0) spectra have a single CO component of width 
$\sim$70~km~s$^{-1}$. This emission may not be co-spatial with that seen 
in the ancillary CO(2-1) and CO(3-2) spectra (see section~\ref{sect:mopra}), 
which are narrower (Table~\ref{table:lines}). 
While H$\alpha$ and [C~{\sc ii}] peak spatially where the current star formation 
episode is taking place and where we detect CO with Mopra, H~{\sc i} peaks 
$\sim$300~pc east \citep{cannon-2004}, where the second peak of [C~{\sc ii}] 
is located and where the APEX and JCMT observations point.

To summarize, the CO line detections are consistent with the velocities 
as obtained from H$\alpha$, H~{\sc i}, and [C~{\sc ii}]. Moreover, for a given 
galaxy, the different CO transitions exhibit the same line center and width 
(except CO(1-0) in NGC\,625) within the uncertainties, showing that they 
originate from the same component. 
This allows us to study CO line ratios (section~\ref{sect:ratios}).

%%%
\subsection{Estimating the total CO emission}
\label{sect:comiss}
%%%
One of the complications in analyzing the various CO observations 
is the difference in beam sizes used for each transition. 
A larger beam may encompass more molecular clouds 
than probed by a smaller beam and thus hamper the 
comparison of the transitions with each other. 
A further complication stems from the possibility of CO emission 
outside the observed area if the galaxy is more extended than 
any of the beams, which influences the results 
when comparing CO properties to global properties (e.g. H~{\sc i} mass).  
Here we investigate how using different beam 
sizes affects the derived intensities. 

%%%
\begin{figure*}
\centering
\includegraphics[clip, trim=.3cm 0 .3cm .5cm,width=6.2cm]{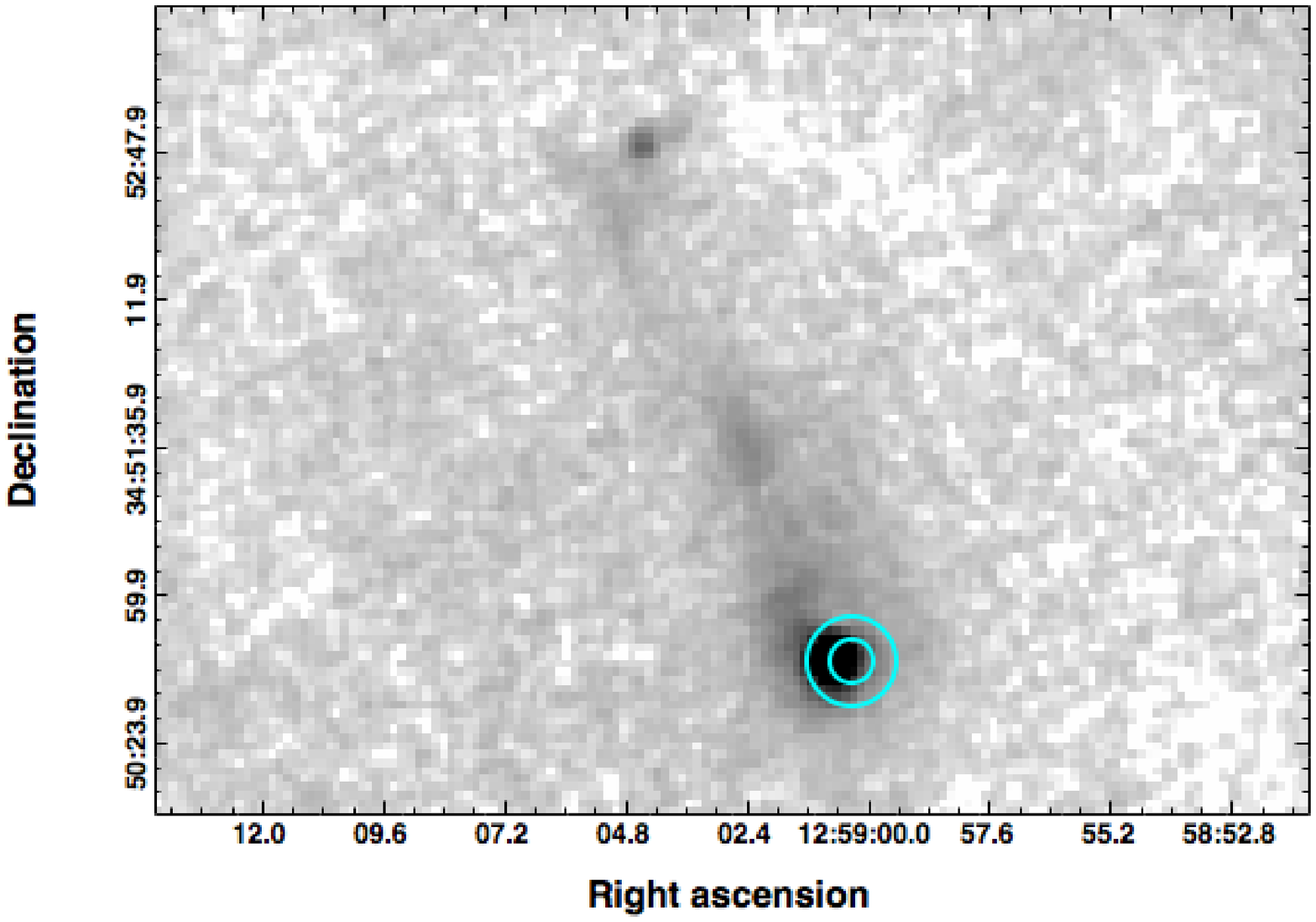}
\includegraphics[clip, trim=.6cm 0 2.2cm .5cm,width=5.8cm]{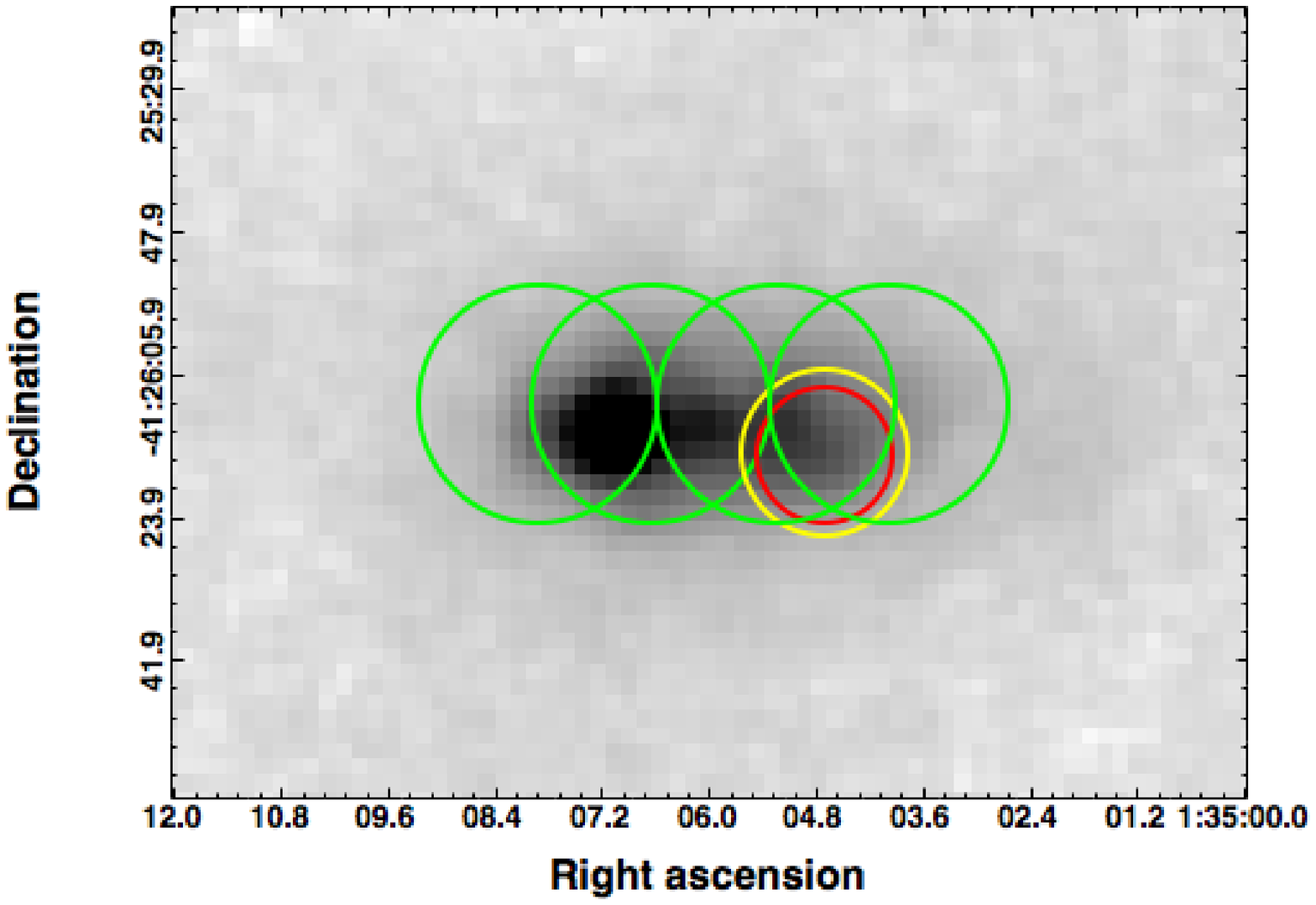}
\includegraphics[clip, trim=.6cm 0 .3cm 1cm,width=6.2cm]{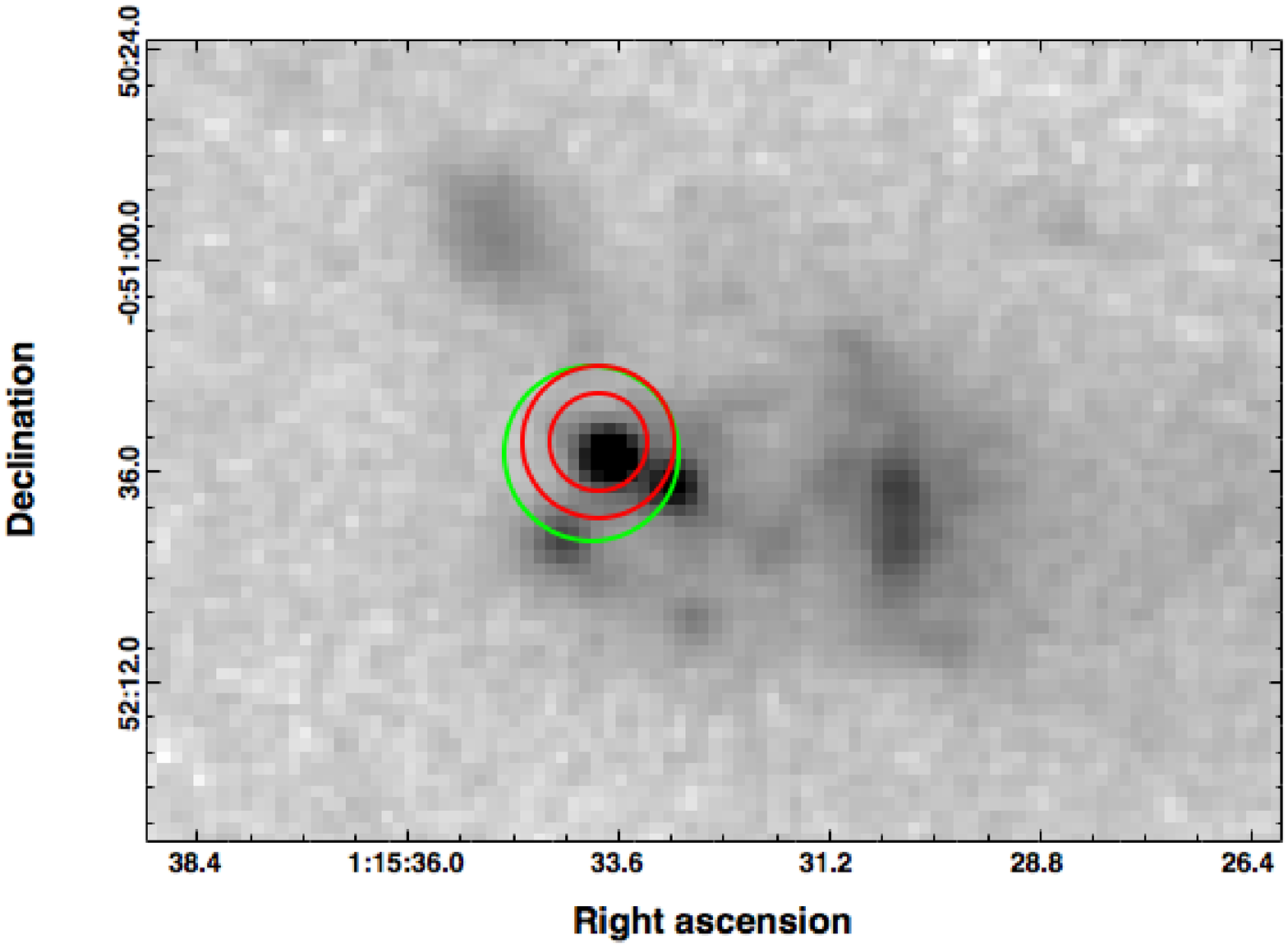}
\caption{
{\it Herschel} PACS 100$\mu$m maps of NGC\,4861 ({\it left}), 
NGC\,625 ({\it middle}), and UM\,311 ({\it right}), with the  
CO beams (Mopra in green, APEX in red, IRAM in blue, and the 
{\it JCMT} in yellow; see Fig.~\ref{fig:cobeam} for details). 
}
\label{fig:dustmaps}
\end{figure*}

%%%
We have at our disposal a complete coverage of the 100$\mu$m 
dust emission from {\it Herschel} for each galaxy 
\citep[Figure~\ref{fig:dustmaps};][]{remy-2013a}. 
If we assume that the distribution of the CO emission follows that 
of the 100$\mu$m dust continuum, we can coarsely estimate 
the fraction of CO emission measured in each CO beam relative to 
the total CO emission. 
In principle, one should first subtract the H~{\sc i} column density 
for such approach. Here, we just assume a flat H~{\sc i} distribution. 
The spatial correlation between CO and the dust may not follow 
the same distribution on small scales in the Galaxy, but on larger scales 
in external galaxies, observations show a correlation between CO 
and IR emission \citep[e.g.][]{tacconi-1987,sanders-1991}. 
The dust emission in the spectral energy distribution (SED) 
of our dwarf galaxies typically peaks between 70 - 100$\mu$m, 
which makes the 100$\mu$m emission a relatively good proxy 
for the dust column density. 
We convolve the 100$\mu$m maps with a 2D Gaussian. 
The position and width of the Gaussian are chosen to correspond to the 
position and width\footnote{The FWHM used equals $\sqrt{FWHM(CO)^2 - FWHM(PACS)^2}$.} 
of the respective CO observations. 
The CO fractions that we estimate this way are listed 
in Table~\ref{table:cofrac}.

For Haro\,11, Mrk\,1089, and Mrk\,930, we have recovered the bulk 
of the CO emission and the CO luminosities that we report in 
Table~\ref{table:lines} can be considered as total values. 
The fractions of the CO emission captured in the different CO beams 
agree within 20\% and comparing the CO intensities from one transition with 
another should be quite reliable (except for Mrk\,930 but we only have 
a limit on its ratio). 
On the other hand, NGC\,4861, NGC\,625, and UM\,311 are more extended 
and with complex shapes, and a large fraction of the total 100$\mu$m flux 
is emitted outside the CO beams. 
For these galaxies, we apply correction factors 
to the observed CO fluxes to estimate the total CO fluxes (Table~\ref{table:lines}). 
For NGC\,4861, in which we targeted only the main H~{\sc ii} region, this correction factor is $3.5$. 
For UM\,311, the correction factor is $5$ because of 100$\mu$m contamination 
from the two neighboring spiral galaxies. In this case, it is unclear whether the 
100$\mu$m emission can be used to calculate beam correction factors since 
these systems may or may not be hosting CO clumps. For the sake 
of consistency, we apply this correction factor and estimate the total 
CO emission for the entire system. 
The CO line ratios are also uncertain by a factor of $\sim$2. 
For NGC\,625, we consider a single Mopra beam at the position 
of the CO(3-2) pointing. The difference between the 100$\mu$m fluxes integrated 
within the individual CO(1-0) and CO(3-2) beams is a factor of 2. 
The reason for this difference in NGC\,625 is because the CO(3-2) 
observation does not point toward the peak of dust emission, and the 
CO(1-0) and CO(3-2) beam sizes are quite different. 
For the total CO emission of NGC\,625, we consider the emission in all 
the CO(1-0) Mopra pointings, which cover the full extent of the galaxy 
seen in the dust continuum. When doing so, only 10\% of the 100$\mu$m 
emission is emitted outside our CO(1-0) coverage. 
%}

%%%
These rough estimates of possible CO emission outside of the respective 
CO beams show that there are uncertainties of {about a factor of 2 when 
comparing CO lines with each other, and a factor of $\sim$3} for NGC\,4861 
and UM\,311 when considering global quantity, such as 
molecular masses.

%%%
\begin{center}
\begin{table}[t!]
  \caption{Estimated CO fractions (in \%).}
  \hfill{}
  \begin{tabular}{c c c c}
    \hline\hline
    \vspace{-8pt}\\
    \multicolumn{1}{c}{Galaxy} & 
    \multicolumn{1}{c}{1-0 / total} &
    \multicolumn{1}{c}{2-1 / total} &
    \multicolumn{1}{c}{3-2 / total} \\
    \hline
    \vspace{-8pt}\\
	{Haro\,11}		& 85	& 81	& 65 \\
	{Mrk\,1089}	& 56	& 63	& 44 \\
	{Mrk\,930}		& 62	& 28	& - \\
	{NGC\,4861}	& 28 	& 10 	& - \\
	{NGC\,625}	& 27 	& 17 	& 11 \\
	{UM\,311}		& 21 	& 18	& 10 \\
    \hline \hline
    \vspace{-8pt}\\
  \end{tabular}
  \hfill{}
  \newline
    \vspace{-1pt}\\
	Fraction of the 100$\mu$m flux from {\it Herschel} measured in each CO beam 
	relative to the total 100$\mu$m flux -- where the distribution of the 100$\mu$m 
	dust emission is used as a proxy for the CO emission. 
  \label{table:cofrac}
\end{table}
\end{center}

%%%%%%%%%%%%%%%%%%%%%%%%
\section{Physical conditions in the molecular clouds}
\label{sect:prop}
%%%%%%%%%%%%%%%%%%%%%%%%
In this section, we investigate the conditions (temperature, density, mass) 
characterizing the CO-emitting gas in our galaxies. 
First we discuss line ratios and derive total molecular masses using two 
values of the CO-to-H$_{\rm 2}$ conversion factor (Galactic and metallicity-scaled), 
and dust measurements. 
Then we compare the observations with the predictions of the code \textsc{RADEX} 
{and contrast our CO results with the analysis of the warm H$_{\rm 2}$ gas.}

%%%%%
\subsection{Empirical diagnostics}
\label{sect:diagnos}
%%%%%
%%%
\subsubsection{Line ratios}
\label{sect:ratios}
%%%
CO line intensity ratios give an insight into the conditions of the gas. 
Typical ratios are CO(2-1)\,/\,CO(1-0)$\sim$0.8 and CO(3-2)\,/\,CO(1-0)$\sim$0.6 
in nearby spiral galaxies \citep{leroy-2008,wilson-2009,warren-2010}, and 
CO(3-2)\,/\,CO(2-1)$\sim$0.8 in Galactic giant molecular clouds \citep{wilson-1999}. 
{In blue compact dwarf galaxies, \cite{sage-1992} find 
an average ratio of CO(2-1)\,/\,CO(1-0)$\sim$0.8, and 
\cite{meier-2001} derive an average CO(3-2)\,/\,CO(1-0) 
ratio of 0.6 in starburst dwarf galaxies.} 
Ratios of CO(2-1)\,/\,CO(1-0)$\ge$1 are found in several starbursting 
compact dwarfs \citep[e.g.][]{israel-2005} and bright regions of the 
Magellanic Clouds \citep{bolatto-2000,bolatto-2003}, and are usually attributed 
to the emission arising from smaller, warmer clumps, where CO(1-0) 
may not trace the total molecular gas.

The line ratios are calculated from the observed CO line intensities $I_{\rm CO}$ 
(K~km~s$^{-1}$), and solid angles of the source ($\Omega_{source}$) and 
of the beam ($\Omega_{beam}$), following \cite{meier-2001}: 
\begin{equation}
{\rm CO}~2-1/1-0 = 
\frac{I_{\rm CO~2-1}}{I_{\rm CO~1-0}} \times 
\frac{\Omega_{source}+\Omega_{beam,~CO~2-1}}{\Omega_{source}+\Omega_{beam,~CO~1-0}}. 
\end{equation}
The source sizes are unknown since we do not resolve individual molecular clouds. 
The assumed spatial distribution of the emission 
influences the derived CO line ratios (see also section~\ref{sect:comiss}). 
Therefore we derive line ratios in two cases: (1)~if the source is point-like 
($\Omega_{source} << \Omega_{beam}$) or (2)~if the source is much larger 
than the beam ($\Omega_{beam} << \Omega_{source}$). 
The latter case also applies if the source is not resolved but the CO emission comes 
from small clumps scattered across the beam and fills the beam (no beam dilution). 
Values of the line ratios are summarized in Table~\ref{table:ratios}. 

For the sample, we find an average CO(3-2)\,/\,CO(2-1) ratio of $\sim$0.6 
(measured on the detections only). Ratios of CO(3-2)\,/\,CO(1-0) and 
CO(2-1)\,/\,CO(1-0) are respectively $\ge$0.2 and $\ge$0.6 across the sample. 
Although there can be significant uncertainties due to the different 
beam sizes (see section~\ref{sect:comiss}) and unknown source distribution, 
overall, we find no evidence for significantly different CO line ratios in our sample 
compared to normal-metallicity environments. 
Unlike resolved low-metallicity environments where CO line ratios 
can be high due to locally warmer gas temperatures, the line ratios 
(hence CO temperature) in our compact dwarf galaxies are moderate.

%%%
\begin{center}
\begin{table}[t!]
  \caption{CO line ratios.}
  \hfill{}
  \begin{tabular}{c c c c c}
    \hline\hline
    \vspace{-8pt}\\
    \multicolumn{2}{c}{Galaxy} & 
    \multicolumn{1}{c}{3-2 / 1-0} &
    \multicolumn{1}{c}{3-2 / 2-1} &
    \multicolumn{1}{c}{2-1 / 1-0} \\
    \hline
    \vspace{-8pt}\\
	\multicolumn{4}{l}{Haro\,11} \\
	 & {\it point-like}		& $>$0.23 	& 0.38 $\pm$ 0.23 	& $>$0.60 \\
	 & {\it extended}	& $>$0.71 	& 0.89 $\pm$ 0.53	& $>$0.79 \\
	\multicolumn{4}{l}{Mrk\,1089} \\
	 & {\it point-like}		& 0.30 $\pm$ 0.13	& 0.30 $\pm$ 0.13	& 1.00 $\pm$ 0.31 \\
	 & {\it extended}	& 0.46 $\pm$ 0.19 	& 0.70 $\pm$ 0.31	& 0.65 $\pm$ 0.20 \\
	\multicolumn{4}{l}{Mrk\,930} \\
	 & {\it point-like}		& -	& - 	& $<$0.69 \\
	 & {\it extended}	& -	& - 	& $<$2.51 \\
	\multicolumn{4}{l}{NGC\,625} \\
	 & {\it point-like}		& $>$0.29 	& 0.64 $\pm$ 0.30 	& $>$0.45 \\
	 & {\it extended}	& $>$0.90 	& 0.88 $\pm$ 0.42	& $>$1.02 \\
	\multicolumn{4}{l}{UM\,311} \\
	 & {\it point-like}		& $>$0.21 	& 0.36 $\pm$ 0.21	& $>$0.58 \\
	 & {\it extended}	& $>$0.66 	& 0.85 $\pm$ 0.49	& $>$0.77 \\
    \hline \hline
    \vspace{-8pt}
  \end{tabular}
  \hfill{}
  \newline
  \label{table:ratios}
\end{table}
\end{center}

%%%
\subsubsection{Molecular gas mass from CO observations}
%%%

\begin{center}
\begin{table*}[!t]\scriptsize
  \caption{H$_{\rm 2}$ masses using CO observations, dust measurements, and models.}
  \hfill{}
  \begin{tabular}{l c c c c c |c |c |c c}
    \hline\hline
    \vspace{-7.8pt}\\
    \multicolumn{1}{l}{Galaxy} & 
    \multicolumn{5}{c}{from CO observations} & 
    \multicolumn{1}{|c}{from dust} &
    \multicolumn{1}{|c}{from LTE} &
    \multicolumn{2}{|c}{from Cloudy} \\  \cline{2-10} 
   \vspace{-7.8pt}\\
    \multicolumn{1}{c}{} & 
    \multicolumn{1}{c}{${X_{\rm CO}}/{X_{\rm CO, gal}} ^{(a)}$} & 
    \multicolumn{4}{c}{M$_{\rm H_2}$ (M$_{\odot}$)}  &
    \multicolumn{1}{|c}{M$_{\rm H_2}$ (M$_{\odot}$)} &
    \multicolumn{1}{|c}{M$_{\rm H_2}^{warm}$ (M$_{\odot}$)} &
    \multicolumn{2}{|c}{M$_{\rm H_2}$ (M$_{\odot}$)}  \\ \cline{3-6}  \cline{9-10} 
    \vspace{-7.8pt}\\
    \multicolumn{1}{c}{} &
    \multicolumn{1}{c}{} & 
    \multicolumn{1}{c}{CO(1-0)} &
    \multicolumn{1}{c}{$R_{21}=0.8 ^{(b)}$} &
    \multicolumn{1}{c}{$R_{31}=0.6 ^{(c)}$} &
    \multicolumn{1}{c}{Average}  &
    \multicolumn{1}{|c}{} & 
    \multicolumn{1}{|c}{} &
    \multicolumn{1}{|c}{CO-free} &
    \multicolumn{1}{c}{CO-traced} \\ 
    \hline
    \vspace{-7.8pt}\\
	Haro\,11 		& 1 	& $<$4.4$\times10^{8}$ 	& 3.3$\times10^{8}$		& 1.7$\times10^{8}$ 		& 2.5$\times10^{8}$ 		& 3.6$\times10^{9}$ & 
							$>$3.6$\times10^{6}$~$<$2.4$\times10^{8}$	& 1.2$\times10^{8}$		& 1.6$\times10^{9}$ \\
	 			& 10 	& $<$4.4$\times10^{9}$	& 3.3$\times10^{9}$ 		& 1.7$\times10^{9}$		& 2.5$\times10^{9}$ 		& - 	&	&	&	\\
	Mrk\,1089 	& 1 	& 3.1$\times10^{8}$  	& 3.9$\times10^{8}$ 		& 1.6$\times10^{8}$ 		& 2.9$\times10^{8}$ 		& 4.0$\times10^{9}$ 	&	&	&	\\
			 	& 16 	& 4.9$\times10^{9}$ 		& 6.2$\times10^{9}$ 		& 2.5$\times10^{9}$ 		& 4.5$\times10^{9}$ 		& - 	&	&	&	\\
	Mrk\,930	 	& 1 	& 3.1$\times10^{7}$  	& $<$9.8$\times10^{7}$	& - 	& 3.1$\times10^{7}$ 		&1.0$\times10^{9}$ 	&	&	&	\\
			 	& 22 	& 6.8$\times10^{8}$	 	& $<$2.1$\times10^{9}$	& - 	& 6.8$\times10^{8}$ 		& - 	&	&	&	\\
	NGC\,4861 	& 1 	& $<$7.3$\times10^{5}$	& $<$1.2$\times10^{6}$	& -	& $<$7.3$\times10^{5}$ 		& - 	&	&	&	\\
			 	& 42 	& $<$3.1$\times10^{7}$ 	& $<$5.1$\times10^{7}$ 	& -	& $<$3.1$\times10^{7}$ 		& - 	&	&	&	\\
	{\it total} \dotfill 	& 1 	& $<$2.6$\times10^{6}$	& -	& -	& $<$2.6$\times10^{6}$ 	& 4.1$\times10^{8}$ 	&	&	&	\\
	NGC\,625 	& 1 	& $<$1.9$\times10^{6}$		& 1.1$\times10^{6}$ 		& 9.0$\times10^{5}$		& 1.0$\times10^{6}$ 		& - 	& 	&	&	\\
			 	& 9 	& $<$1.7$\times10^{7}$ 		& 9.7$\times10^{6}$	 	& 8.2$\times10^{6}$		& 9.0$\times10^{6}$ 		& - 	&	&	&	\\
	{\it total} \dotfill 	& 1 	& 5.0$\times10^{6}$			& -	& -	& 5.0$\times10^{6}$	 	& 1.5$\times10^{8}$ 	& 
							$>$6.0$\times10^{4}$~$<$5.9$\times10^{6}$	&	&	\\
	UM\,311 		& 1 	& $<$2.7$\times10^{7}$ 	& 2.0$\times10^{7}$ 		& 9.4$\times10^{6}$ 		& 1.5$\times10^{7}$ 		& - 	&	&	&	\\
		 		& 4 	& $<$1.2$\times10^{8}$ 	& 8.5$\times10^{7}$	 	& 4.1$\times10^{7}$	 	& 6.3$\times10^{7}$ 		& - 	&	&	&	\\
	{\it total} \dotfill 	& 1 	& $<$1.3$\times10^{8}$	& 9.8$\times10^{7}$		& -	& 9.8$\times10^{7}$ 	& 4.5$\times10^{9}$ 	&	&	&	\\
    \hline \hline
  \end{tabular}
  \hfill{}
    \\
  \newline
  The H$_{\rm 2}$ masses from CO observations are measured within the CO beams. 
  For extended galaxies, we also consider the estimated total CO emission 
  (see section~\ref{sect:comiss} for details). 
  $(a)$~Ratio of our adopted $X_{\rm CO}$ value to the Galactic value $X_{\rm CO, gal}$. 
  To bracket the range of molecular gas masses, there are two lines for each galaxy 
  corresponding to the molecular masses derived with two values of the $X_{\rm CO}$ factor: 
  $X_{\rm CO} = X_{\rm CO, gal}$ (lower mass case) and 
  $X_{\rm CO} = X_{\rm CO, Z} \propto Z^{-2}$ (upper mass case). 
  $(b)$~$R_{21}$ is the CO(2-1)\,/\,CO(1-0) ratio. 
  $(c)$~$R_{31}$ is the CO(3-2)\,/\,CO(1-0) ratio. 
  \label{table:mass}
\end{table*}
\end{center}

The molecular gas mass is usually derived from CO observations 
as a tracer of the cold H$_{\rm 2}$ using the CO-to-H$_{\rm 2}$ conversion 
factor $X_{\rm CO}$\footnote{
\cite{dame-2001} report an average value in the Galaxy of 
$X_{\rm CO, gal} = 1.8 \times 10^{20}~\rm{cm^{-2}~(K~km~s^{-1})^{-1}}$, 
and {\it Fermi} measurements give 
$X_{\rm CO, gal} \simeq 2 \times 10^{20}~\rm{cm^{-2}~(K~km~s^{-1})^{-1}}$ 
for large-scale complexes \citep[e.g.][]{ackermann-2011}. 
In the following, we adopt 
$X_{\rm CO, gal} = 2 \times 10^{20}~\rm{cm^{-2}~(K~km~s^{-1})^{-1}}$. 
Note that these variations in $X_{\rm CO, gal}$ are insignificant here 
compared to the orders of magnitude variations that may prevail 
in our low-metallicity sample. 
}
which relates the observed CO intensity to the column 
density of H$_{\rm 2}$: 
\begin{equation}
X_{\rm CO} = N({\rm H_2})/I_{\rm CO}~\rm{[cm^{-2}~(K~km~s^{-1})^{-1}]}. 
\end{equation}
%
%%%%%
The extent to which the global conversion from CO to cold H$_{\rm 2}$ reservoir 
is affected by metallicity depends crucially on the relative proportion of H$_{\rm 2}$ 
residing at moderate A$_V$ compared to the amount of enshrouded molecular gas. 
Even though both simulations and observations support a change 
in the CO-to-H$_{\rm 2}$ conversion factor at low metallicities, the exact 
dependence of $X_{\rm CO}$ on metallicity is not empirically well-established. 
Many physical characteristics other than metallicity, such as the 
topology of the molecular clouds and the distribution of massive 
ionizing stars, are also at play. 
{We refer the reader to \cite{bolatto-2013} for a review on the topic.} 
%
%%%%%%
Therefore we consider two cases: one where $X_{\rm CO}$ is equal 
to the Galactic value $X_{\rm CO, gal}$, and one where $X_{\rm CO}$ scales 
with metallicity (noted  $X_{\rm CO, Z}$) following \cite{schruba-2012}: 
$X_{\rm CO, Z} \propto Z^{-2}$. 
Although the $X_{\rm CO}$ conversion factor is calibrated for the 
CO(1-0) transition, we also estimate the molecular mass from 
the CO(2-1) and CO(3-2) observations, using the values 
{of 0.8 and 0.6 for the CO(2-1)\,/\,CO(1-0) and CO(3-2)\,/\,CO(1-0) 
ratios respectively (section~\ref{sect:ratios})}. 
For each CO transition, the source size is taken as its corresponding beam size. 
For galaxies observed with IRAM (Mrk\,930 and NGC\,4861), we assume 
the source size to be that of the CO(1-0) beam since only the 
CO(1-0) observations in Mrk\,930 leads to a detection. 
{The resulting molecular gas masses of the 6 galaxies are reported 
in Table~\ref{table:mass}.}

%%%%%
For Haro\,11, {our CO detections yield a molecular gas mass of} 
$\rm{M(H_{2}) \simeq 2.5 \times 10^{8}~M_{\odot}}$ (with $X_{\rm CO, gal}$). 
This value is higher than the previous upper limit of \cite{bergvall-2000}, where 
they estimated $\rm{M(H_{2}) \le 10^{8}~M_{\odot}}$. 
For Mrk\,1089, \cite{leon-1998} found $\rm{M(H_{2}) = 4.4 \times 10^{8}~M_{\odot}}$, 
This value is $1.5$ times larger than that we derive with $X_{\rm CO, gal}$ 
due to a beam (hence assumed source size) difference, and is below the 
value that we derive with $X_{\rm CO, Z}$. 
There are no other molecular mass estimates for Mrk\,930, NGC\,625, NGC\,4861, 
and UM\,311 in the literature to which we can compare our estimates. 
The molecular gas masses that we find with $X_{\rm CO, gal}$ (denoted 
M(H$_{\rm 2 ,gal}$)) are particularly low compared to, e.g., the IR luminosity 
or H~{\sc i} reservoir of our galaxies. 
We find an average ratio (with minimum and maximum values) 
of M(H$_{\rm 2, gal}$)/M(H{\sc i})$\sim$12\,$^{50} _{1.0}$\% 
and M(H$_{\rm 2 ,gal}$)/L$_{TIR}$$\sim$1.0\,$^{1.9} _{0.1}$\% 
(typically 10\% in isolated galaxies; \citealt{solomon-1988}).  
This stresses how deficient our galaxies are in CO.

%%%
\subsubsection{Molecular gas mass from dust measurements}
\label{sect:xcodust}
%%%
Other methods based on dust measurements are also used to quantify 
the total H$_{\rm 2}$ gas, assuming that 
the gas mass is proportional to the dust mass 
via a fixed dust-to-gas mass ratio \citep[e.g.][]{leroy-2011,sandstrom-2013}: 
\begin{equation}
{\rm M(H_2)~=~M_{dust} \times [D/G]^{-1}~-~[M(HI)+M(HII)]}. 
\end{equation}
{We consider quantities within apertures defined from the dust emission 
and we use aperture-corrected H~{\sc i} masses from \citealt{remy-2013c} 
(Table~\ref{table:ksparams}). 
Except for Haro\,11 \citep[H~{\sc ii} mass from][]{cormier-2012}, we ignore M(H~{\sc ii}) from the equation 
since there is no ionized gas mass reported in the literature for our galaxies.} 
We note that M(H~{\sc ii}) might be negligible in normal galaxies but contribute 
more to the mass budget of dwarf galaxies. 
For the dust-to-gas mass ratio (D/G), we choose a D/G that scales 
linearly with metallicity \citep[e.g.][]{edmunds-2001} and 
adopt the value $\rm{D/G}=1/150$ for $O/H = 8.7$ \citep{zubko-2004}. 
The dust masses are measured in \cite{remy-2013b} from a full dust SED 
modeling \citep{galliano-2011} up to 500$\mu$m with Galactic dust opacities. 
\cite{remy-2013b} show that the thus derived dust masses are not 
significantly affected by the presence of a possible submm excess. 
Uncertainties on the dust masses due to the submm excess or 
a change of grain properties in low-metallicity environments are 
of a factor 2-3 \citep{galametz-2011,galliano-2011}. 
We provide the D/G and dust masses in Table~\ref{table:xcodust}, and 
the derived H$_{\rm 2}$ masses in Table~\ref{table:mass}. 
We find that the H$_{\rm 2}$ masses resulting from this method are 
systematically larger by one or two orders of magnitude (except in Mrk\,1089) 
compared to those determined from CO and $X_{\rm CO, gal}$. 

If we assume that the H$_{\rm 2}$ mass derived from the dust method 
is the true molecular mass, we can use the observed CO intensities and calculate 
the corresponding $X_{\rm CO}$ value, $X_{\rm CO, dust}$. 
The full extent of the CO emission being unknown for the extended galaxies, 
we consider the fraction of the H$_{\rm 2}$ mass that would fall in the 
CO(1-0) beam (section~\ref{sect:comiss}). We re-estimate the 
CO(1-0) intensities based on the average H$_{\rm 2}$ masses of the 
three CO transitions (Table~\ref{table:mass}) to derive $X_{\rm CO, dust}$ 
(Table~\ref{table:xcodust}). 
The $X_{\rm CO, dust}$ values are 10-50 times higher than $X_{\rm CO, gal}$, 
and sometimes even higher than $X_{\rm CO, Z}$. 
They scale with metallicity with power-law index $\simeq -2.6$: 
$X_{\rm CO, dust} \propto (O/H)^{-2.6}$. 
This relation with metallicity is steeper than what is found in \cite{leroy-2011} 
for Local Group galaxies (index of $-1.7$), and similar to that found 
in \cite{israel-2000} (index of $-2.5$) and in \cite{schruba-2012} 
for their complete sample of dwarf galaxies (index of $-2.8$).

Note that the numbers from this method are given as an indication 
since they bear large uncertainties. 
We have done the same exercise on 7 additional compact galaxies of the DGS 
(names in Table~\ref{table:ksparams}). 
These galaxies have $7.75<O/H$ and are selected with total detected CO 
emission (within 50\%, comparing the [C~{\sc ii}] emission inside and outside 
the CO aperture, \citealt{cormier-2014}). 
However, the total gas masses predicted from the dust are often lower than 
the H~{\sc i} masses (and not only for the lowest metallicity objects). 
This shows the limitation of this method for compact low-metallicity galaxies 
which are H~{\sc i} dominated and for which the D/G scales 
more steeply than linearly with metallicity, although its value is strongly variable 
as it depends on the star formation history of each individual galaxy \citep{remy-2013c}. 

To conclude, the use of different methods ($X_{\rm CO, gal}$, $X_{\rm CO, Z}$ or dust) 
to measure the molecular gas reservoir in our galaxies results in very different 
mass estimates. This highlights the large uncertainties when using one method 
or the other. $X_{\rm CO, gal}$ gives very low molecular gas masses which 
probably under-estimate the true H$_{\rm 2}$ gas in these low-metallicity galaxies. 
More realistic molecular masses (accounting for a CO-dark reservoir) are obtained 
with higher conversion factors from the dust method or from the scaling of 
$X_{\rm CO}$ with metallicity.

%%%
\begin{table}[t!]\footnotesize
  \caption{Dust-derived $X_{\rm CO}$ factors.}
  \hfill{}
  \begin{tabular}{l c c c c}
    \hline\hline
    \vspace{-8pt}\\
    \multicolumn{1}{l}{Galaxy} & 
    \multicolumn{1}{c}{M$_{dust}$ [M$_{\odot}$]} &
    \multicolumn{1}{c}{D/G} &
    \multicolumn{1}{c}{$X_{\rm CO, dust}$/$X_{\rm CO, gal}$} &
    \multicolumn{1}{c}{$X_{\rm CO, dust}$/$X_{\rm CO, Z}$} \\
    \hline
    \vspace{-8pt}\\
	Haro\,11	 	& 9.9 $\times$ 10$^6$	& 10$^{-2.7}$	& 14	 	& 1.4 \\ %41 \\ %17 \\
	Mrk\,1089 	& 2.6 $\times$ 10$^7$ 	& 10$^{-2.8}$	& 2.8		& 0.2 \\ %89 \\ %130 \\
	Mrk\,930	 	& 6.0 $\times$ 10$^6$ 	& 10$^{-2.8}$	& 33 		& 1.5 \\ %85 \\ %148 \\
	NGC\,4861 	& 6.0 $\times$ 10$^5$ 	& 10$^{-3.0}$	& 159	 & 3.8 \\ %81 \\ %431 \\
	NGC\,625 	& 4.5 $\times$ 10$^5$ 	& 10$^{-2.7}$	& 55 		& 6.0 \\ %306 \\ %152 \\
	UM\,311	 	& 2.0 $\times$ 10$^7$ 	& 10$^{-2.5}$	& 60 		& 14 \\ %10$^{3}$ \\ %464 \\
    \hline \hline
    \vspace{-8pt}\\
  \end{tabular}
  \hfill{}
  \newline
  Dust masses are from \cite{remy-2013b}. 
  \label{table:xcodust}
\end{table}

%%%%%
\subsection{Modeling of the CO emission with \textsc{RADEX}}
\label{sect:radex}
\begin{figure}[!t]
\begin{minipage}{8.8cm}
\centering
\includegraphics[clip,trim=1cm 0 0 0,width=8.8cm]{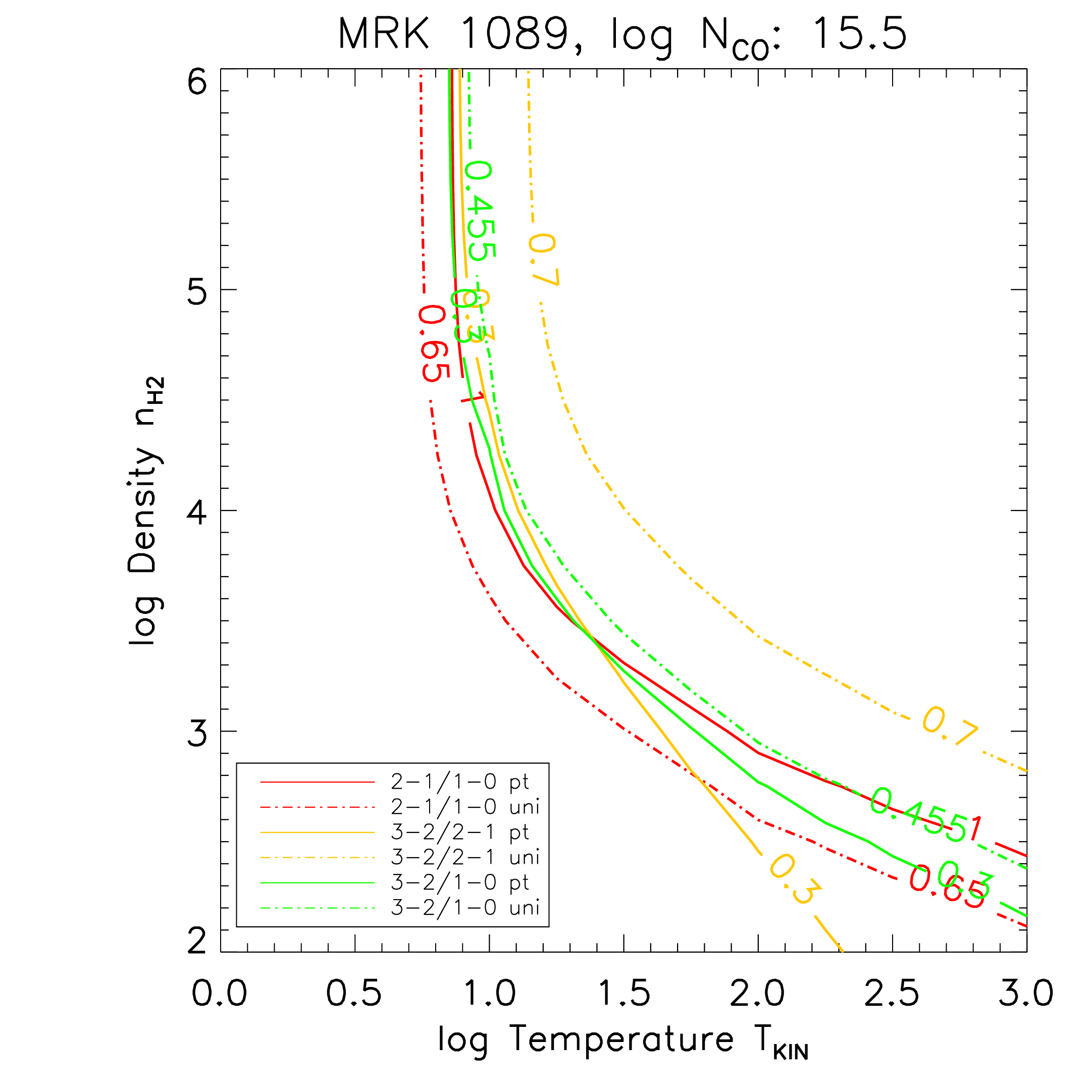}
\caption[]{{\small
Results from \textsc{RADEX} for Mrk\,1089: Contour plot in kinetic temperature 
($T_{kin}$) -- density (n$_{H2}$) parameter space (logarithmic scale), 
for a column density of $N({\rm ^{12}CO}) = 3 \times 10^{15}$~cm$^{-2}$. 
The colored lines correspond to the observed line ratios 
of CO(2-1)\,/\,CO(1-0) (red), 
CO(3-2)\,/\,CO(1-0) (green), 
and CO(3-2)\,/\,CO(2-1) (orange), 
for both uniform filling (dashed-dotted) and point-like (plain) limits. 
}}
\label{fig:radex}
\end{minipage}
\end{figure}

%%%%%
%%%
We use the non-Local Thermodynamic Equilibrium (non-LTE) code \textsc{RADEX} 
\citep{vandertak-2007} to analyze the physical conditions of the low-J CO-emitting gas.  
\textsc{RADEX} predicts the expected line intensity of a molecular cloud defined by constant 
column density ($N(^{\rm 12}$CO)), gas density (n$_{\rm H_2}$), and kinetic temperature (T$_{kin}$). 
In Figure~\ref{fig:radex}, we draw contour plots of CO line ratios 
{for Mrk\,1089 (similar plots are obtained for Haro\,11, NGC\,625, and UM\,311)} 
in the n$_{\rm H_2}$ -- T$_{kin}$ space for a given value of 
$N({\rm ^{12}CO}) = 3 \times 10^{15}$~cm$^{-2}$ 
(typical of subsolar metallicity clouds; \citealt{shetty-2011a}). 
The behavior of each line ratio considered is similar in the n$_{\rm H_2}$ -- T$_{kin}$ 
space and relatively insensitive to $N(^{\rm 12}$CO). 
Given the small number of observational constraints, no precise estimate 
on the physical conditions of the gas can be reached, 
except that T$_{kin} \ge 10$~K. 
To break the degeneracy shown in Figure~\ref{fig:radex}, and particularly 
constrain better the gas column density, one would need 
optically thin line diagnostics with, e.g., $^{13}$CO measurements. 
When $^{13}$CO measurements are available, they however usually 
preclude models with only a single molecular gas phase, and require 
at least two different phases \citep{meier-2001,israel-2005}.

%%%

%%%%%
\subsection{Excitation diagrams of the H$\rm{_{2}}$ molecule}
\label{sect:lte}
\begin{figure}[t]
\centering
\includegraphics[clip,trim=.5cm 0 .45cm 0,width=4.66cm]{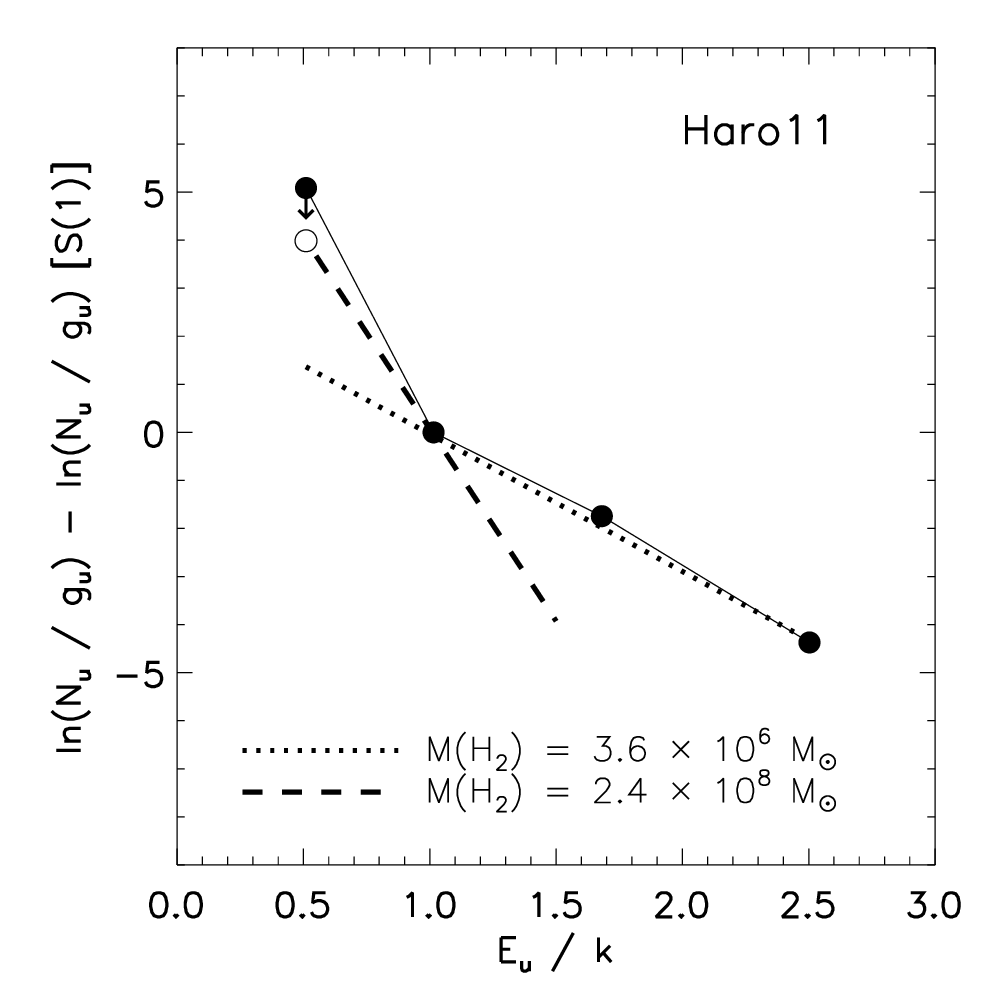}
\hspace{-0.2cm}
\includegraphics[clip,trim=1.75cm 0 0 0,width=4.25cm]{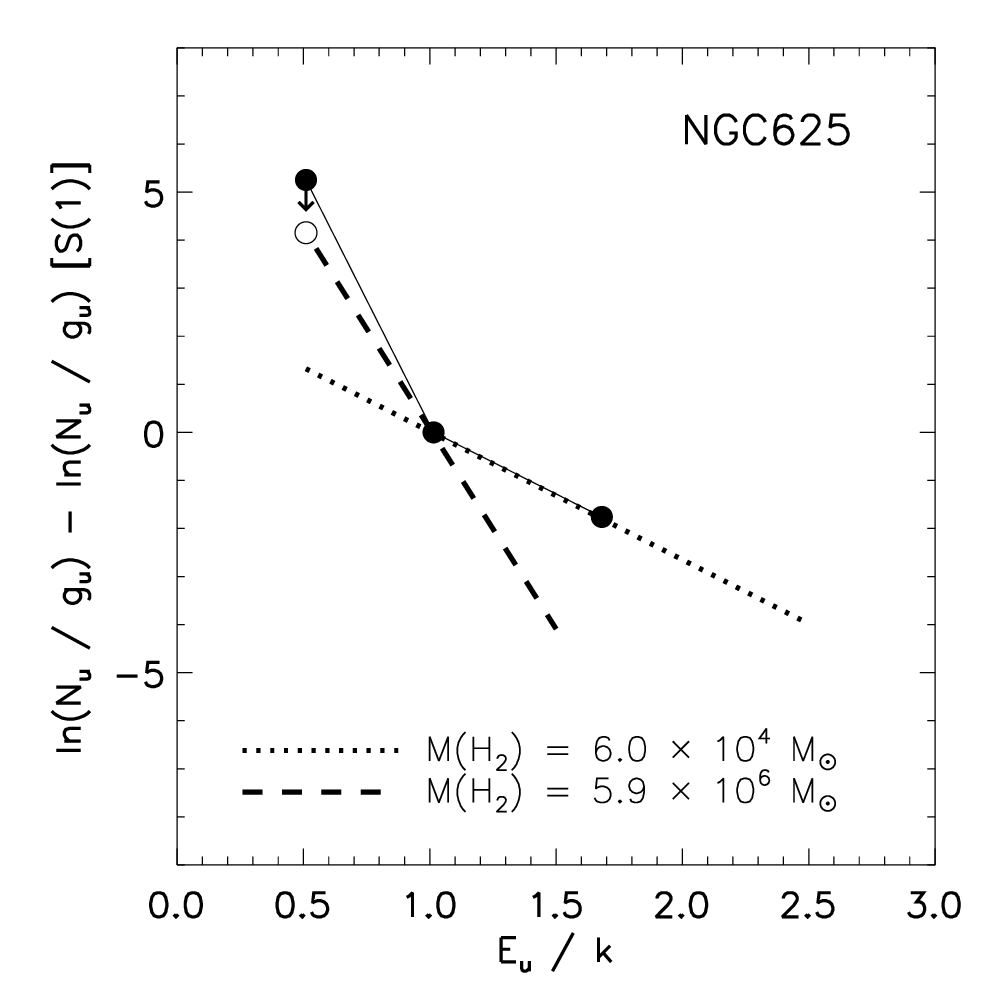}
\caption{
Excitation diagrams of the S(0) to S(3) rotational H$\rm{_{2}}$ lines 
for Haro\,11 ({\it left}) and NGC\,625 ({\it right}). 
The $N_u/g_u$ ratios are normalized by the S(1) transition. 
The observations are shown with filled circles. 
The dotted lines indicate the best fit to the S(1) to S(3) transitions. 
The dashed line indicates the fit to the S(0) 1-$\sigma$ limit (open circle) 
and the S(1) transition. 
}
\label{fig:h2diag}
\end{figure}

%%%%%
Rotational transitions of molecular hydrogen are visible in the MIR 
with {\it Spitzer} and originate from a warm gas phase (temperature 
greater than a few hundreds of~K), therefore putting constraints on 
the PDR properties of star-forming regions (and in particular the mass) 
rather than the cold molecular phase \citep[e.g,][]{kaufman-2006}. 
As a first step in characterizing the conditions of the PDR, 
we assume local thermal equilibrium (LTE) and build 
an excitation diagram to derive the population of the H$\rm{_{2}}$ levels and 
estimate the temperature, column density, and mass of the warm molecular gas. 
We conduct this analysis only for the two galaxies Haro\,11 and NGC\,625, 
which have H$\rm{_{2}}$ lines detected in their IRS spectra. 
We consider the total fluxes in Table~\ref{table:h2flux}. 
The H$\rm{_{2}}$ emission is assumed optically thin. 
This approximation is justified by the fact that low-metallicity galaxies 
have, on average, lower A$_V$, and clumpy material of high optical depth 
while the warm H$_{\rm 2}$ should arise from regions with larger filling factors. 
Also, the effects of extinction on the MIR H$\rm{_{2}}$ lines are not important 
in normal galaxies \citep{roussel-2007} and ULIRGs \citep{higdon-2006}. 
We follow the method from \cite{roussel-2007}, and assume a single temperature 
to match the S(1) to S(3) lines. This choice is biased toward higher temperatures 
(and lower masses) due to the non-detection of the S(0) line, as mentioned in 
\cite{higdon-2006} and \cite{roussel-2007}. 
The source sizes are taken as the CO(1-0) Mopra coverage.  
The excitation diagrams are shown in Figure~\ref{fig:h2diag}. 
On the y-axis, the observed intensities are converted to column densities 
in the upper levels $N_u$ normalized by the statistical weights $g_u$. 
For Haro\,11 ({\it left} panel), the best fit to the S(1), S(2), and S(3) lines 
is represented by the dotted line and agrees with a temperature of 350~K 
and column density of $\rm{1.1 \times 10^{18}~cm^{-2}}$. 
The resulting H$\rm{_{2}}$ mass is $\rm{M(H_2) > 3.6 \times 10^{6}~M_{\odot}}$, 
which can be regarded as a lower limit since the S(0) line is not detected 
and hence not used as a constraint. 
Fitting the 1-$\sigma$ limit on the S(0) line and the S(1) line (dashed line on 
Fig.~\ref{fig:h2diag}) agrees with a temperature of 100~K and column density 
of $\rm{7.5 \times 10^{19}~cm^{-2}}$. This yields an upper limit of 
$\rm{M(H_2) = 2.4 \times 10^{8}~M_{\odot}}$. 
The exact warm molecular gas mass is therefore in between the two limits: 
$3.6 \times 10^{6}~M_{\odot} < \rm{M(H_{2, warm}) < 2.4 \times 10^{8}~M_{\odot}}$. 
For NGC\,625 ({\it right} panel), the fit to the S(1) and 
S(2) lines (dotted line) yields a temperature of 380~K, and a column density of 
$\rm{2.7 \times 10^{17}~cm^{-2}}$. This results in a lower limit on the mass of 
$\rm{M(H_2) > 6.0 \times 10^{4}~M_{\odot}}$. 
Fitting the 1-$\sigma$ limit on the S(0) line and the S(1) line (dashed line) agrees 
with a temperature of 120~K and column density of $\rm{2.6 \times 10^{19}~cm^{-2}}$.
This gives a warm H$_{\rm 2}$ mass of $\rm{M(H_2) < 5.9 \times 10^{6}~M_{\odot}}$. 

The limits that we obtain on $\rm{M(H_{2, warm})}$ indicate that the warm 
molecular gas mass in Haro\,11 and NGC\,625 is lower than their cold gas mass 
(Table~\ref{table:mass}). \cite{roussel-2007} find an average warm-to-cold ratio 
of $\rm{M(H_{2, warm})/M(H_{2, cold})} \sim$10\% in star-forming galaxies. 
The temperatures and column densities of the warm H$_{\rm 2}$ gas that 
we find are globally consistent with previous studies of nearby galaxies. 
In BCDs, \cite{hunt-2010} find an average temperature for the warm gas 
of $\sim$245~K, and of $\sim$100~K for Mrk\,996, which is their only galaxy 
detected in the S(0) transition. However, their column densities 
are somewhat larger (N$(H_2) \sim \rm{3 \times 10^{21}~cm^{-2}}$) than ours. 
In more metal-rich galaxies, \cite{roussel-2007} find a median temperature 
of $\sim$150~K for the colder H$_{\rm 2}$ component (fitting the S(0) transition), 
and a temperature $\geq$400~K for the warmer component. 
Their column densities (N$(H_2) \sim \rm{2 \times 10^{20}~cm^{-2}}$) are 
in better agreement with ours.

%%%

%%%%%%%%%%%%%%%%%%%%%%
\section{Full radiative transfer modeling of Haro\,11}
\label{sect:cloudy}
%%%%%%%%%%%%%%%%%%%%%%
%%%
In this section, we investigate a radiative transfer model of Haro\,11 
using the spectral synthesis code v13.01 Cloudy \citep{ferland-2013,abel-2005}, 
for a more complete picture of the conditions in the PDR and molecular cloud. 
\cite{cormier-2012} analyzed the emission of $\sim$17 MIR and FIR 
fine-structure lines from {\it Spitzer} and {\it Herschel} in Haro\,11 with Cloudy. 
The model consists of a starburst surrounded by a spherical cloud. 
It first solves for the photoionized gas, which is constrained by $\sim$9 ionic lines, 
and then for the PDR, which is constrained by the FIR [O~{\sc i}] lines, assuming 
pressure equilibrium between the phases. 
A Galactic grain composition with enhanced abundance of small size grains, 
as often found in dwarf galaxies, is adopted (see \citealt{cormier-2012} for details). 
We extend this model in light of the molecular data presented. 
%

%%%
\subsection{PDR properties}
\label{sect:pdr}
%%%
The PDR model of \cite{cormier-2012} is described by a covering factor 
$\sim$10\% and parameters $\rm{n_H = 10^{5.1}~cm^{-3}}$, $\rm{G_0 = 10^{3}~Habing}$.
This model is stopped at a visual extinction A$_V$ of 3~mag, just before CO starts to form, 
to quantify the amount of dark gas untraced by CO, and because no CO detections 
had been reported before. 
The emission of the {\it Spitzer} warm H$_{\rm 2}$ lines is produced by this PDR 
which has an average temperature of $\sim$200~K. The total gas mass of the PDR 
derived with the model is $\rm{M(PDR) = 1.2 \times 10^{8}~M_{\odot}}$ 
\citep{cormier-2012}, and is mainly molecular: 90\% is in the form of H$_{\rm 2}$ 
(dark gas component), and only 10\% is in the form of H~{\sc i}. 
This molecular gas originates from a warm phase and its mass falls in between 
the limits on the warm molecular gas mass set in section~\ref{sect:lte}. 
The H~{\sc i} mass of the PDR is lower than the \cite{bergvall-2000} limit, 
which was the value adopted in \cite{cormier-2012}. Thus the higher H~{\sc i} 
mass found by \cite{machattie-2013} would not affect our PDR model but 
rather the diffuse ionised/neutral model component of \cite{cormier-2012}.

We find that the intensity of the H$_{\rm 2}$ S(1) transition 
is reproduced by the model within 10\%, while the intensities of the S(2) 
and S(3) transitions are under-predicted by a factor of two. The S(0) 
prediction is below its upper limit. 
PDR models indicate that the H$\rm{_{2}}$ observations, and particularly 
the S(2)/S(1)~(0.6) and S(3)/S(1)~(0.7) ratios, could be better matched if another PDR 
phase of higher density or lower radiation field were considered. 
However, other factors such as uncertainties in the excitation physics of H$\rm{_{2}}$, 
along with uncertainties in UV and density, can also play a role \citep{kaufman-2006}. 
Reproducing lines coming from J$>$3 have inherent uncertainties in the microphysics, 
and inconsistencies can arise from trying to simultaneously reproduce these lines 
at the same time as reproducing [C~{\sc ii}] and [O~{\sc i}].

%%%
\subsection{Molecular cloud properties}
\label{sect:mol}
%%%
To model the molecular cloud (A$_V \ge 3$~mag, where CO starts to form) 
with Cloudy, we continue the PDR model described above to larger A$_V$. 
This way, we assume that the observed CO emission is linked to the 
densest star-forming cores rather than a diffuse phase. 
Since we have no observations of molecular optically thin tracer or denser 
gas tracer than CO, we use the cold dust emission to set our stopping criterion 
for the models and determine the A$_V$ appropriate to reproduce the 
FIR/submm emission of the spectral energy distribution of Haro\,11. 
We find that the {\it Herschel} PACS and SPIRE photometry is best 
reproduced (within 30\%) for A$_V$$\sim$30~mag and we take this value 
as our stopping criterion. 

\begin{figure}
\includegraphics[clip,trim=1cm .5cm .5cm .5cm,width=9cm]{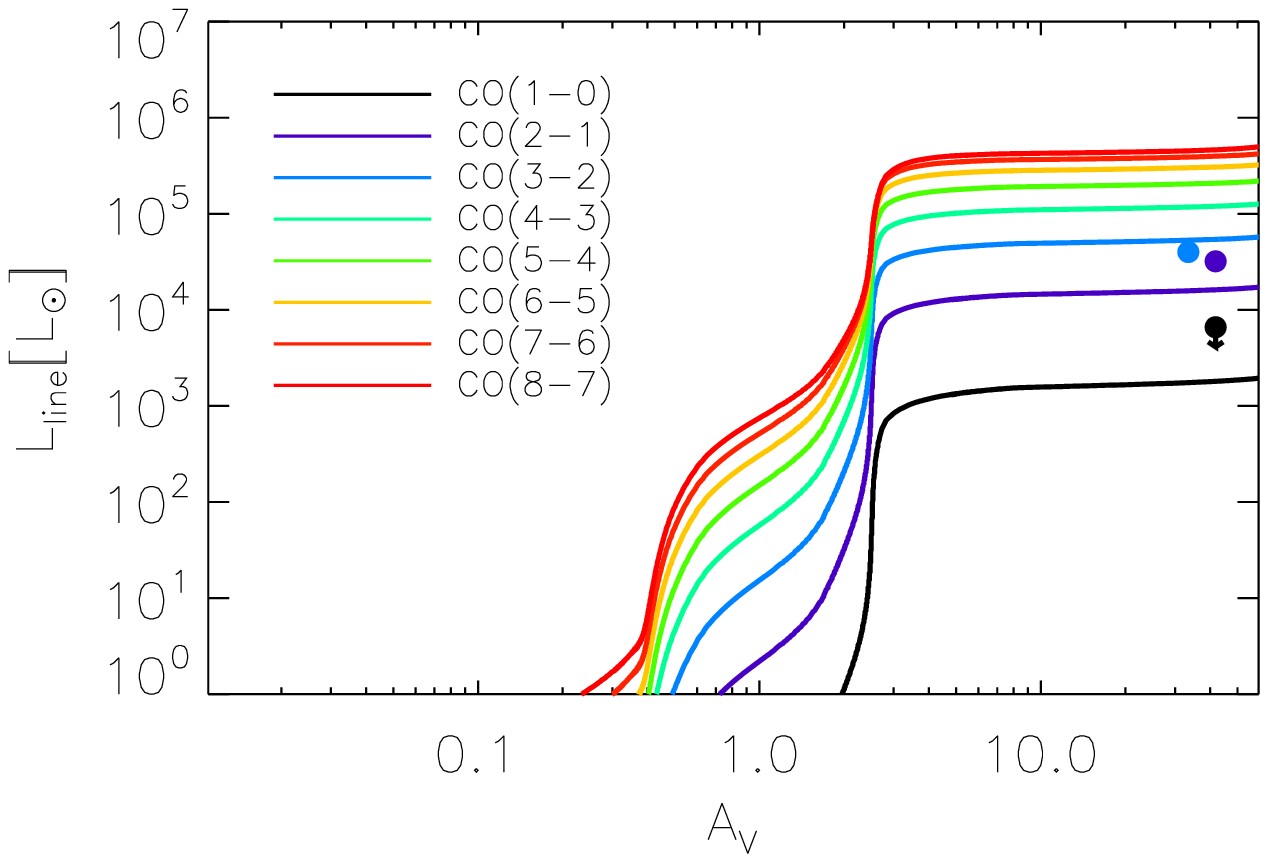}
\includegraphics[clip,trim=1cm .5cm .5cm .5cm,width=9cm]{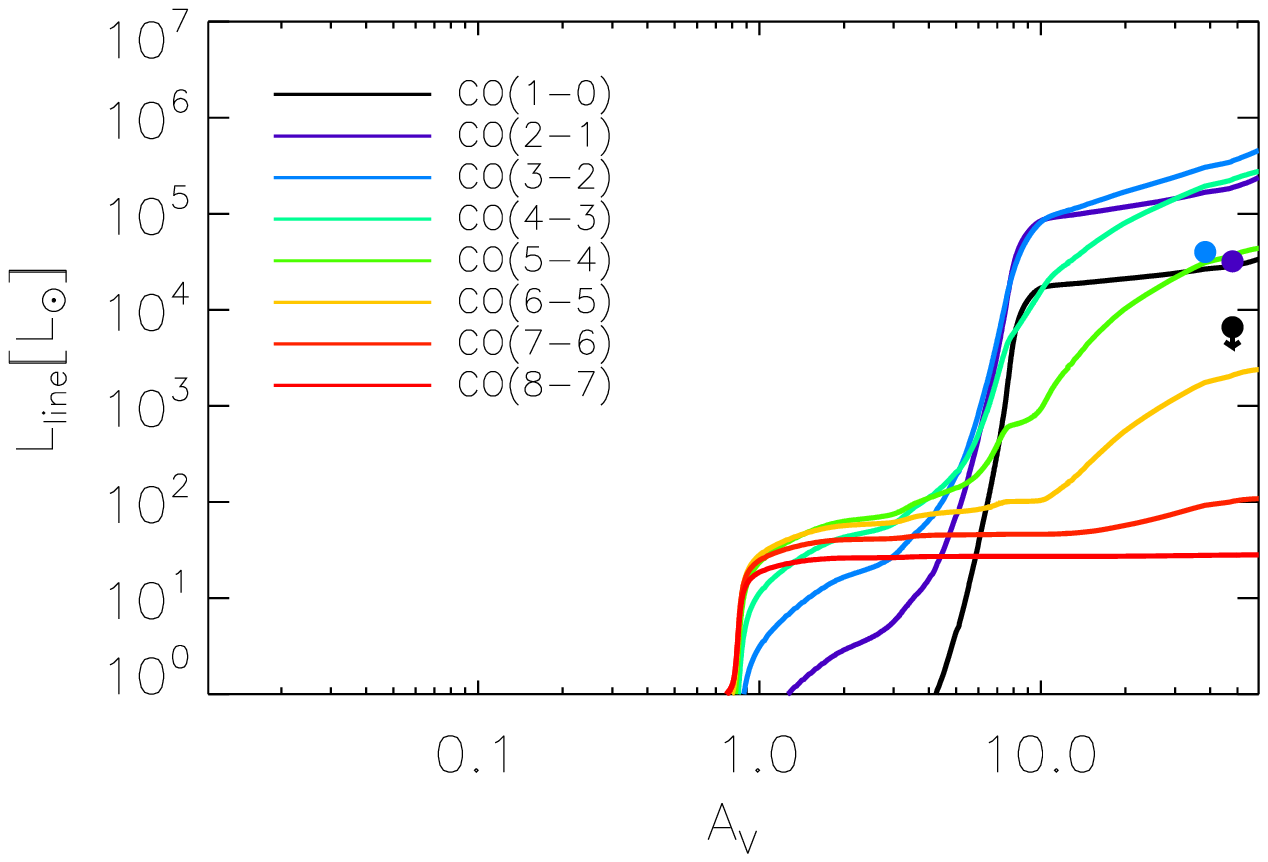}
\caption{
Cloudy model predictions for the CO line luminosities 
as a function of A$_V$. The models are stopped at A$_V$ of 30~mag, 
and the covering factor is set to $\sim$5\%. 
Observations of the CO(1-0) (upper limit), 
CO(2-1), and CO(3-2) lines 
are represented by filled circles on the right hand side of the plots. 
{\it Top:} 
No turbulent velocity included. 
{\it Bottom:} 
Including a turbulent velocity $u_{turb}$ of 30~km~s$^{-1}$. 
}
\label{fig:cocloudy}
\end{figure}

We find that the CO emission is essentially produced at 3$<$A$_V$$<$6~mag, 
where CO dominates the cooling of the gas. For larger A$_V$, the 
gas temperature stabilizes around 30~K, and denser gas 
tracers such as HNC, CN, and CS take over the cooling. 
Densities in the molecular cloud reach 10$^{6.3}$~cm$^{-3}$. 
For A$_V > 10$~mag, $\sim$30\% of the carbon is in the form of CO 
and the condensation of CO onto grain surfaces reaches $\sim$10\%. 
The photoelectric effect on grains is mostly responsible for the heating 
of the gas at the surface of the cloud, and grain collisions are the main 
heating source deeper into the cloud. 
Figure~\ref{fig:cocloudy} ({\it top}) shows the Cloudy model predictions for the 
CO line luminosities (from CO(1-0) to CO(8-7)) as a function 
of A$_V$ (models are stopped at A$_V = $30~mag). 
The observed CO(2-1) and CO(3-2) luminosities 
are matched by this model within a factor of 2, and the CO(1-0) model 
prediction falls below the observed upper limit. 
The predicted CO(1-0) intensity is only 3 times lower than 
our observed upper limit, indicating that higher-performance telescopes 
(such as ALMA) should be able to detect this transition. 
The CO(3-2)/CO(2-1) ratio is over-predicted by 
a factor of 3 compared to the observed uniform-filling ratio. 

The modeled total mass depends strongly on the assumed geometry 
and on the SED fitting, which is has an additional complication due to 
an excess of submm emission \citep{galametz-2011}. 
We select two values of maximum A$_V$, one at which the shielding is sufficient 
for CO to completely form and the SED is well fit (within 30\%) up to 350~$\mu$m 
(A$_{V, max1}$=15~mag), and one at which the SED is fit up to 500~$\mu$m 
(A$_{V, max2}$=30~mag). 
The molecular gas masses corresponding to those A$_V$ are 
M(H$_{\rm 2}$)=7.2$\times$10$^8$~M$_{\odot}$ and 
M(H$_{\rm 2}$)=1.6$\times$10$^9$~M$_{\odot}$ respectively. 
We adopt the latter value in the remainder of this paper, but remind that 
this mass estimate is uncertain by a factor of $\sim$2. 
This is the mass in the molecular phase associated with CO emission. 
The molecular gas mass in the CO-dark PDR layer is then $\sim$10\% that 
of the CO-emitting core (section~\ref{sect:pdr}), hence the warm-to-cold molecular gas mass 
in Haro\,11 is very close to the values found in \cite{roussel-2007}. 
The total molecular mass in Haro\,11 (CO-dark gas+CO core) is 
$\rm{1.7 \times 10^9\,M_{\odot}}$, close to the value derived from $X_{\rm CO, Z}$. 
The resulting $\alpha_{\rm CO, model}$ is $\sim50\,{\rm M_{\odot}\,(K\,km\,s^{-1}\,pc^2)^{-1}}$ 
and is about 10 times $\alpha_{\rm CO, gal} = 4.3\,{\rm M_{\odot}\,(K\,km\,s^{-1}\,pc^2)^{-1}}$. 
The total (H~{\sc ii}+H~{\sc i}+H$_{\rm 2}$, including He) model gas mass is 
$\rm{3.3 \times 10^9~M_{\odot}}$, and the total gas mass to stellar mass 
($\rm{10^{10}~M_{\odot}}$) ratio is $\sim$35\%. 
Despite active photodissociation and low CO luminosities, the overall 
molecular gas content of Haro\,11 is not particularly low. The mass budget of Haro\,11 
is dominated by its molecular gas, followed by the ionized and neutral gas phases.

Note that modeling the molecular cloud to A$_V$ of 30~mag has implications 
on the model predictions for the lines originating from other media given 
the geometry adopted for the model. 
Even though the ionic lines are little affected, the PDR H$_{\rm 2}$ lines 
can be subject to dust extinction and the [O~{\sc i}] lines become optically thick, 
their emission being reduced by a factor of two at most. 
If true, than we would need an additional component in the model, another 
PDR phase or other excitation mechanisms, such as shocks 
\citep[e.g.][]{hollenbach-1989} to reproduce the observations. 
Spatially resolved observations of, e.g., [C~{\sc i}] for the PDR, as well as optically 
thin molecular tracers, are needed to confirm this PDR/molecular structure.

%%%

%%%
\subsection{Influence of micro-turbulence}
\label{sect:vturb}
%%%
We discuss the effects of adding a turbulent velocity 
in the Cloudy models, in particular to the predicted CO line ratios, 
which are sensitive to the temperature in the high density regime. 
Turbulence may be an important factor in setting the ISM conditions 
in Haro\,11 since it has been found dynamically unrelaxed 
\citep{ostlin-2001,james-2013}. 
The microphysical turbulence is linked to the observed CO FWHM 
($\sim$60\,km~s$^{-1}$) via $u_{obs} = \sqrt{u_{thermal} ^2 + u_{turb} ^2} = FWHM/\sqrt{4ln(2)}$. 
Turbulence is included in the pressure balance, with the term 
$P_{turb} \propto n_{H} u_{turb} ^2$, hence dissipating 
some energy away from the cloud. 

\begin{figure}
\centering
\includegraphics[clip,width=8.8cm]{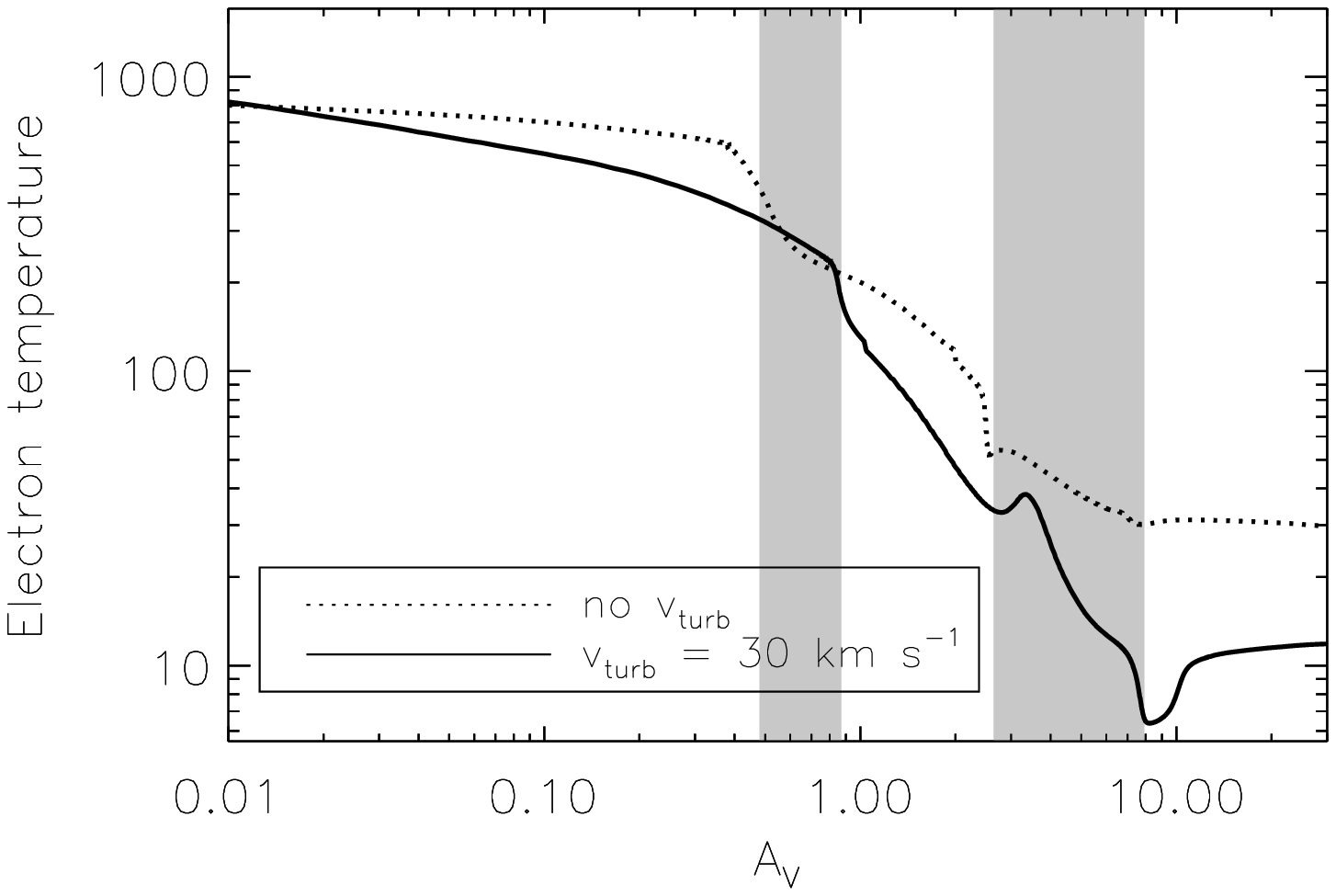}
\includegraphics[clip,width=8.4cm]{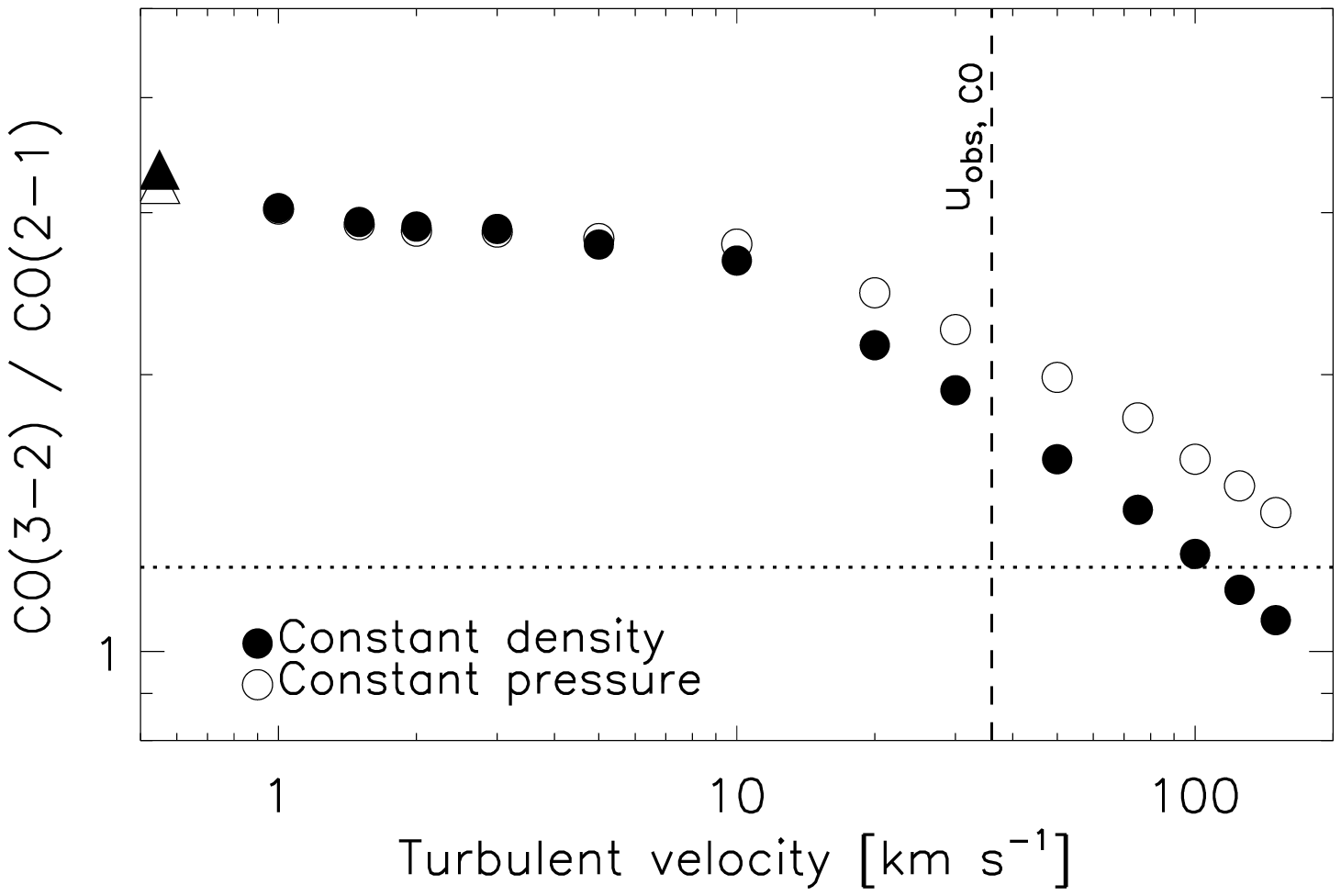}
\caption{
{\it Top:} 
Temperature profile in the cloud. 
The dotted profile is without turbulence, and the plain profile 
with a turbulent velocity of 30~km~s$^{-1}$. 
The first vertical grey band indicates the transition into 
the molecular cloud (H$_{\rm 2} > $H~{\sc i}), and the second 
grey band when CO starts to form. 
{\it Bottom:} 
Cloudy model predictions for the CO(3-2)/CO(2-1) 
ratio as a function of turbulent velocity. The two cases of constant 
density (open symbols) and constant pressure (filled symbols) are considered. 
The horizontal dotted line indicates the observed CO ratio 
and the vertical dashed line the observed CO velocity $u_{obs} = FWHM/\sqrt{4ln(2)}$. 
Triangles correspond to the model without turbulent velocity. 
}
\label{fig:temp-vturb}
\end{figure}

We show the temperature profile in the cloud in Figure~\ref{fig:temp-vturb} 
({\it top} panel). {We have chosen to display the results for a turbulent velocity 
of 30~km~s$^{-1}$ to illustrate clearly the effect of turbulence on the profiles}. 
The gas temperature decreases more quickly with A$_V$ in the molecular cloud, 
from 200~K down to 7~K, and stabilizes around 10~K in the CO-emitting region. 
The density stays around that of the PDR (10$^{5.2}$~cm$^{-3}$) hence 
similar results are found in the case of a constant density. 
Adding turbulence has a profound effect on the CO ladder, where high-J lines 
are emitted in a thinner region close to the cloud surface compared to 
low-J CO lines (Figure~\ref{fig:cocloudy}, {\it bottom}). 
With turbulence, the low-J CO lines start to form at A$_V$ of 5-6~mag. 
The [C~{\sc i}]~370 and 610$\mu$m lines are actually enhanced by an order 
of magnitude between A$_V$ of 2-5~mag, and then the lowest CO transitions 
(lowest $T_{exc}$) are enhanced. As a consequence, ratios of CO(3-2)/CO(2-1) 
decrease with increasing turbulent velocity {(Figure~\ref{fig:temp-vturb}, {\it bottom})}. 
The observed CO(3-2)/CO(2-1) ratio is better 
matched by the models when including a turbulent velocity $> 10$~km~s$^{-1}$.
Observations of [C~{\sc i}] and higher-J ($J>4$) CO lines would help validate 
the importance of turbulence, and investigate possible excitation 
mechanisms in low-metallicity galaxies.

%%%
%

%%%%%%%%%%%%%%%%%%%%%%%%%%%
\section{Star formation and total gas reservoir}
\label{sect:sfr}
%%%%%%%%%%%%%%%%%%%%%%%%%%%
In this section, we analyze the global star formation properties of 
our low-metallicity galaxies and relate their gas reservoir 
to the star formation activity in order to contrast what we find for SFE when 
we use a {\it tracer} of a specific ISM phase as a proxy for the total gas reservoir.  
%

%%%%%
\subsection{[C~{\sc ii}], [O~{\sc i}], and CO}
\begin{figure}
\includegraphics[clip,trim=0 .5cm 0 0,width=8.8cm]{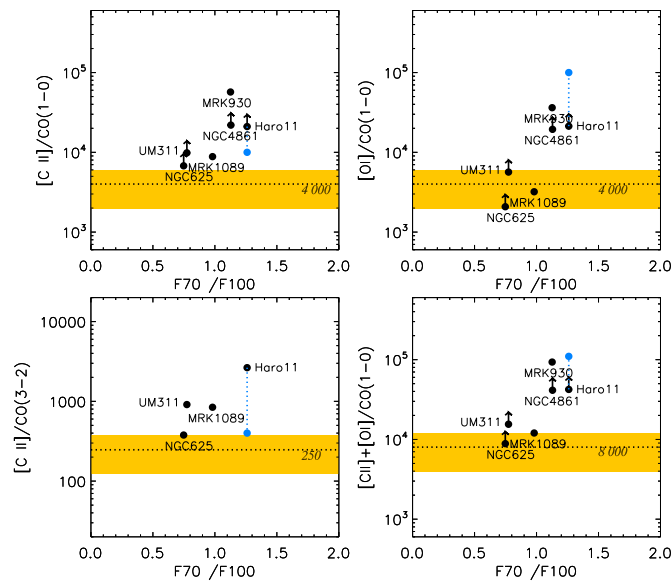}
\caption{
[C~{\sc ii}] and [O~{\sc i}]-to-CO ratios versus 70$\mu$m/100$\mu$m. 
The horizontal orange band indicates the average ratio 
[C~{\sc ii}]/CO(1-0) of 2\,000-6\,000 observed in 
star-forming regions of the Galaxy and galaxies \citep{stacey-1991}. 
For the [C~{\sc ii}]/CO(3-2) horizontal band, 
a ratio $R_{31} = 0.6$ is considered. 
The blue circles correspond to the model predictions for Haro\,11. 
}
\label{fig:cii-co-colors}
\end{figure}

%%%%%
The [C~{\sc ii}]-to-CO(1-0) ratio can be used as a diagnostic of 
{the star formation activity} 
in nearby normal galaxies \citep{stacey-1991} and high-redshift galaxies 
\citep[z$\sim$1-2;][]{stacey-2010}, with average values between 2\,000 and 6\,000 
(for star-forming galaxies). 
High [C~{\sc ii}]/CO ratios are observed in low-metallicity dwarf galaxies 
\citep{madden-2000} because of the different cloud structure, 
with smaller CO cores and larger PDR envelopes, and also because of 
increased photoelectric heating efficiencies in the PDR, caused by dilution 
of the UV radiation field which can enhance the [C~{\sc ii}] cooling \citep{israel-2011}. 
For our sample, we also find high ratios with [C~{\sc ii}]/CO(1-0)$\ge$7\,000 
(Figure~\ref{fig:cii-co-colors}, {\it top left}). 
This indicate that most of the gas is exposed to intense radiation fields 
and that the star-forming regions dominate the global emission of our galaxies.

%%%

C$^0$ has an ionization potential of 11.26~eV, and critical densities 
of 50~cm$^{-3}$ for collisions with electrons and of $\rm{3 \times 10^3~cm^{-3}}$ 
for collisions with hydrogen atoms. Thus, C$^+$ can arise from the surface 
layers of PDRs but also from diffuse ionized gas as well as 
diffuse neutral gas. 
In Figure~\ref{fig:cii-co-colors}, we also show line ratios with [O~{\sc i}], 
which could be a more accurate tracer of the PDR envelope given 
the possible origins of the [C~{\sc ii}] emission, and with CO(3-2), 
for which we have more detections than CO(1-0). 
[O~{\sc i}]/CO(1-0) ratios are globally lower than the [C~{\sc ii}]/CO(1-0) 
ratios because of the high [C~{\sc ii}]/[O~{\sc i}] ratios found in our dwarf galaxies. 
However, the ratios remain high when using [O~{\sc i}]. 
Although there are only 4 data points, the scatter in the [C~{\sc ii}]/CO(3-2) 
values is slightly reduced. 
The color ratio of 70$\mu$m/100$\mu$m is an indicator of the dust temperature 
and thus of the relative contribution of the H~{\sc ii} region and PDR 
to the global emission. Hence one might expect larger line-to-CO ratios 
for higher 70$\mu$m/100$\mu$m ratios. 
We do not see a direct correlation between the observed line-to-CO ratios 
and 70$\mu$m/100$\mu$m color or metallicity, but maybe a slight increase 
of [C~{\sc ii}]/CO with 70$\mu$m/100$\mu$m.

%%%%%
\subsection{Star formation law}
\label{sect:sflaw}
\begin{figure}
\centering
\includegraphics[clip,trim=.2cm .2cm 0 0,width=8.8cm]{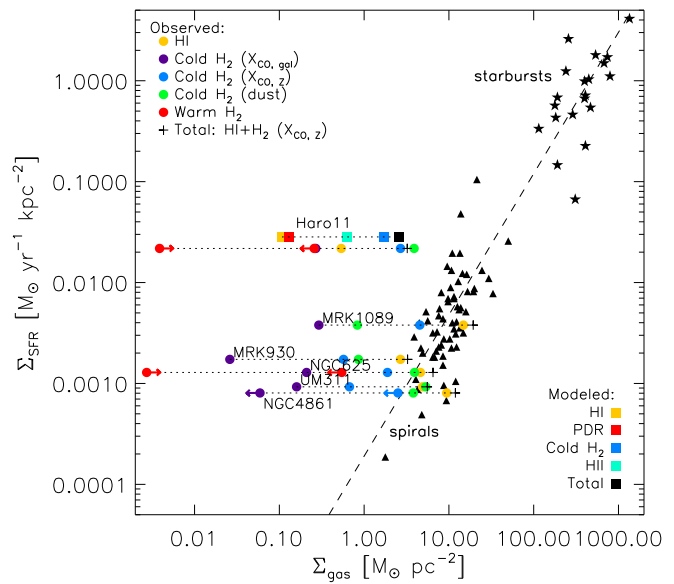}
\caption{
Star formation rate surface density ($\Sigma_{\rm SFR}$) versus 
gas surface density ($\Sigma_{\rm gas}$). 
Our dwarf galaxies are represented by filled circles, and the colors 
show the contribution from the separate observed phases. 
Cloudy models for Haro\,11 correspond to the filled rectangles 
(section~\ref{sect:mol}). 
The starbursts and spirals from \cite{kennicutt-1998} 
are represented by stars and triangles respectively. 
The diagonal dotted line indicates a fit to the Schmidt-Kennicutt law, 
with power index $1.4$ to starbursts and spirals \citep{kennicutt-1998}. 
}
\label{fig:ksobs}
\end{figure}

\begin{figure}
\centering
\includegraphics[clip,trim=0 .2cm 0 0,width=8.8cm]{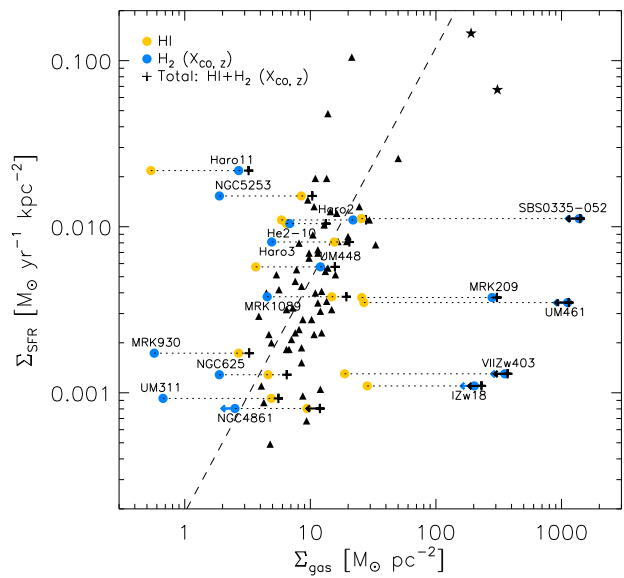}
\caption{
Same as Figure~\ref{fig:ksobs} for all the compact dwarf galaxies 
of the DGS with CO observations. 
}
\label{fig:ksobsall}
\end{figure}

%%%%%
%
We investigate how the different gas reservoirs of H~{\sc i} and H$_{\rm 2}$ 
are related to the star formation activity measured in our galaxies, and 
compare them to the dataset of starburst and spiral galaxies from \cite{kennicutt-1998}, 
{as well as the other compact dwarf galaxies from the DGS with 
CO(1-0) measurements}. 
We look at the SFR surface density ($\Sigma_{\rm SFR}$) and gas surface 
density $\Sigma_{\rm gas} = \Sigma_{HI} + \Sigma_{H_{\rm 2}}$. 
{
For $\Sigma_{\rm SFR}$, we use the $\rm{L_{TIR}}$-derived SFRs from 
the formula of \cite{murphy-2011}, divided by the source size, which is taken 
as the dust aperture from \cite{remy-2013a}. 
For the hydrogen masses, we consider the H~{\sc i} within the dust apertures 
from \cite{remy-2013c}, and the H${\rm _2}$ derived from total CO intensities 
with $X_{\rm CO, gal}$, $X_{\rm CO, Z}$, and from dust measurements 
(Table~\ref{table:mass}) for our sample (Fig.~\ref{fig:ksobs}), and with 
$X_{\rm CO, Z}$ and CO intensities from the literature 
for the DGS galaxies (Fig.~\ref{fig:ksobsall}). Those galaxies have 
metallicities between $1/40$ and 1~Z$_{\odot}$ \citep{madden-2013}. 
We summarize those parameters in Table~\ref{table:ksparams}.} 
The hydrogen masses are multiplied by a factor $1.36$ on the figures to include helium. 
%

%%%

\begin{table}[t!]\tiny
  \caption{Parameters for the star formation law.}
  \hfill{}
  \begin{tabular}{l c c c c c}
    \hline\hline
    \vspace{-8pt}\\
    \multicolumn{1}{l}{Galaxy} & 
    \multicolumn{1}{c}{Metal.} &
    \multicolumn{1}{c}{M(H~{\sc i})} &
    \multicolumn{1}{c}{M(H$_{\rm 2}$)} &
    \multicolumn{1}{c}{SFR} &
    \multicolumn{1}{c}{Area} \\
     \multicolumn{1}{l}{} &
    \multicolumn{1}{c}{[O/H]} &
    \multicolumn{1}{c}{[log M$_{\odot}$]} &
    \multicolumn{1}{c}{[log M$_{\odot}$]} &
    \multicolumn{1}{c}{[M$_{\odot}$~yr$^{-1}$]} &
    \multicolumn{1}{c}{[kpc$^2$]} \\
    \multicolumn{1}{l}{(1)} &
    \multicolumn{1}{c}{(2)} &
    \multicolumn{1}{c}{(3)} &
    \multicolumn{1}{c}{(4)} &
    \multicolumn{1}{c}{(5)} &
    \multicolumn{1}{c}{(6)} \\
    \hline
    \vspace{-8pt}\\
	Haro\,11	 	& 8.20	& 8.70	& 9.40 	& 27.6 	& 1266 \\ %45 \\ 
	Mrk\,1089 	& 8.10	& 10.2	&  9.65	& 5.11 	& 1350 \\ %75 \\ 
	Mrk\,930	 	& 8.03 	& 9.50	& 8.83	& 2.80 	& 1617 \\ %60 \\ 
	NGC\,4861 	& 7.89 	& 8.62	& $<$8.04	& 0.048 	& 60 \\ %120 \\ 
	NGC\,625 	& 8.22	& 8.04	&  7.65	& 0.042 	& 32 \\ %170 \\ c
	UM\,311	 	& 8.38	& 9.48	&  8.61	& 0.77 	& 833 \\ %140 \\ c
    \hline
    \vspace{-8pt}\\
    \multicolumn{6}{c}{Additional DGS compact galaxies} \\
    \hline
    \vspace{-8pt}\\
	Haro\,2	 	& 8.23	& 8.58	& 9.04	& 0.95	& 87 \\ %50 \\ 
	Haro\,3	 	& 8.28	& 9.05	& 8.55	& 0.80	& 99 \\ %60 \\ 
	He\,2-10	 	& 8.43	& 8.49	& 8.52	& 0.68	& 65 \\ %108 \\ 
	I\,Zw\,18	 	& 7.14	& 8.00	& $<$8.85	& 0.005	& 4.8 \\ %10 \\ 
	Mrk\,209		& 7.74	& 7.43	& 8.47	& 0.005	& 1.4 \\ %24 \\ 
	NGC\,5253	& 8.25	& 8.02	& 7.37	& 0.26	& 17 \\ %120 \\ 
	SBS\,0335-052	& 7.25	& 8.64	& $<$10.4	& 0.26	& 23 \\ %14 \\ 
	UM\,448	 	& 8.32	& 9.78	& 10.3	& 13.4	& 2332 \\ %64 \\ 
	UM\,461	 	& 7.73	& 7.86	& $<$9.49	& 0.014	& 3.7 \\ %17 \\ 
	VII\,Zw\,403	& 7.66	& 7.52	& $<$8.79	& 0.003	& 2.4 \\ %40 \\ 
    \hline \hline
    \vspace{-8pt}\\
  \end{tabular}
  \hfill{}
  \newline
  The columns are as follow: (1)~Galaxy name; (2)~Metallicities from \cite{madden-2013}; 
  (3)~Aperture-corrected H~{\sc i} masses assuming an exponential distribution, from \cite{remy-2013c}; 
  (4)~H$_{\rm 2}$ masses using $X_{\rm CO, Z}$. 
  CO data are from: 
\cite{sage-1992} for Haro\,2, UM\,448, and UM\,461; 
\cite{tacconi-1987} for Haro\,3 and Mrk\,209; 
\cite{taylor-1998} for NGC\,5253; \cite{kobulnicky-1995} for He\,2-10; 
\cite{leroy-2005} for I\,Zw\,18 and VII\,Zw\,403; 
and \cite{dale-2001} for SBS0335-052.
  (5)~Star formation rate using L$_{\rm TIR}$ from \cite{remy-2013b} and the formula from \cite{murphy-2011}; 
  (6)~Area defined from the {\it Herschel} photometry apertures \citep{remy-2013a} and 
  using distances from \cite{madden-2013}. 
  \label{table:ksparams}
\end{table}

In Figure~\ref{fig:ksobs} and \ref{fig:ksobsall}, we see that most low-metallicity 
galaxies fall close to the Schmidt-Kennicutt relation. The less active galaxies are 
generally dominated by their H~{\sc i} gas while some of the most active galaxies, 
such as Haro\,11, are dominated by their H$_{\rm 2}$ gas. Several galaxies stand further 
to the right of the relation (I\,Zw\,18, Mrk\,209, NGC\,4861, UM\,461, VII\,Zw\,403) and hence appear 
either more gas-rich, with a part of their H~{\sc i} gas probably less connected 
to the star formation activity, or more quiescent than expected, where the TIR 
luminosity might under-estimate slightly the star formation rate \citep{kennicutt-2012}. 
Mrk\,209 shows a strikingly high molecular surface density, which would be an indication 
that it is on the verge to form stars. However, it is known to already host young super 
star clusters and WR features \citep[e.g.][]{thuan-2005b}. Moreover, 
its warm molecular component inferred from {\it Spitzer} data is only 
$\sim 6 \times 10^3$~M$_{\odot}$ \citep{hunt-2010}, which is orders of magnitude 
lower than the estimated cold molecular gas. These indications combined probably 
mean that we are over-estimating $\Sigma_{H_{\rm 2}}$ with the high conversion factor 
$X_{\rm CO, Z}$ used (and because of uncertainties on the metallicity). 
Two galaxies (Haro\,11 and, more moderately, NGC\,5253), which have the highest $\Sigma_{\rm SFR}$, 
lie to the left of the standard Schmidt-Kennicutt relation, even when 
considering high $X_{\rm CO}$ values. Both are active blue compact galaxies. 
The presence of young super star clusters in NGC\,5253 and clear lack of CO 
suggest that the starburst was triggered by accretion of gas from outside the 
galaxy and formed with high efficiency \citep{meier-2002}. 
Haro\,11 is also known to have undergone a merger event \citep{ostlin-1999} 
and behaves like starbursting mergers \citep{daddi-2010}. 
We note that the position of our low-metallicity galaxies coincide with the position of spiral 
galaxies on Figure~\ref{fig:ksobsall}. We have used apertures that are large enough 
to encompass the total FIR continuum emission and are $\ge$1.5$\times$r$_{25}$ 
\citep{remy-2013a}. If we were to zoom on the star-forming regions of our galaxies, 
those would generally move closer towards the position of the starburst galaxies.

%%%%%
The gas depletion time, i.e. the time needed 
to consume the existing gas reservoir given the SFR, is expressed 
as the inverse of the star formation efficiency: 
$\tau_{dep} = \Sigma_{\rm gas}/\Sigma_{\rm SFR}$\,[yr]. 
In nearby, predominantly late-type disk galaxies, the average molecular 
depletion time found by \cite{bigiel-2008} is around 2~Gyr. 
For our sample, taking the cold H${\rm _2}$ ($X_{\rm CO, Z}$) mass values 
from Table~\ref{table:mass} and including helium, we get molecular depletion times 
of $\sim$0.1~Gyr (for Haro\,11) - 0.3~Gyr (for Mrk\,930) up to 3~Gyr (for NGC\,4861), 
{and a median value of 0.7~Gyr for the compact DGS sample}. 
\cite{sage-1992} find similarly low values ($\sim$0.1~Gyr) for their sample of BCDs, 
although they use relatively lower conversion factors, of 2-3 times $X_{\rm CO, gal}$. 
Therefore, even when using a metallicity-scaled $X_{\rm CO}$ factor, 
we find that our systems have short molecular gas depletion times, typical 
of interacting systems, starbursts, and ULIRGs. 
If we are accounting for all the gas related to the star formation, such short depletion 
times indicate that our dwarf galaxies are efficient in forming stars. {This 
can be understood by the fact that several super star clusters are observed 
in those galaxies, and BCGs do have higher cluster formation efficiencies than 
normal galaxies \citep{adamo-2011b}.} 
If these galaxies keep forming stars at the current rate, they would consume 
all their gas quickly. Therefore they might be caught in a bursty episode 
(as they are selected bright in the IR). 
When adding the H~{\sc i} data, the average total gas depletion time over the sample 
is $\tau_{dep}$$\sim$5~Gyr. This value is significantly higher than the molecular 
depletion times since most galaxies are H~{\sc i}-dominated.  
{This H~{\sc i} gas may also play a role in regulating 
the star formation, as a direct fuel at very low metallicities \citep{krumholz-2012b}, 
or as future fuel since part of the H~{\sc i} we measure may be residing in the 
envelopes of the molecular clouds. Regardless of the role of the H~{\sc i}, the 
picture that emerges is a very rapid and efficient conversion of molecular gas 
into stars in low-metallicity dwarf galaxies.} 
%

%%%%%%%%%%%%%
\section{Conclusions}
%%%%%%%%%%%%%
The state and abundance of the molecular gas, observed through the molecule CO, 
is not well known in dwarf galaxies. To address this issue, we have obtained new 
observations of the CO(1-0), CO(2-1), and CO(3-2) lines in a subsample of 
6 low-metallicity galaxies of the {\it Herschel} Dwarf Galaxy Survey (Haro\,11, Mrk\,930, 
Mrk\,1089, NGC\,4861, NGC\,625, and UM\,311), which we combine 
with existing broad band and atomic fine-structure line observations. 
We summarize our results as follow:

\begin{itemize}

\item
Out of the 6 targets, 5 are detected in at least one CO transitions, 
but NGC\,4861 remains undetected in CO. Three of the targets, 
Haro\,11, Mrk\,930 (marginal detection), and UM\,311, {with 
metallicities Z$\sim$1/3, 1/5, and 1/2~Z$_{\odot}$ respectively}, were never 
detected in CO before, and no CO data are reported in the literature 
for NGC\,625. 

\item
CO is found to be faint in the 6 dwarf galaxies observed, 
especially relative to the [C~{\sc ii}]~157$\mu$m line, as we derive 
[C~{\sc ii}]/CO(1-0) ratios $\ge$7\,000. In particular, this hints at 
low filling factor of the CO-emitting material {and high 
photoelectric [C~{\sc ii}] heating efficiency}, as has been 
previously suggested \citep{israel-1996,madden-2000}. 

\item
Molecular gas mass estimates are quite uncertain and vary significantly 
depending on the method used ($X_{\rm CO}$, dust measurements, etc.). 
A Galactic conversion factor predicts unrealistically low masses while 
the dust method, a metallicity-scaled $X_{\rm CO}$, or radiative transfer 
modeling predict larger molecular masses with accuracy of a few. 
Even when considering high $X_{\rm CO}$ conversion factors for our sample of 
low-metallicity galaxies, the molecular content {make a small contribution 
to the total ISM mass budget compared to} the H~{\sc i} gas (except for Haro\,11). 

\item
We have compared the star formation activity to the gas reservoir for 
a sample extended to the compact low-metallicity galaxies of the DGS. 
Despite their low molecular content, all galaxies are in relative agreement 
with the (total gas) Schmidt-Kennicutt relation. 
{The BCGs NGC\,5253 and Haro\,11, which has a high 
molecular-to-atomic gas fraction, are exceptions and may be intrinsically 
more efficient at forming stars.} 
Even with a metallicity-scaled $X_{\rm CO}$ factor, the molecular gas 
depletion times of our low-metallicity galaxies are short. 
Our results imply either that: 
1)~the molecular gas forms stars with a different efficiency in the dwarf 
galaxies {of our sample} than in normal spirals or 
2)~if a universal {\it total} (as opposed to molecular) gas depletion time 
applies, then the atomic and possibly the ionized gas are 
participating in star formation in {those} galaxies.

\end{itemize}

What the role of the different ISM gas phases is, and whether star formation 
efficiencies are truly higher in some dwarf galaxies, is still unclear. 
Our interpretations are limited by the different antenna beam sizes, 
and the lack of information on the source size and distribution. 
In particular, we cannot conclude whether the cold ISM in the dwarf galaxies 
that we have studied is a result 
of enhanced photodissociation of molecular clouds having a mass 
distribution similar to clouds in normal metallicity galaxies, or 
points to a fundamentally altered structure of the low-metallicity ISM with 
molecular clouds having a different mass distribution compared to H$_{\rm 2}$ 
clouds in normal metallicity galaxies. 
The analysis presented here could be taken to the next level with 
interferometry observations with the SMA, ALMA, or the JVLA to map 
the distribution and properties of the gas reservoirs on smaller scales.

\begin{acknowledgements}
This research was supported by Sonderforschungsbereich SFB 881 
"The Milky Way System" of the German Research Foundation (DFG), 
and by the Agence Nationale de la Recherche (ANR) through the programme 
SYMPATICO (Program Blanc Projet ANR-11-BS56-0023). 
V. L. is supported by a CEA/Marie Curie Eurotalents fellowship. 
This research has made use of the NASA/IPAC Extragalactic Database (NED) 
which is operated by the Jet Propulsion Laboratory, California Institute of Technology, 
under contract with the National Aeronautics and Space Administration. 
This work is based in part on observations made with the Spitzer Space Telescope, 
which is operated by the Jet Propulsion Laboratory, California Institute of Technology under a contract with NASA.
  The Mopra radio telescope is part of the Australia Telescope National Facility 
  which is funded by the Commonwealth of Australia for operation as a 
  National Facility managed by CSIRO. 
   The University of New South Wales Digital Filter Bank used 
   for the observations with the Mopra Telescope was provided 
   with support from the Australian Research Council.
We thank the IRAM-30m staff for their help with the observations. 
This publication is based on data acquired with the Atacama Pathfinder Experiment (APEX). 
APEX is a collaboration between the Max-Planck-Institut fur Radioastronomie, 
the European Southern Observatory, and the Onsala Space Observatory.

\end{acknowledgements}

\bibliographystyle{aa}
\bibliography{../../BIB/references}

\end{document}